%% file: main.tex
\newif\ifanon
\anonfalse

\documentclass[11pt]{article}
\pdfoutput=1

\usepackage[T1]{fontenc}
\usepackage[utf8]{inputenc}
\usepackage[american]{babel}
\usepackage{multicol}
\usepackage{amsfonts}
\usepackage{amsmath}
\usepackage{amssymb}
\usepackage{amthm}
\usepackage{complexity} 
\usepackage{hyphenat} 

\usepackage[letterpaper,margin=1in]{geometry}

\makeatletter
\let\c@author\relax
\makeatother
\usepackage{ambib}
\let\H\relax

\usepackage{ammacros}

\DeclareMathOperator{\id}{id}

\makeatletter
\newlang{\@col}{col}
\newcommand*{\kcol}{k{\mhyph}\@col}
\newcommand*{\twocol}{2{\mhyph}\@col}
\makeatother

\newcommand*{\alvgd}{\mathcal{V}(\mathcal{D},n)}

\newcommand*{\aviewsd}{\mathsf{Aviews}(\mathcal{D},n)}
\newcommand*{\aviews}{\mathsf{Aviews}(\mathcal{D})}
\newcommand*{\bad}{\mathrm{bad}}
\newcommand*{\base}{\mathrm{base}}
\newcommand*{\raviewsd}{\mathsf{AuthenticAviews}(\mathcal{D},n)}
\newcommand*{\cent}{\mathsf{center}}
\newcommand*{\ids}{\mathsf{Id}}
\newcommand*{\kcolg}{\mathcal{G}(\kcol)}

\newcommand*{\ports}{\mathsf{prt}}

\newcommand*{\twocolg}{\mathcal{G}(\twocol)}

\newcommand*{\view}{\mathsf{view}}

\newcommand*{\cldiameq}[1]{\mathcal{H}_{\diam = #1}}
\newcommand*{\cldiamle}[1]{\mathcal{H}_{\diam \le #1}}
\newcommand*{\clbipdiameq}[1]{\cldiameq{#1}^\mathrm{bip}}
\newcommand*{\clbipdiamle}[1]{\cldiamle{#1}^\mathrm{bip}}
\newcommand*{\clwise}{\mathcal{H}_\mathrm{w}}

\usepackage{subcaption}
\usepackage{stmaryrd}
\usepackage{tikz}
\usetikzlibrary{arrows.meta,calc,fit,shapes.geometric,positioning}
\tikzset{>=stealth}

\usepackage{standalone}

\usepackage{xcolor}
\definecolor{stgoblue}{RGB}{74,144,226}
\definecolor{stgogreen}{RGB}{80,227,194}
\definecolor{stgored}{RGB}{255,69,0}
\definecolor{stgoorange}{RGB}{255,165,0}

\usepackage{hyperref}
\hypersetup{
  colorlinks=true,
  linkcolor=black,
  citecolor=black,
  filecolor=black,
  urlcolor=black,
}
\newcommand{\email}[1]{\href{mailto:#1}{\texttt{#1}}}

\usepackage{microtype}
\usepackage{csquotes}
\usepackage[capitalize]{cleveref}
\crefname{claim}{Claim}{Claims}

\usepackage[draft]{fixme}
\fxusetheme{color}
\FXRegisterAuthor{am}{anam}{\footnotesize\color{blue}Augusto}
\FXRegisterAuthor{pm}{apm}{\footnotesize\color{orange}Pedro}
\FXRegisterAuthor{mrw}{amrw}{\footnotesize\color{red}Martin}
\FXRegisterAuthor{bj}{abj}{\footnotesize\color{purple}Benja}

\title{Strong and Hiding Distributed Certification of Bipartiteness%
  \ifanon\else\thanks{%
    Parts of this work previously appeared as a brief announcement at PODC 2025
    \cite{modanese25_brief_podc}.
  }\fi 
}
\date{}
\author{%
  \ifanon 
  Anonymous Authors
  \else
  Benjamin Jauregui\thanks{Universidad de Chile, Chile and Université Paris
  Cité, France.
    Email: \email{jauregui@irif.com}}
  \and Augusto Modanese\thanks{CISPA Helmholtz Center for Information Security,
    Germany.
    Email: \email{augusto.modanese@aalto.fi}}
  \and Pedro Montealegre\thanks{Universidad Adolfo Ibáñez, Chile.
    Email: \email{p.montealegre@uai.cl}}
  \and Martín Ríos-Wilson\thanks{Universidad Adolfo Ibáñez, Chile.
    Email: \email{martin.rios@uai.cl}}
  \fi 
}

\begin{document}
    
\maketitle
\thispagestyle{empty}

\begin{abstract}
Distributed certification is a framework in distributed computing where nodes in a network collaboratively verify whether the entire graph satisfies a given property. Within this framework, a \emph{locally checkable proof (LCP)} is a non-deterministic distributed algorithm designed to verify global properties of a graph $G$. It involves two key components: a \emph{prover} and a \emph{distributed verifier}. The prover is an all-powerful computational entity capable of performing any Turing-computable operation instantaneously. Its role is to convince the distributed verifier, composed of the graph's nodes, that $G$ satisfies a particular property $\Pi$.  

An LCP is correct if it satisfies \emph{completeness} and \emph{soundness}. Completeness ensures that, for every graph $G$ satisfying $\Pi$, there exists a certificate assignment that is accepted by all nodes. Soundness guarantees that for every graph $G$ that does not satisfy $\Pi$, at least one node rejects the assignment for any given certificate assignment.  

In this paper, we study the problem of certifying whether a graph is
\emph{bipartite} (i.e. \emph{$2$-colorable}) with an LCP that is able to \emph{hide} a $2$-coloring
from the verifier.
More precisely, we say an LCP for $2$-coloring is \emph{hiding} if, in a
yes-instance, it is possible to assign certificates to nodes \emph{without
revealing an explicit $2$-coloring}. 
Motivated by the search for promise-free separations of extensions of the LOCAL
model in the context of locally checkable labeling (LCL) problems, we also
require the LCPs to satisfy what we refer to as the \emph{strong soundness}
property.
This is a strengthening of soundness that requires that, in a no-instance (i.e.,
a non-$2$-colorable graph) and for every certificate assignment, the subset of
accepting nodes must induce a $2$-colorable subgraph.
An LCP satisfying completeness, soundness, hiding, and strong soundness is
called a \emph{strong and hiding LCP}. 

We show that strong and hiding LCPs for $2$-coloring exist in specific graph
classes and requiring only $O(\log n)$-sized certificates.  
Furthermore, when the input is promised to be a cycle or contains a node of
degree $1$, we show the existence of strong and hiding LCPs even in an 
\emph{anonymous} network and with \emph{constant-size} certificates. 

Despite these upper bounds, we prove that there are no strong and hiding LCPs for $2$-coloring in general, unless the algorithm has access to node identifiers and uses certificates of size~$\omega(1)$. Furthermore, in anonymous networks, the lower bound holds regardless of the certificate size.  The proof relies on a Ramsey-type result as well as an  argument about
the realizability of subgraphs of the neighborhood graph consisting of the
accepting views of an LCP.
Along the way, we also give a characterization of the hiding property for the
general $k$-coloring problem that appears to be a key component for future
investigations in this context.

\end{abstract}

\clearpage
\pagenumbering{arabic} 
\thispagestyle{plain}

\section{Introduction}

Distributed certification is a framework in distributed computing where nodes in
a network collaboratively verify that the entire network (graph) satisfies a
given property. Within this framework, a \emph{locally checkable proof (LCP)} is
a non-deterministic distributed algorithm designed to verify global properties
of a graph \( G \). It involves two components: a \emph{prover} and a
\emph{distributed verifier}. The prover is an omnipotent computational entity,
capable of performing any (even uncomputable) operation instantaneously. Its
role is to convince the distributed verifier, which consists of a collection of
nodes within the graph, that \( G \) satisfies a particular property \(
\mathcal{P} \).
The LCP must satisfy the properties of \emph{completeness} and \emph{soundness},
which are defined in a manner akin to the classical proof systems from
computational complexity theory.
(See \cref{sec:defs} for precise definitions.)

In this paper, we focus on the fundamental \emph{$k$-coloring} problem, which
consists of determining whether the input graph $G$ is $k$-colorable. 
This problem admits a simple single-round LCP with certificates consisting of
$O(\log k)$ bits, representing the color of the node in a proper $k$-coloring
of $G$.
We ask whether this is the only genuine strategy possible.
\begin{quote}
  \textbf{Question.}
  Is it possible to certify $k$-coloring \emph{by other means than fully
  revealing} a proper $k$-coloring to the nodes?
\end{quote}
To explore this question, we introduce and explore the notion of \emph{hiding}
locally checkable proofs.
Informally, an LCP for $k$-coloring is said to be hiding if there exist
certificates that convince the verifier on every $k$-colorable graph, but there
is no deterministic local algorithm that can \emph{fully} extract a $k$-coloring
from these certificates.
In particular, if we have any distributed network executing such an algorithm,
then \emph{at least one node} fails to produce a correct color.
We postpone a more in-depth discussion of this new notion to
\cref{sec:intro-model}.
For now, the reader should keep in mind that it is \emph{very different} from
cryptographic concepts as, for example, zero-knowledge proofs; see
\cref{sec:related-work} for a more extensive discussion.

In addition to the hiding property, we also require our LCPs to satisfy a
stronger but still very natural variant of soundness, which we accordingly dub
\emph{strong soundness}.
The main reason for using this version instead of the usual one is best
understood in the context of a connection to \emph{promise-free separations of
extensions of the LOCAL model}, which we discuss next.

\paragraph{Main motivation: application to promise-free separations.}
\emph{Locally checkable labeling} (LCL) problems are central to the study of
the LOCAL model of computation
\cite{naor95_what_siamjc,linial92_locality_siamjc}.
These are problems in which one must label the nodes of a graph according to
a set of constraints that can be locally verified.
By far, the most studied variant of LOCAL is randomized LOCAL
\cite{ghaffari18_derandomizing_focs,balliu21_lower_jacm,balliu20_much_podc,%
brandt16_lower_stoc,chang19_exponential_siamjc,chang19_time_siamjc,%
fischer17_sublogarithmic_disc,rozhon20_polylogarithmic_stoc}, but recently
interest has extended to several other augmentations of the LOCAL model,
including access to shared randomness \cite{balliu25_shared_icalp}, quantum
computing
\cite{legall19_quantum_stacs,coiteux-roy24_no_stoc,balliu25_distributed_stoc},
and global symmetry breaking
\cite{akbari23_locality_icalp,chang24_tight_podc,ghaffari17_complexity_stoc,%
ghaffari18_derandomizing_focs}.
See also \cite{akbari25_online_stoc} for a broader discussion of this landscape.

Due to this wide range of extended models, a basic question is whether we can
prove separation results between them.
Here it cannot be stressed enough that we are looking for separations that hold
in a \emph{promise-free} setting, that is, without additional assumptions on the
input graph.
The reason for this is that the bulk of the theory on LCL problems is
promise-free, and assumptions on the input graph may have severe effects on the
complexity landscape.
For instance, in trees some complexity classes turn out to collapse across a
wide range of models \cite{dhar24_local_opodis}.
Nevertheless, proving promise-free separations is usually quite an ordeal and
only \emph{ad hoc} constructions are known \cite{balliu25_shared_icalp}.

Let us focus on the case of separating the SLOCAL and online LOCAL models, which
we mention here explicitly for concreteness.
To appreciate our paper and its contribution, it is not necessary to be
acquainted with these two models---other than to know that they are extensions
of LOCAL---, and hence we skip their definitions. 
The interested reader is referred to, e.g., \cite{akbari23_locality_icalp}.

There is an exponential separation between these SLOCAL and online LOCAL for the
problem of $3$-coloring---but only when restricted to bipartite graphs.
More precisely, there is an $O(\log n)$ algorithm for $3$-coloring bipartite
graphs in online LOCAL, whereas in SLOCAL we have a polynomial lower bound for
$3$-coloring that holds even in grids \cite{akbari23_locality_icalp}.

Can this separation be made promise-free?
One simple approach would be to encode in the LCL input a certificate of
$2$-colorability while asking for a $3$-coloring (i.e., the nodes must label
themselves with a proper $3$-coloring).
Since we must allow arbitrary input graphs, the certificate cannot be valid in
all of them, so we require the nodes to output a correct $3$-coloring \emph{only
on the parts of the graph where the $2$-coloring certificate is valid}, which is
a standard approach in the context of LCLs (see also, e.g., the aforementioned
paper \cite{balliu25_shared_icalp}).
In addition, for it to be useful, the certificate of $2$-colorability must be
\emph{hiding} in the sense that every SLOCAL algorithm with $O(\log n)$
locality fails to extract a $3$-coloring from it.

Let us refer to this LCL problem, that is, $3$-coloring under the presence of a
certificate of $2$-colorability as $\Pi$.
Assuming such a hiding certificate exists, we immediately get a
\emph{promise-free separation} between SLOCAL and online LOCAL:
SLOCAL fails to solve $\Pi$ with $O(\log n)$ locality since it cannot extract
a $3$-coloring from the certificate; meanwhile online LOCAL can simply execute
the existing $O(\log n)$ coloring algorithm on the parts of the graph where
the certificate is valid, labeling the rest of the nodes arbitrarily.

Alas, the only way of certifying bipartiteness that is known is to reveal a
$2$-coloring, which renders the $3$-coloring problem trivial.
Hence it would be highly desirable to have a certificate for bipartiteness that
\emph{hides} a $3$-coloring.
In addition, this certificate must be resilient against being used to label
graphs that are not bipartite.
That is, for any graph $G$ where someone tries to place the certificate labels
on, the parts of $G$ where the certificate is valid is indeed $2$-colorable, and
hence the online LOCAL algorithm succeeds in $3$-coloring these parts of $G$.
This motivates the stronger variant of soundness that we introduce.
We discuss these two notions in more detail next.

\subsection{The model}
\label{sec:intro-model}

Before we continue, we introduce our model in more detail but still in the form
of a high-level discussion.
The reader is invited to consult \cref{sec:defs} for the formal definitions and
notation when needed.

As mentioned above, the goal of the prover in the LCP model is to convince the
distributed verifier that the underlying graph $G$ satisfies a certain property
$\mathcal{P}$.
To achieve this, the prover assigns a \emph{certificate} \( \ell(v) \) to each
node \( v \). Once these certificates are distributed, each node \( v \)
receives the certificates of its vicinity. Using this local information---its
own certificate, the certificates of its neighbors, and the graph structure
visible to it---each node independently decides whether to accept or reject the
assertion that \( G \) satisfies \( \mathcal{P} \). 

We consider LCPs where the verifier is allowed to run for $r$ rounds, for a
fixed positive integer $r$, which we refer to as $r$-round LCPs.
(For $r=1$, we call these single-round LCPs.)
Hence every node \enquote{sees} the graph structure and all certificates up to
$r$ hops away from it.
Every node is also given a unique identifier that is smaller than $N$, where $N$
scales polynomially with the number of nodes and is known by the nodes of the
graph.
If the LCP operates independently of the values of the identifiers (which is
desirable but may limit its power), then we refer to it as an anonymous LCP.

We say that an LCP is \emph{correct} if it satisfies \emph{completeness} and \emph{soundness}. Roughly, soundness guarantees that if all nodes accept the assigned certificates (i.e., the proof is locally verified as correct at each node), then the underlying structure genuinely satisfies the intended property. In other words, no incorrect configurations are falsely accepted. Completeness, on the other hand, ensures that whenever the structure does satisfy the intended property, there is indeed a way to assign certificates such that local verification at every node passes. Thus, if the structure is valid, it will not be mistakenly rejected. 

\paragraph{Strong soundness.}
In the standard version of soundness, we require that, if $G$ does not satisfy
$\mathcal{P}$, then, for any assignment of certificates that a malicious prover
attempts to distribute, there will be at least one node in $G$ that rejects.
We are interested in strengthening this to the following:
If $G$ does not satisfy $\mathcal{P}$, then, for any assignment of certificates,
\emph{the subgraph induced by accepting nodes of $G$} must satisfy the property
$\mathcal{P}$.
Moreover, similar notions have been studied in the context of local certification. For instance, in \cite{feuilloley22_error_jpdc}, the authors study a variant of the proof labeling scheme model in which the number of rejecting nodes is proportional to the Hamming distance between the current state and the set of accepting states. Enforcing this condition guarantees that instances that are far from satisfying the desired property are detected by many nodes in the network.

It is also worth noting that this distinction between the two notions of soundness is intrinsic to distributed proof systems; in particular, there is no analogous distinction in classical proof systems from computational complexity theory.

This strengthening is non-trivial even for hereditary graph classes for which compact proof-labeling schemes are already known. For instance, the proof-labeling scheme of \cite{feuilloley2021compact} for planarity, as well as the ones for various $H$-minor-free classes \cite{bousquet2024local}, are designed to satisfy the standard notion of soundness only: on a non-planar (or non-$H$-minor-free) graph, every certificate assignment forces at least one node to reject, but there is no guarantee that the subgraph induced by the accepting nodes remains planar or $H$-minor-free under arbitrary, possibly corrupted, certificates. A similar phenomenon appears in the recent certifications for geometric graph classes \cite{jauregui2025compact} such as interval, chordal or permutation graphs, where the constructions again only ensure that some node rejects whenever the whole graph violates the property. Hence strong soundness does not come for free, even for hereditary properties that already admit compact PLSs.

\paragraph{Hiding LCPs.}
Recall the central novel notion of our interest is that of a \emph{hiding} LCP.
Focusing on the case of coloring, an $r$-round LCP for $k$-coloring is hiding on
a particular $k$-colorable graph $G$ if there is a certificate $\ell$ we can
place on the nodes of $G$ that is correct (i.e., it is accepted unanimously by
the verifier), but there is no $r$-round local algorithm that, given $\ell$, can
extract a $k$-coloring of $G$.
Note that, in order for an algorithm to fail in extracting a $k$-coloring from
$\ell$, it suffices if a \emph{single} node fails to output its color.
Hence this is a purely algorithmic notion and is far from having cryptographical
applications---after all, the model considered in this paper uses a
deterministic verifier.
For further discussion, please consult the next section
(\cref{sec:related-work}).

The above gives a notion of a hiding LCP on a fixed graph $G$.
We extend this to any graph class $\mathcal{G}$ (containing only $k$-colorable
graphs) by saying an LCP is hiding on $\mathcal{G}$ if, for any local algorithm
$\mathcal{A}$, there is $G \in \mathcal{G}$ and a (valid) certificate $\ell$ of
$k$-colorability on $G$ such that $\mathcal{A}$ fails to extract a $k$-coloring
from $\ell$.
It is important to notice that we cannot hope to have a hiding LCP on every
graph class.
For instance, consider the language of star graphs; that is, we would like to
certify that stars are $2$-colorable but without revealing a $2$-coloring.
Since the nodes know that the input graph is a star, they can always output a
color: 
If they have degree 1, they choose color 1; otherwise, they choose color 0.
This contrasts with classical LCPs, where it is known that 
every Turing-computable graph property $\mathcal{P}$ admits an LCP with
certificates of size $O(n^2)$ (i.e., simply provide the entire adjacency
matrix of the input graph to every vertex, along with their corresponding node
identifiers).

\subsection{Related work}
\label{sec:related-work}

\paragraph{Distributed zero-knowledge.}

The notion of hiding LCPs is reminiscent of zero-knowledge proof systems, which
have also been studied in the distributed setting
\cite{bick22_distributed_soda,grilo25_distributed_arxiv}.
Zero-knowledge is a different notion, mainly motivated by cryptographic applications. Informally, it requires that the verifier learns ``nothing'' about the solution. More formally, one fixes a class of algorithms $\mathcal C$, and a protocol is zero knowledge with respect to the class $\mathcal C$ if whatever the nodes can infer from the interaction with the prover can be obtained by running some algorithm of $\mathcal C$. In contrast, the hiding property is more lenient, requiring only that the solution not be revealed everywhere in the network.

\paragraph{Search vs.~decision.} Still in a complexity-theoretical vein, the
question of whether there is a hiding LCP for a problem $\Pi$ can also be
intrpreted as asking whether the \emph{decision} and \emph{search} versions of
$\Pi$ are equal in complexity in the context of local certification.
That is, when is certifying that a solution to $\Pi$ exists on a graph $G$ as
complex as producing a concrete solution from the same certificates?
Comparing the decision and search versions of a problem is a natural
complexity-theoretic question \cite{goldreich08_computational_book} and has been
studied in some sense in the distributed setting by comparing the complexity of
verifying and producing a solution \cite{sarma12_distributed_siamjc}.
To the best of our knowledge, however, this angle has not yet been considered in
the context of local certification.

\paragraph{Resilient labeling schemes.}
Another related model is the \emph{resilient labeling scheme}, a type of
distributed proof system (specifically, a proof-labeling scheme) that considers
the scenario where the certificates of up to \( f \) nodes are erased. This
model was introduced by \textcite{fischer2022explicit} in the context of
studying trade-offs between certificate sizes and verification rounds. 
(See also the recent paper by \textcite{censor2025near}.) 
The key idea is to design a distributed algorithm in which nodes can reconstruct
their original labels from the remaining ones, even after some certificates have
been erased. 
This can be achieved using error-correcting codes and  techniques based on graph
partitions.
Unlike our model, resilient labeling schemes impose a requirement on
\emph{completeness} rather than \emph{soundness}; specifically, in
\emph{yes-instances}, all nodes must accept the verification, even when a subset
of nodes has lost their certificates. However, no constraints are imposed on 
\emph{soundness} as we propose in this article. 

\paragraph{\boldmath $k$-colorability.}
Certain previous articles are concerned with proving lower bounds for
$k$-coloring, where for standard LCPs we have a trivial upper bound of
$\ceil{\log k}$ bits (i.e., give each node its color in a proper $k$-coloring).
\textcite{goos2016locally} show that any LCP deciding whether the chromatic
number of a graph is strictly greater than $3$ requires certificates of size
$\tilde{\Omega}(n^2)$. 
Meanwhile, \textcite{bousquet24_local_stacs} showed that, in the $d$-round LCP
model, certificates of size $\Omega((\log k)/d)$ are necessary. 
Furthermore, for the anonymous model with single-round verification, the authors
show that $\ceil{\log k}$ bits are necessary, which matches the trivial upper
bound.

In this work, we ask whether any distributed proof that a graph is $k$-colorable \emph{necessarily leaks} a proper $k$-coloring of the graph. This question may be related to the minimum certificate size required to prove that a graph is $k$-colorable, but it is not the same. In fact, it may be the case that achieving strong and hiding LCPs for $k$-coloring requires certificates that are strictly larger than those needed to certify the property without any additional conditions, as we will see in our main results.

\subsection{Main results}

Our results are mainly focused on the most fundamental case, that is, on
certifying $2$-coloring in single-round LCPs.
Some of our results do extend beyond this case; for example, we also prove
impossibility results beyond the single-round case and the framework there is
also applicable to $k$-coloring. 
That being said, the main tenor is to provide a comprehensive atlas of the
landscape for $2$-coloring.
After all, this is precisely what we need in order to address the questions in
our main motivation regarding promise-free separations between extensions of the
LOCAL model.

Since the technical foundations of each result do not have a significant
overlap, we integrate the discussion of the proof strategy with that of the
results themselves.

\subsubsection{Upper bound with constant-size certificates}
Our first result is the cornerstone case where the LCP is anonymous and we wish
to certify a $2$-coloring using only constant-size certificates.
We prove it is possible to certify the $2$-coloring without revealing it
provided we have access to graphs with minimum degree one or even cycles (or 
both).
In the statement below, $\twocol_\mathcal{G}$ refers to the promise version of
the $2$-coloring problem where we allow only input graphs from the class
$\mathcal{G}$.

\begin{restatable}{theorem}{restateThmDegOneOrEvenCycle}
  \label{thm:deg-one-or-even-cycle}
  Let $\mathcal{H} = \mathcal{H}_1 \cup \mathcal{H}_2$, where $\mathcal{H}_1$ is
  the class of graphs $G$ for which $\delta(G) = 1$ and $\mathcal{H}_2$ is the
  class of even cycles.
  Then, for any class of bipartite graphs $\mathcal{G}$ that has non-empty
  intersection with $\mathcal{H}$, there is a strong and hiding single-round
  anonymous LCP for $\twocol_\mathcal{G}$ using constant-size certificates.
\end{restatable}

As already explained in \cref{sec:intro-model}, our requirement for hiding is
that, for any local algorithm $\mathcal{A}$ that attempts to extract a
$2$-coloring from our certificate, there is a graph $G \in \mathcal{G}$ together
with a certificate of $G$ on which $\mathcal{A}$ fails.
This is why in the statement we demand that $\mathcal{H}$ is contained in
$\mathcal{G}$---it is in this restricted \emph{hardcore set} of instances that
we force $\mathcal{A}$ to fail.
If we can achieve this in $\mathcal{H}$, then we can afford to use trivial
certificates (i.e., revealing a $2$-coloring everywhere) on any graph of
$\mathcal{G} \setminus \mathcal{H}$, as this will still fulfill the requirements
for hiding. In fact, it suffices that there exists a single instance in which it is possible to hide the coloring. This observation can be presented in an equivalent way: if there exists a class of graphs $\mathcal{G}$ such that, for some non-empty subclass $\mathcal{H} \subseteq \mathcal{G}$, there exists a protocol $\mathcal{D}$ such that the coloring cannot be extracted in every graph in $\mathcal{H}$, then there exists a hiding protocol $\overline{\mathcal{D}}$ extending $\mathcal{D}$ to $\mathcal{G}$ such that $\overline{\mathcal{D}}$ coincides with $\mathcal{D}$ on $\mathcal{H}$ and is the trivial protocol on the graphs in $\mathcal{G} \setminus \mathcal{H}$. 

Regarding the certification scheme itself, we use two distinct strategies for
graphs coming from the classes $\mathcal{H}_1$ and $\mathcal{H}_2$.
If the input graph has minimum degree one, we can reveal the coloring everywhere except for a degree-one node of our choice and its unique neighbor. Note that the unique neighbor can still infer the correct coloring from its other neighbors.
Conversely, if we are in a cycle, we can use a $2$-\emph{edge}-coloring, which
hides the $2$-coloring \emph{from all the nodes}.

\subsubsection{Impossibility results}
\label{sec:intro-results-impossibility}
Next, we discuss our lower bound results. 
First, we show that \Cref{thm:deg-one-or-even-cycle} characterizes the set of
graphs that admit a hiding and strong single-round anonymous LCP for
$2$-coloring.
In fact, the same impossibility argument extends beyond this class to any
\emph{non-anonymous} LCP that uses constant-size certificates.

\begin{restatable}{theorem}{restateThmDegOneOrEvenCycleLB}\label{thm:deg-one-or-even-cycleLB}
  Let $\mathcal{H} = \mathcal{H}_1 \cup \mathcal{H}_2$ be as in
  \cref{thm:deg-one-or-even-cycle}, that is, $\mathcal{H}$ is the union of is
  the class $\mathcal{H}_1$ of graphs $G$ such that $\delta(G) = 1$ and the
  class $\mathcal{H}_2$ of even cycles. 
  Then the following holds for any class $\mathcal{G}$ of graphs that has empty
  intersection with $\mathcal{H}$:
  \begin{enumerate}
    \item In anonymous networks, there is no hiding single-round LCP for
    $\twocol_\mathcal{G}$ that is strongly sound, regardless of the certificate
    size.
    \item In non-anonymous networks, there is no hiding single-round LCP for
    $\twocol_\mathcal{G}$ that uses constant-size certificates and is strongly
    sound.
  \end{enumerate}
\end{restatable}

Furthermore, we can extend this result to $r$-round LCPs, for each positive
integer $r$, thought at the cost of restricting the graph class.
Namely, for larger values of $r$, our lower bound applies only to graphs
satisfying a specific connectivity property that we dub
\emph{$r$-forgetfulness}.
The general idea is that, in an $r$-forgetful graph, if we have nodes $u$ and
$v$ that share an edge, then there is a path $P = (v_0 = v, \dots, v_r)$
starting at $v$ that allows us to \enquote{forget} nodes that are nearer to $u$
than $v$ in the sense that they no longer appear in our radius-$r$ view.
More precisely, if $w$ is a node in the $r$-neighborhood of $v$ that has
distance $d = \dist(w,u)$ to $u$, then, as we visit the nodes $v_0, v_1, \dots$
of $P$, the distance to $w$ monotonically increases until it exceeds $r$ (namely
at $v_{r-d}$).
There are natural graphs classes that satisfy this property, a couple of
examples being large enough (even) cycles and toroidal grids.
See \cref{fig:forgetful-torus} for an example and \cref{sec:proof-oi-lb} for the
precise definition.
Note that $1$-forgetfulness is equivalent to having minimum degree at least two.

\begin{figure}
  \centering
  \includestandalone{figs/forgetful_torus}
  \caption{Example of the $r$-forgetful property for (bipartite) tori.
    Here we are considering $r = 3$.
    The shaded area indicates the intersection $N^3(u) \cap N^3(v)$ of the
    $3$-neighborhoods of $u$ and $v$.
    The orange nodes (plus $u$ itself) are those in that set that have smaller
    distance to $u$ than to $v$.
 Along the path $(v, v_1, v_2, v_3)$, the distance to every node $w$ among these orange nodes (including $u$) increases monotonically, and by the time we reach $v_3$ we have $\dist(v_3, w) > r$.}
  \label{fig:forgetful-torus}
\end{figure}

Extending \cref{thm:deg-one-or-even-cycleLB}, our main impossibility result is
that it is impossible for $r$-round LCPs to hide a $2$-coloring in $r$-forgetful
graphs except when we are in one of the two cases from
\cref{thm:deg-one-or-even-cycle}.
Again, this holds even if the LCP is allowed to use the values of the
identifiers it sees arbitrarily (i.e., even if it is not anonymous).

\begin{restatable}{theorem}{restateThmFullBipImpossibility}
  \label{thm:full-bip-impossibility}
  Let $r$ be a positive integer, and let $\mathcal{G}$ be a class of
  $r$-forgetful bipartite graphs that has empty intersection with $\mathcal{H}$,
  where $\mathcal{H}$ is as in
  \cref{thm:deg-one-or-even-cycle,thm:deg-one-or-even-cycleLB} (i.e., the union
  of all graphs with minimum degree $1$ and even cycles).
  Then the same conclusion as in \cref{thm:deg-one-or-even-cycleLB} holds for
  $r$-round LCPs, that is:
  \begin{enumerate}
    \item In anonymous networks, there is no hiding $r$-round LCP for
    $\twocol_\mathcal{G}$ that is strongly sound, regardless of the certificate
    size.
    \item In non-anonymous networks, there is no hiding $r$-round LCP for
    $\twocol_\mathcal{G}$ that uses constant-size certificates and is strongly
    sound.
  \end{enumerate}
\end{restatable}

The proof relies on our main technical contribution: a characterization of
hiding LCPs for $k$-coloring based on a graph of \emph{accepting local views},
where nodes represent accepting local views, and two views are adjacent if they
appear in the same yes-instance.
We refer to this graph as the \emph{accepting neighborhood graph} (see
\cref{sec:alvg}). 
Namely, an LCP for $k$-coloring is hiding if and only if the neighborhood graph
itself is \emph{not} $k$-colorable (\cref{lem:hiding-char}).
Besides the lower bound proof, this technical result is useful for proving the
hiding property of our upper bound LCPs as well.
We discuss the details further below.

Incidentally, note this lemma also explains why we touched the point of hiding a
$3$-coloring while certifying a $2$-coloring when discussing the main
motivation, but did not define the notion of an LCP for \enquote{hiding a
$K$-coloring while certifying a $k$-coloring}, where $K > k$.
This is because an LCP that hides a $K$-coloring (i.e., from which we cannot
extract a $K$-coloring) must have a neighborhood graph that is not $K$-colorable
and thus in particular is not $k$-colorable; hence the LCP also hides a
$k$-coloring.
Taken in the contrapositive, the non-existence of a hiding LCP for $k$-coloring
(in the sense that we define, i.e., hiding a $k$-coloring) implies the
non-existence of an LCP for $k$-coloring that hides a $K$-coloring, for
\emph{every} value of $K \ge k$.
If this sounds counter-intuitive, note it is still possible to have, say, an
LCP for $k$-coloring that hides a $k$-coloring while there is no LCP for
$k$-coloring that is able to hide a $K$-coloring, where $K > k$.
Indeed, the latter always holds trivially in the non-anonymous setting, as the identifier assignment can be seen as a proper $K$-coloring with $K$ equal to the number of nodes in the graph.

\paragraph{Proving \cref{thm:full-bip-impossibility}.}
The first step is to reduce the general non-anonymous case to that of
\emph{order-invariant} LCPs.
As the name suggests, these are LCPs where the verifier's behavior is only
sensitive to the \emph{order} and not the actual value of the identifiers in the
neighborhood of a node.
That is, we can replace the identifiers around a node without altering the
node's output as long as we preserve the relative order of identifiers. 
(Clearly, this setting is still stronger than the anonymous one.)
This reduction as shown in \cref{sec:ramsey} relies on constant-size
certificates and is based on Ramsey theory. 
In fact, similar arguments have already been considered in the literature; we
follow a related construction \cite{balliu24_local_podc}, albeit our context is
different.

Hence, having accounted for this reduction, the \enquote{true form} of
\cref{thm:full-bip-impossibility} that we prove in \cref{sec:oi-lb} is the
following:

\begin{restatable}{theorem}{restateThmOIBipImpossibility}
  \label{thm:oi-bip-impossibility}
  Let $r$ be a positive integer, and let $\mathcal{H}$ be the union 
  of all graphs with minimum degree $1$ and even cycles.
  In addition, let $\mathcal{G}$ be a class of $r$-forgetful bipartite graphs
  having empty intersection with $\mathcal{H}$.
  Then there is no \emph{order-invariant} hiding $r$-round LCP for
  $\twocol_\mathcal{G}$ that is strongly sound, \emph{no matter the certificate
  size}. 
\end{restatable}

Let us address the main technical aspects behind the proof of
\cref{thm:oi-bip-impossibility}.
As already mentioned above, the key realization is that the hiding property can
be fully characterized in terms of the \emph{neighborhood graph} of
yes-instances of an LCP.
For an $r$-round LCP $\mathcal{D}$, we denote the respective neighborhood graph
for instances with at most $n$ nodes by $\alvgd$.
The characterization we prove (\cref{lem:hiding-char}) is that, for any positive
integer $k$, $\mathcal{D}$ hides a $k$-coloring if and only if $\alvgd$ is
\emph{not} $k$-colorable.

The proof of the lemma is by showing how we can use a $k$-colorable $\alvgd$ to
extract a coloring for $\mathcal{D}$.
For this idea to go through, it is imperative that the nodes have an upper bound
on the number of nodes in the input graph.
The case where such a bound is not given remains an interesting open question.
See \cref{sec:alvg} for details.

The characterization of \cref{lem:hiding-char} suggests using non-$k$-colorable
subgraphs of $\alvgd$ in order to prove \cref{thm:oi-bip-impossibility}.
That is, assuming that $\mathcal{D}$ hides $2$-colorings, we would like to take
a non-bipartite subgraph $H$ in $\alvgd$ and turn it into a real instance.
Since we are in the special case of $2$-colorability, it suffices if $H$ is an
odd cycle---or, more generally, an odd closed walk---in $\alvgd$.

The main challenge is that, although $\alvgd$ consists of views that can be
found in instances accepted by $\mathcal{D}$, there is no guarantee that such an
arbitrary $H$ corresponds to views of a real instance.
This problem is evident if $H$ is a closed walk but not a cycle, but even then
$H$ may contain views that are inherently incompatible.
For instance, we may have distinct views of $H$ in which a node with the same
identifier has a different number of neighbors.
Indeed, if $H$ can be realized directly, then it is \emph{immediately obvious}
that $\mathcal{D}$ violates strong soundness.

\paragraph{\boldmath Realizing subgraphs of $\alvgd$.}
We introduce and develop a notion of \emph{realizability} of subgraphs of
$\alvgd$ with which we are able to gradually manipulate an odd cycle of $\alvgd$
into a real instance.
To this end it is imperative to assume some properties of valid yes-instances
(e.g., the existence of more than one cycle as well as the property of
$r$-forgetfulness we defined previously) for the proof to go through.
The manipulations needed are fairly technical since we must ensure the resulting
views all agree with each other.
To make the construction more digestible, we give the proof as a series of steps
where one can exactly keep track of where each of the properties assumed in the
theorem statement finds its use.

In retrospect, the question of when subgraphs of the neighborhood graph can be
realized as a proper graph is an interesting blend of graph theory and
distributed computing that may be of independent interest.
In its most general form, what we are asking in this context is the following:
\emph{Given a neighborhood graph $\alvgd$ and assuming certain properties of the
original class of graphs, when can we realize a subgraph $H$ of $\alvgd$ as a
real instance accepted by $\mathcal{D}$?}
From this angle, our proof of \cref{thm:oi-bip-impossibility} may also be seen
as an attempt to answer to this question in the case where $H$ is an odd cycle.
Our proof shows that, although not every odd cycle $H$ can be realized, (under
the assumptions of the theorem) it is possible to manipulate $H$ and obtain a
realizable subgraph of $\alvgd$ with the property that we want (i.e.,
non-bipartiteness).

\subsubsection{From soundness to strong soundness in hereditary languages}

The results just discussed show that, in anonymous networks with constant-size certificates, one cannot in general hope for protocols that are simultaneously hiding and strongly sound. This naturally raises the question of what can be guaranteed if we only require the hiding property together with \emph{ordinary soundness}. We show that, for a very broad class of languages, this distinction essentially disappears: for every \emph{hereditary} distributed language $\mathcal{L}$, the existence of an anonymous single-round LCP with constant-size certificates on graphs of maximum degree~$\Delta$ already implies the existence of such a protocol that is \emph{strongly sound}. Moreover, for languages of network pairs $(G,x)$, if the original protocol is hiding, then the transformation can be carried out while preserving hiding.

Recall that a graph language $\mathcal{L}$ is \emph{hereditary} if, for every graph $G$ and every subset of vertices $U \subseteq V(G)$, whenever $G$ is a yes-instance of $\mathcal{L}$, the induced subgraph $G[U]$ is also a yes-instance of $\mathcal{L}$.

\begin{restatable}{theorem}{restateSoundImpliesStrong}

\label{thm:sound-implies-strong}
Let $\mathcal{L}$ be a hereditary distributed language.
If $\mathcal{L}$ admits an anonymous single-round LCP $\mathcal{D}$ with constant-size certificates on graphs of maximum degree~$\Delta$, then it also admits such a protocol that is strongly sound.
Furthermore, if $\mathcal{L}$ is a language of network pairs $(G,x)$ and, in addition, $\mathcal{D}$ is hiding for $\mathcal{L}$, then $\mathcal{L}$ admits an anonymous single-round LCP with constant-size certificates that is both hiding and strongly sound.
\end{restatable}

To prove \Cref{thm:sound-implies-strong}, we proceed in two conceptual steps.

\paragraph{Step 1: Normalization via realizable accepting views.}
A single-round decoder may accept local views that are \enquote{locally consistent} but cannot occur in any globally accepting yes-instance: if such a view appears, some node must reject somewhere else. Since these views play no meaningful role, we define an accepting view $\mu$ to be \emph{authentic} if it appears in a labeled yes-instance in which \emph{every} node accepts, and we \emph{normalize} the decoder by restricting acceptance to realizable views only. The normalized decoder differs from the original one only by rejecting additional views; hence it preserves completeness and soundness, and it also preserves hiding, because it reveals no more information than the original decoder.

\paragraph{Step 2: An algebraic model for authenticity and a bounded completion argument.}
To make normalization effective and to later \enquote{complete} partially
accepting configurations, we develop an algebraic representation of local views.
Fix the certificate alphabet $\Sigma=\{0,1\}^c$. For each accepting view $\mu$,
we associate an antisymmetric integer \emph{demand vector}
$x(\mu)\in\mathbb{Z}^{\Sigma\times\Sigma}$ encoding, for each ordered pair
$(A,B)$, how many neighbors of certificate $B$ are required by a center of
certificate $A$ (with the antisymmetry ensuring that actual edges contribute
opposite signs at their endpoints). For a multiset $\mathcal{M}$ of views, the
sum $X(\mathcal{M})=\sum_{\mu\in\mathcal{M}} m_\mu x(\mu)$ captures the net
imbalance of cross-certificate requirements. This imbalance admits a concrete
\enquote{stub} interpretation: each unit of demand corresponds to a half-edge of
type $(A,B)$, and matching compatible stubs $(A,B)$ with $(B,A)$ corresponds to
creating an edge between certificates $A$ and $B$.

A key lemma  shows that \emph{any} multiset $\mathcal{M}$ of accepting views can be realized, up to a bounded blow-up factor $k=2^{O(\Delta)}$, by a labeled instance whose remaining unmatched stubs encode exactly $k\cdot X(\mathcal{M})$. In particular, if $X(\mathcal{M})=0$, then $\mathcal{M}$ can be realized with no unmatched stubs.

We then show that the witness of an authentic view admits a \emph{uniform size bound}: if an accepting view $\mu$ is realizable at all, then it can be realized in a labeled yes-instance whose size is bounded by a function $F(\Delta,c)$ depending only on the maximum degree and the certificate size. This bound is independent of the size $n$ of the original network. 

This size bound is crucial in the proof that soundness implies strong soundness. Given an arbitrary labeled instance, let $S$ be the set of accepting nodes and let $H=G[S]$ be the induced subgraph. The external demands created by cutting $H$ away from the rest of the graph are captured by the demand vector $X(\mathcal{M}_S)$. Since the decoder is normalized, every accepting view appearing in $S$ is authentic, and hence admits a bounded complement whose total demand cancels that of the view. Taking the disjoint union of these complements yields a multiset $\overline{\mathcal{M}_S}$ whose realization requires only a bounded number of nodes per accepting node in $S$. Applying the global realizability construction to $\overline{\mathcal{M}_S}$ therefore produces a completion gadget whose size is linear in $\abs{S}$, with a multiplicative constant depending only on $\Delta$ and $c$.

 The resulting graph $H'$ is fully accepting and has a size linear on the size of $G$, which fits within the polynomial upper bounds on the network size assumed by the protocol. By soundness, this graph must be a yes-instance of $\mathcal{L}$. Since $H$ appears as an induced subgraph and $\mathcal{L}$ is hereditary, we conclude that $H\in\mathcal{L}$, establishing strong soundness.

\subsubsection{Upper bounds with variable-size certificates}
\label{sec:results-variable-cert}

Our final set of results concern obtaining hiding and strongly sound LCPs where
we are not restricted to the anonymous setting and may use variable-size
certificates.
Our main result in this part shows that there is a single-round protocol with logarithmic certificate
complexity that achieves this.

\begin{restatable}{theorem}{restateThmUpperBoundVariableSize}
  \label{thm:upper-bound-variable-size}
  There is a hiding and strongly sound single-round LCP for $\twocol$ with
  certificate size $O(\log n)$.
\end{restatable}

Recall our discussion following \cref{thm:deg-one-or-even-cycle}; that is, for
our protocol to be hiding, it suffices if there is a labeling $\ell$ of \emph{at
least one} yes-instance $G$ such that, given $\ell$, no local algorithm can
extract a $2$-coloring of $G$ from $\ell$.
In particular, this means that we may focus on constructing such a hiding
labeling for a $G$ with properties of our choice and simply use the trivial
labeling (i.e., a $2$-coloring of $G$) in the remaining cases.
In the proof, we use a graph $G$ that has diameter equal to four.
It is simplest to state the proof for a fixed graph $G$, but we should note we
do not have any requirement on $G$ besides $\diam(G) \ge 4$ and that one cannot
already trivially extract a $2$-coloring of $G$ without even looking at the
certificates. This situation might occur, for instance, if the degree sequence of the nodes directly leaks its color, as it happens in the example of the language of star graphs we discussed above.
The labeling itself is similar to our anonymous LCP for $\delta(G) = 1$ from
\cref{thm:deg-one-or-even-cycle}:
We hide the $2$-coloring at a designated node $v$ and reveal it everywhere
outside $N[v]$.
The difference here is that we also reveal $\id(v)$ to all the nodes so as to
prevent having more than one node that does not know its color (which would
obviously allow odd cycles to be built).
In addition, nodes outside $N[v]$ are also given the color of the neighbors of
$v$.
This information is necessary as otherwise there could be neighbors of $v$ with
distinct colors and then an odd cycle could be closed at $v$ without any of the
nodes noticing it.

Since the hiding property of the protocol of
\Cref{thm:upper-bound-variable-size} relies on the diameter of the graph being
at least four, we exhibit another protocol where we restrict our attention to
graphs with diameter at most three.
Here, maintaining the hiding property comes at the cost of increasing the
certificate size.

\begin{restatable}{theorem}{restateThmUpperBoundDiamThree}
  \label{thm:upper-bound-diam-three}
  Let $\clbipdiamle{3}$ denote the class of bipartite graphs with diameter at
  most three.
  Then there is a hiding and strongly sound single-round LCP for
  $\twocol_{\clbipdiamle{3}}$ with certificate size $O(n \log n)$.
\end{restatable}

Even protocol used to prove \Cref{thm:upper-bound-variable-size} still guarantees completeness and strong soundness restricted to the class $\clbipdiamle{3}$, this protocol does not hold the hiding property, given that our hiding construction relies in graphs of diameter at least $4$. In fact, when dealing with graphs of diameter at most $3$, the strategy used to prove \Cref{thm:upper-bound-variable-size} no longer applies, because this protocol essentially consists in revealing the $2$-coloring to all nodes at distance at least $2$ from a selected node $v$, whose identity is known to all nodes in the graph. 
This is sufficient to verify the correctness of the coloring while preventing node $v$ from reconstructing it.

However, when we restrict our attention to graphs of diameter at most $3$, we cannot reveal the identity of a unique node $v$ to \emph{all} nodes in the graph. 
In this setting, there exists a trivial way to reconstruct a valid $2$-coloring using only the identifier of a single node. 
Specifically, node $v$ assigns itself color $0$; its neighbors assign themselves color $1$; and any remaining node assigns color $0$ or $1$, depending if they have a neighbor that was not assigned to any color. This procedure allows all nodes to recover a valid $2$-coloring.

Therefore, to obtain \Cref{thm:upper-bound-diam-three}, we develop other protocol that satisfies the hiding property for graphs within this class. The idea is to reveal to each node $v$ which side of the bipartition $V = A \cup B$ it belongs to. Each node must then verify that all its neighbors lie on the opposite side. In order to prove the hiding property, while still ensuring the completeness and strong soundness of our protocol, we restrict this new protocol to graphs that we call $2$-wise graphs, which roughly, are graphs such that there exists two nodes $a$ and $b$ such that $N(a)$ and $N(b)$ form a partition of the nodes. We refer to $a$ and $b$ are {\it wise nodes}. We prove (\Cref{lem:relations-wise-graphs}) that the $2$-wise graphs are contained in $\clbipdiamle{3}$, and all the bipartite graphs with diameter $2$ are $2$-wise graphs (see \Cref{fig:2wise} for a graphical example). Then, the new consists in, if the input graph $G$ has diameter $3$, then it is the same protocol of \Cref{thm:upper-bound-diam-three}, and if $G$ has diameter $2$, protocol selects in each side a wise node $a\in A$ and $b\in B$ and sends $\id(a)$ to all the nodes in $B$ and $\id(b)$ to all the nodes in $A$. Afterwards, select two nodes $d_a\in A$ and $d_b\in B$ (neither of which is a wise node) that will not receive any further information. For the rest of the nodes in $A$ and $B$ (different from $d_a$ and $d_b$, respectively), the prover reveals to each node $v$ the $\id$ of all the nodes on the same side as $v$, which is encoded in $\mathcal O(n\log n)$ bits. 

Completeness and strong soundness for the graphs with diameter $3$ are directly given in \Cref{thm:upper-bound-diam-three}, and we prove that these properties hold in the new protocol defined for $2$-wise graphs, and moreover, we prove that this new protocol satisfies the hiding property for $2$-wise graphs.

The certificate size of $O(n\log n)$ of \Cref{thm:upper-bound-diam-three} is due to the new protocol defined for $2$-wise graphs, where the prover sends to some nodes the complete list of $\id$'s of its side. Given that this new protocol is required to satisfy the hiding property in the class $\clbipdiamle{3}$, it remains an interesting open question whether we can obtain $O(\log n)$ as in
\cref{thm:upper-bound-variable-size} for any class of bipartite graphs with
small diameter (where the $2$-coloring problem is not trivial).
This problem appears to be hard since, when the diameter is small, nodes have a great deal of information at their disposal.
In particular, strategies based on hiding the coloring at a single, distinguished node $v$ apparently fails since any node can determine its distance
from $v$ and then use that to color itself.

\subsection{Key future directions}
\label{sec:open-problems}

We return to the motivation we posed at the beginning of this introduction, that
is, removing the promise from the separation of two models $\textbf{A}$ and
$\textbf{B}$ where $3$-coloring is \enquote{easy} in $\textbf{A}$ and
\enquote{hard} in $\textbf{B}$ \emph{when restricted to bipartite graphs}.
Using the terminology that we have introduced, in order to obtain such a
separation, we would like to obtain a certification scheme that achieves the
following:
\begin{enumerate}
  \item The scheme is a \emph{constant}-radius, strongly sound LCP for
  $2$-coloring.
  (Constant radius is necessary since we are working in the realm of LCL
  problems.)
  Due to strong soundness, it is admissible if the scheme only works in a
  restricted class of bipartite graphs where we can prove that $3$-coloring is
  hard for $\textbf{B}$ (e.g., grids for the concrete case of separating
  online-LOCAL from SLOCAL).
  \item The scheme is not only hiding in the sense that we introduced, that is,
  such that no local algorithm with \emph{constant} radius can extract a
  \emph{$2$-coloring}---but such that no \emph{variable}-radius algorithm in the
  model $\textbf{B}$ (which is potentially much stronger than the standard
  LOCAL model!) fails to extract a \emph{$3$-coloring}.
\end{enumerate}
As one can see, our results are still far from clarifying whether this is
possible.
Next we discuss each of the extensions that will be necessary to consider moving
forward.
Along the way we also mention other interesting questions that arise from them
even if they are not directly related to this goal.

\paragraph{\boldmath Hiding beyond $r = 1$.} 
The first extension is considering hiding certification of $2$-coloring for
larger radii.
It is likely that the $r$-forgetfulness property or generalizations thereof will
play a major role in this context.
One evidence for this is that it is apparently possible to come up with an
constant-size hiding and strongly sound anonymous LCP on grids if $r = 2$. 
(This does not contradict \cref{thm:deg-one-or-even-cycleLB} since grids are
\emph{not} $r$-forgetful.)
Namely, one can use a similar construction as in our
\cref{thm:deg-one-or-even-cycle} for the case of graphs with minimum degree $1$,
\enquote{hiding} the coloring at one of the corners of the grid and using the
overlap between the neighborhoods of its two neighbors to guarantee strong
soundness.

\paragraph{Stricter versions of hiding.}
The second extension is strengthening our hiding notion so that we can hide the
$2$-coloring even against models that are \emph{strictly stronger} than our
verifier.
Concretely, can we, say, certify a $2$-coloring with (constant) radius $r$ but
in such a way that our certification scheme is hiding against local algorithms
with radius $R > r$ (where potentially $R = \omega(1)$)?
Reconsidering our two constructions for constant radius from
\cref{thm:deg-one-or-even-cycle} under this aspect, there are reasons for both
optimism and pessimism:
While the scheme for graphs with minimum degree one definitely fails as it leaks
the coloring to algorithms with radius $2$, the one for cycles is very robust
and is able to tolerate even $R = o(n)$.

Beyond our main motivation of promise-free separations, there are another two
possible strengthenings of our notion of hiding that would be interesting to
consider:
\begin{enumerate}
  \item The first is a \enquote{fractional} one where we require, say, that at
  least a constant fraction of nodes fail to extract a correct color.
  This variant gives a more robust notion where, for example, hiding in a single
  node is not sufficient. 
  This fractional notion could have connections to areas such as distributed
  property testing \cite{censor-hillel19_fast_dc}.
  \item The second one refers to the order of quantifiers in the definition of
  hiding.
  Namely one could consider a stronger version where, instead of it being
  sufficient to hide the coloring in a single graph (for every possible
  algorithm that tries to extract it from the certificates), we impose the
  stronger requirement of fooling \emph{any} extraction algorithm on \emph{any}
  graph in the class that we are interested in.
  (Using the terminology from the discussion after
  \cref{thm:deg-one-or-even-cycle}, this would mean that the subclass of graphs
  we are certifying bipartiteness for and the hardcore set of our scheme are the
  same.)
  Exploring this notion would shed more light into what structure (in a more
  abstract sense) is required in order for a hiding strategy to be successful.
\end{enumerate}

\paragraph{\boldmath $3$-coloring.}
Finally, notice that it is not enough to hide a $2$-coloring but actually a
$3$-coloring from model $\mathbf{B}$.
If we wish to prove a certification scheme as outlined above does not exist,
then this is perhaps not as important:
As we point out in the discussion following \cref{thm:full-bip-impossibility}, a
certification scheme that hides a $K$-coloring must simultaneously be hiding a
$k$-coloring, for any $k < K$.
Hence it is well possible that we can already rule out a scheme that is hiding a
$2$-coloring, let alone a $3$-coloring.
At the same time, proving a scheme is able to hide a $3$-coloring based on our
\cref{lem:hiding-char} is a significantly more challenging task.
This would require showing the accepting neighborhood graph of the LCP is not
$3$-colorable, which is much harder to obtain than the $2$-coloring case as
there is no clear obstruction to be constructed---as opposed to the proofs we
pursue here, where we \enquote{only} have to show the existence an odd cycle.

\subsection{Organization}

The rest of the paper is organized as follows:
We start in \cref{sec:defs} by introducing basic definitions and the concept of
hiding LCPs.
\Cref{sec:alvg} introduces our main technical tool, the neighborhood graph of
accepting views, and proves the characterization for hiding that was mentioned
in the discussion after \cref{thm:oi-bip-impossibility}.
In \cref{sec:anonymous-ub} we show \cref{thm:deg-one-or-even-cycle}, which is
our upper bound in the anonymous case.
Following that, \cref{sec:stgo-postulate} presents the proof of
\cref{thm:sound-implies-strong}.
Then in \cref{sec:oi-lb,sec:ramsey} we prove our lower bound results
\cref{thm:oi-bip-impossibility,thm:full-bip-impossibility}.
Finally, in \cref{sec:non-anon-ubs} we develop our upper bound results
\cref{thm:upper-bound-variable-size,thm:upper-bound-diam-three}.

\section{Basic definitions}
\label{sec:defs}

Given two integers $s$ and $t$, $[s, t]$ denotes the set $\{k \in \mathbb{Z} : s
\leq k \leq t\}$. 
We also let $[t] = [1, t]$. 
For a function $f\colon A \to B$, we write $f|_{A'}$ for the restriction of $f$
to $A' \subseteq A$.
We write $A + B$ for the set that results from the disjoint union of $A$ and
$B$.

\subsection{Graphs}
\label{sec:def-graphs}

We only consider undirected, simple graphs.
Given a graph $G$, we denote the set of nodes and edges of $G$, respectively, by
$V(G)$ and $E(G)$.
We use the shorthand $uv$ to denote an edge $\{ u, v \}$.
For a set of nodes $U\subseteq V(G)$, we write $G[U]$ for the subgraph of $G$
induced by $U$.

Given $v \in V(G)$, the \emph{set of neighbors} of $v$ is $N(v) = \{u \in V(G) :
\{u,v\} \in E\}$, and $N[v] = N(v) \cup \{v\}$ is the \emph{closed neighborhood
of $v$}. 
Furthermore, we denote by $N^k(v)$ the set of nodes at distance at most $k$ from
$v$. 
Abusing notation, we occasionally also use $N(v)$, $N[v]$, and $N^k(v)$ to refer
to the \emph{subgraph} in $G$ induced by the respective set of nodes.
In case we are considering multiple neighborhoods from distinct graphs, we
denote the graph in question by a subscript (e.g., $N_G(v)$ for the neighborhood
of $v$ in $G$).

The \emph{degree of $v$}, denoted $d(v)$, is defined as the cardinality of
$N(v)$, and the minimum and maximum degree of $G$ are denoted $\delta(G)$ and
$\Delta(G)$, respectively.
We denote the diameter of $G$ by $\diam(G)$ and the distance between two nodes
$u,v \in V(G)$ by $\dist(u,v)$.

\subsection{Languages and models}

We consider \emph{distributed languages} \(\mathcal{L}\) as sets of pairs \((G,
x)\), where \(G = (V, E)\) is a simple, finite, undirected graph, and \(x\colon
V \rightarrow \{0, 1\}^*\) is an \emph{input} function, sometimes called
\emph{witness}. 
For instance, in this article, we primarily focus on the language 
\[
\textsf{k-col} = \left\{ (G, x) \mid x\colon V(G) \to \{1, \dots, k\}, \forall
\{u, v\} \in E(G), x(u) \neq x(v)\right\},
\]
which is the language that contains all pairs \((G, x)\) such that \(x\) is a proper \(k\)-coloring of \(G\).

Associated with a distributed language \(\mathcal{L}\), we define the set of graphs 
\[
\mathcal{G}(\mathcal{L}) = \left\{G \mid \exists x\colon V(G) \to \{0,1\}^*, (G,
x) \in \mathcal{L}\right\}.
\]
containing all graph that admit a witness. For instance, \(\mathcal{G}(\textsf{2-col})\) is the set of bipartite graphs. 

\subsubsection{Distributed certification of distributed problems}

\paragraph{Ports, identifiers, and labelings.} Given an $n$-node graph $G$, a
\emph{port assignment} is a function 
\[\ports_G\colon V(G) \times E(G) \to [\Delta(G)]\] 
that satisfies the following conditions for each $v \in V(G)$: (1)
For every edge $e$ incident to $v$, $\ports_G(v, e) \leq d(v)$. (2) For every pair
of edges $e_1, e_2$ incident to $v$, $\ports_G(v, e_1) \neq \ports(v, e_2)$. 
We omit the subscript when $G$ is clear from the context.
We slightly abuse notation and call $\ports(v)$ the function $\ports(v,\cdot)$.

Given an $n$-node graph $G$, and $N = \poly(n)$, an \emph{identifier assignment}
is an injective function 
\[
  \ids_G: V(G) \rightarrow [N].
\]
Observe that for each $v \in V(G)$, $\ids(v)$ can be encoded in $\lceil \log(N)
\rceil = O(\log n)$ bits. 
We assume that all identifiers are encoded using the same number of bits
(possibly padded with zeros), so it is possible to compute a polynomial upper
bound of $n$ from the number of bits used to encode $\id(v)$, for each $v \in
V(G)$. 

For a given graph $G$ and a positive integer $c$, a \emph{labeling of $G$ of
size $c$} is a function 
\[
  \ell_G: V(G) \rightarrow \{0, 1\}^c.
\]
A labeling plays the role of a certificate in a non-deterministic distributed
algorithm. 

In all the definitions above, we may drop the subscript when $G$ is clear from
the context.

\paragraph{Views.}
Fix an $n$-node graph $G$, a port assignment $\ports$, an identifier assignment
$\ids$, an \emph{input function} $I\colon V(G) \rightarrow \{0,1\}^*$, and a
positive integer $r$.
Each node $v$ in $G$ has a \emph{view} of radius $r$, which we denote by
$\view_r(G,\ports,\ids,I)(v)$.
If $r$, $\ports$, $\ids$, and $I$ are clear from the context, then we simply
write $\view(G)(v)$ instead; if the same can be said for $G$, then we shorten
this further to $\view(v)$.
Concretely, $\view(v)$ is the tuple 
\[
  \left(G_v^r, \ports|_{N^r(v)}, \ids|_{N^r(v)}, I|_{N^r}(v)\right)
\] 
where $G_v^r$ is the subgraph of $G$ induced by the union of all paths of length
at most $r$ starting from $v$ and $\ports|_{N^r(v)}$, $\ids|_{N^r(v)}$, and
$I|_{N^r}(v)$ are the maps induced by restricting $\ports$, $\ids$, and $I$ to
$N^r(v)$, respectively. 
Notice that $G_v^r$ has $N^r(v)$ as its node set and contains the full structure
of $G$ up to $r-1$ hops away from $v$ but not any connections between nodes that
are at $r$ hops away from $v$.
Hence $G_v^r$ may be distinct from the subgraph of $G$ induced by $N^r(v)$; see
\Cref{fig:basic-view} for an illustration.
Given a view $\mu = \view(v)$, we refer to $v$ as the \emph{center node} of
$\mu$.

\begin{figure}
  \centering
  \includestandalone[width=\textwidth]{figs/basic_view}
  \caption{A graph $G$ and the subgraph $G_v^r$ associated with the view of a
  node $v$ in $G$, where $r = 2$.
  Notice how the edge between the nodes $x$ and $y$ is missing in $G_v^r$.}\label{fig:basic-view} 
\end{figure}

When convenient, we may also refer to $\view(v)$ as the graph $G_v^r$ itself
labeled with the values of respective restrictions of $\ports$, $\ids$, and $I$.
We let $\mathcal{V}_r$ denote the set of all views of radius $r$ (in every
possible graph).

\paragraph{Local algorithms and decoders.} 
An $r$-round \emph{local algorithm} $\mathcal{A}$ is a computable map
$\mathcal{V}_r \to O$, where $O$ is some set of outputs.
We can imagine that $\mathcal{A}$ is the result of the nodes broadcasting to
their neighbors everything they know for $r$ rounds in succession, followed by
the execution of an internal unbounded (but computable) procedure at every node.
When running $\mathcal{A}$ on fixed $G$, $\ports$, $\ids$, and $I$, we write
$\mathcal{A}(G,\ports,\ids,I)(v)$ for $\mathcal{A}(\view(v))$; again, if $G$,
$\ports$, $\ids$, and $I$ are clear from the context, then we simply write
$\mathcal{A}(v)$.

A local algorithm $\mathcal{A}$ is \emph{anonymous} if its output is independent
of the identifier assignment; that is, for any two identifier assignments $\ids$
and $\ids'$, we have $\mathcal{A}(G,\ports,\ids,I)(v) =
\mathcal{A}(G,\ports,\ids',I)(v)$ for every $G$, $\ports$, $I$, and $v \in
V(G)$. 
Furthermore, the algorithm $\mathcal{A}$ is \emph{order-invariant} if its output
stays unchanged as long as the ordering of the identifiers is preserved; that
is, for any two identifier assignments $\ids$ and $\ids'$ such that $\ids(u) <
\ids(v)$ if and only if $\ids'(u) < \ids(v')$ for any $u,v \in V(G)$, we have
$\mathcal{A}(G,\ports,\ids,I)(v) = \mathcal{A}(G,\ports,\ids',I)(v)$ for every
$G$, $\ports$, $I$, and $v \in V(G)$. 
Notice that an anonymous algorithm is necessarily order-invariant, but the
converse may not be true.

A \emph{decoder} $\mathcal{D}$ is a $r$-round local algorithm that runs on
graphs where the input function $I$ of every node $v$ is a pair $I(v) = (N,
\ell(v))$ where $N = \poly(n)$ is a positive integer and $\ell$ is a label
assignment of $G$. 
A \emph{binary} decoder is a decoder whose outputs are $0$ or $1$.

\paragraph{Locally checkable proofs.} We are now ready to define the
\emph{locally checkable proofs} (LCP) model. 
We say that a distributed language $\mathcal{L}$ admits an $r$-round locally
checkable proof with certificates of size $f(n)$ if there exists an $r$-round
binary decoder $\mathcal{D}$ that satisfies the following two conditions:
\begin{description}
  \item[Completeness:] For every $n$-node $G \in \mathcal{G}(\mathcal{L})$, every port assignment $\ports$, and every identifier assignment $\ids$, there exists a labeling $\ell$ of $G$ of size $f(n)$ such that, for every $v \in V(G)$,~$\mathcal{D}(v) =~1$.
  \item[Soundness:] For every $n$-node $G \notin \mathcal{G}(\mathcal{L})$, every port assignment $\ports$, every identifier assignment $\ids$, and every labeling $\ell$ of $G$ of size $f(n)$, there exists a node $v \in V(G)$ such that~$\mathcal{D}(v) = 0$.
\end{description}
An LCP is called \emph{anonymous} if the binary decoder that defines it is
anonymous.

\subsubsection{Strong locally checkable proofs}

We introduce our first variant of the LCP model. We say that a language $\mathcal{L}$ admits a \emph{strong LCP} if, in addition to the completeness and soundness properties, the decoder $\mathcal{D}$ satisfies the \emph{strong soundness} property:
\begin{description}
  \item[Strong soundness:] For every $n$-node graph $G$, every port assignment
  $\ports$, every identifier assignment $\ids$, and every labeling $\ell$ of $G$
  of size $f(n)$, the subgraph induced by the set of nodes $\{v \in V(G) :
  \mathcal{D}(v) = 1\}$ is in $\mathcal{G}(\mathcal{L})$.
\end{description}
Observe that \emph{strong soundness implies soundness}. 
Indeed, in a no-instance \( G \), the set \( V(G) \) does not induce a
yes-instance; hence, \( \mathcal{D}(v) = 0 \) for at least one node \( v \in
V(G) \).

\subsubsection{Hiding locally checkable proofs}
\label{sec:hiding-def}

Remember that we defined distributed languages $\mathcal{L}$ as a set of pairs
$(G, x)$, where $x: V(G) \rightarrow \{0,1\}^*$. 
A \emph{hiding LCP} for $\mathcal{L}$ is one whose $r$-round decoder
$\mathcal{D}$ satisfies the following additional condition:
\begin{description}
  \item[Hiding:] For every $r$-round decoder $\mathcal{A}$, there is an $n$-node $G \in
  \mathcal{P}$, a labeling $\ell$ of $G$ of size $f(n)$, a port assignment
  $\ports$, and an identifier assignment $\ids$ for which the following holds:
  \begin{itemize}
    \item For every $v \in V$, $\mathcal{D}(v) = 1$.
    \item $\mathcal{A}$ fails to extract a valid witness for $G$, that is,
    $(G,(\mathcal{A}(v))_{v\in V(G)})\notin \mathcal{L}$.
  \end{itemize}
\end{description}
Intuitively, a hiding LCP is a distributed certification algorithm that
certifies $G \in \mathcal{G}(\mathcal{L})$ \emph{without (fully) revealing} a
witness $x$ such that $(G, x) \in \mathcal{L}$. 
Note that in our definition, we allow the witness $x$ to be hidden in a single
node. 
In other words, we do not assume any condition on the number of nodes where the
witness is hidden.

An LCP can satisfy any of the previously defined properties simultaneously. For
instance, an anonymous, strong, and hiding LCP is an LCP that satisfies the
anonymous, hiding, and strong properties simultaneously. 
In particular, a hiding LCP is said to be \emph{anonymous} when all the decoders
involved in its definition (i.e., $\mathcal{D}$ and $\mathcal{A}$) are
anonymous. 

\subsubsection{Locally checkable proofs for promise problems} 
\label{sec:def-promise-problems}

In some of our results, we consider strong and hiding LCPs that should only work
on a subset of graphs having a certain property.
For that reason, it is convenient to work with \emph{promise problems}, where we
assume that yes-instances belong to a certain class of graphs $\mathcal{H}$.
Membership in $\mathcal{H}$ may not be easy to certify; nevertheless, we assume
(without having to verify it) that the input instance belongs to it.

Formally, let $\mathcal{L}$ be a distributed language, and let $\mathcal{H}
\subseteq \mathcal{G}(\mathcal{L})$ be a class of graphs. 
We define the \emph{$\mathcal{H}$-restricted problem} $\mathcal{L}_\mathcal{H}$
as the pair $(\mathcal{L}_\mathcal{H}^Y, \mathcal{L}_\mathcal{H}^N)$ where:
\begin{itemize}
  \item $\mathcal{L}_\mathcal{H}^Y$ is the set of yes-instances, that is,
  $\mathcal{L}_\mathcal{H}^Y = \mathcal{H}$.
  \item $\mathcal{L}_\mathcal{H}^N$ is the set of no-instances, namely the set
  of graphs $G$ such that $G \notin \mathcal{G}(\mathcal{L})$.
\end{itemize}
For graphs that are neither yes- nor no-instances, one may exhibit arbitrary
behavior.

The definitions of completeness, soundness, anonymity, and hiding are directly
adapted when considering LCPs for promise problems.
That being said, there are two details that need to be revised, which we discuss
next.

The first of this is the remark that, if we have a class of graphs $\mathcal{H}$
and an LCP $\mathcal{D}$ that is hiding with respect to the promise problem
induced by $\mathcal{H}$, then in general $\mathcal{D}$ will \emph{not} be also
hiding for every $\mathcal{H}' \subseteq \mathcal{H}$.
The reason for this lies in the choice of quantifiers in the definition of
hiding:
If there is a certain decoder $\mathcal{A}$ that uses an extraction strategy
that fails only on $\mathcal{H} \setminus \mathcal{H}'$, then $\mathcal{A}$
succeeds on $\mathcal{H}'$; hence taking the restriction to $\mathcal{H}'$
suddenly breaks the hiding property as we now have a decoder that extracts it.

In addition to the above, we also need to be more precise in the case of the
strong property. 
For $\mathcal{L}_\mathcal{H}$, we say that a LCP is strong if it satisfies the
following condition:
\begin{description}
  \item[Strong promise soundness:] For every $n$-node $G \in \mathcal{L}_\mathcal{H}^N$, every port assignment $\ports$, every identifier assignment $\ids$, and every labeling $\ell$ of $G$ of size $f(n)$, the subgraph induced by the set of nodes $\{v \in V \mid \mathcal{D}(v) = 1\}$ belongs to $\mathcal{G}(\mathcal{L})$.
\end{description}
In full words, when considering strong LCP for a promise problem, we require that the subset of nodes that accept induces a graph in $\mathcal{G}(\mathcal{L})$, but not necessarily to $\mathcal{H}$.  
For instance, consider the problem of deciding whether a graph is bipartite,
with the promise that the input graph is a grid. 
For this problem, the goal is to define a certification algorithm that accepts
when the input graph is a grid and rejects when the input graph is not
bipartite. 
The above condition requires that, in no-instances, the subset of nodes that
accept induces \emph{some bipartite graph but not necessarily a grid}.

\section{The accepting neighborhood graph}
\label{sec:alvg}

In this section we introduce a couple of technical tools that will be used
throughout the paper.
These are the accepting neighborhood graph and the characterization of hiding
based on it (\cref{lem:hiding-char}).

Let $\mathcal{L}$ be a distributed language, and $\mathcal{D}$ be an $r$-round binary decoder for $\mathcal{L}$. Let an arbitrary graph $G$ be given along with port, identifier, and label assignments $\ports$, $\ids$, and $\ell$.
We refer to the tuple $(G,\ports, \ids, \ell)$ as a \emph{labeled instance}. When $G\in \mathcal{G}(\mathcal{L})$ and $\mathcal{D}(v) =1$ for all $v\in V$, then the tuple is called \emph{labeled yes-instance}.

Let $n$ be a positive integer. We define the set $\aviewsd$ as the set of all possible \emph{accepting views} of $\mathcal{D}$ on graphs of at most $n$ nodes. Formally, a view $\mu \in \mathcal{V}_r$ belongs to $\aviewsd$ if there exists a labeled instance $(G,\ports,\ids,\ell)$ and a node $v \in V(G)$ such that $\view(v) = \mu$ and $\mathcal{D}(v) = 1$.

Two views $\mu_1, \mu_2 \in \aviewsd$ are said to be \emph{yes-instance-compatible} if there exists a labeled yes-instance $(G,\ports,\ids,\ell)$ and an edge $\{u,v\} \in E(G)$ such that $\mu_1 = \view(u)$ and $\mu_2 = \view(v)$.

The \emph{accepting neighborhood graph} of $\mathcal{D}$, denoted $\alvgd$, is
the graph obtained from taking $\aviewsd$ as the node set and drawing edges
between every pair of yes-instance-compatible views. 
The reason for restricting the neighborhood graph to graphs of size bounded by
$n$ is its finiteness (and hence computability):

\begin{lemma}
  \label{lem:alg-construct-alvg}
  Let $\mathcal{L}$ be a distributed language and $r$ and $n$ be positive
  integers. 
  Consider an LCP of size at most $f(n)$ for $\mathcal{L}$ with decoder $\mathcal{D}$, for some computable function $f$.
  There exists a sequential algorithm that computes $\alvgd$.
\end{lemma}

\begin{proof}
 The algorithm iterates over all possible labeled yes-instances $(G,\ports,
 \ids, \ell)$ such that $G$ is of size at most $n$. 
 On each iteration, the algorithm adds to the vertex set of \(\alvgd\) the set
 $\{\view(v)\mid v\in V(G)\}$ and to the edge set all pairs
 \[
  \{\{\view(u), \view(v)\}\mid \{u, v\} \in E(G)\}.
 \]
  Since all sets are finite, the algorithm halts in a finite number of steps. 
\end{proof}

The following lemma gives us a characterization of hiding in terms of the local
view graph $\alvgd$:

\begin{lemma}
  \label{lem:hiding-char}
  Let $\mathcal{D}$ be an $r$-round LCP for $\kcol_\mathcal{H}$. 
  Then $\mathcal{D}$ is hiding if and only if there is $n$ such that $\alvgd
  \notin \kcolg$.
\end{lemma}


\begin{proof}
  We prove the claim by showing that $\mathcal{D}$ is not hiding if and only if
  $\alvgd \in \kcolg$ for every $n$.

  First we prove the forward implication.
  If $\mathcal{D}$ is not hiding, then we have a decoder $\mathcal{A}$ such
  that, for every labeled yes-instance $(G,\ports,\ids,\ell)$,
  $(\mathcal{A}(v))_{v \in V(G)}$ is a witness for $G$ in $\kcolg$.
  Then, for any $n$, (the restriction of) $\mathcal{A}$ is a proper
  $2$-coloring of $\alvgd$ as a map from $\aviewsd$ to $[k]$:
  By definition of $\alvgd$, for any two neighboring views $\mu_1$ and $\mu_2$
  in $\alvgd$, there is a $k$-colorable graph $G$ (on less than $n$ nodes) with
  nodes $u$ and $v$ connected by an edge in $G$ and such that $\mu_1 = \view(u)$
  and $\mu_2 = \view(v)$.
  Since $\mathcal{A}$ outputs a correct witness to $G \in \kcolg$, we have
  $\mathcal{A}(\mu_1) \neq \mathcal{A}(\mu_2)$.

  To prove the converse, we assume $\alvgd \in \kcolg$ for every $n$ and show
  how to construct a decoder $\mathcal{A}$ that extracts a $k$-coloring from
  the certificate for $\mathcal{D}$.
  Let $n$ be given.
  Using \cref{lem:alg-construct-alvg}, we have an algorithm $\mathcal{P}$ that
  constructs $\alvgd$ 
  We fix an arbitrary proper $k$-coloring $c$ of $\alvgd$ that can be computed
  deterministically from $\alvgd$ (e.g., $c$ is the lexicographically first
  coloring of $\alvgd$ where nodes are ordered as they appear in the output of
  $\mathcal{A}$).

  Now let $G$ be an input graph on $n$ nodes, and let $\ell$ be a labeling of
  $G$ that is accepted by $\mathcal{D}$. Recall that the input function of each node in $G$ includes some integer $N=\text{poly}(n).$
  The decoder $\mathcal{A}$ has every node $u \in V(G)$ execute the following
  procedure:
  \begin{enumerate}
    \item Execute $\mathcal{P}$ to construct $\mathcal{V}(\mathcal{D},N)$.
    \item Compute $c$.
    \item Determine own view $\mu_u = \view(u)$ in $\mathcal{V}(\mathcal{D},N)$.
    \item Output $c(\mu_u)$.
  \end{enumerate}
  This procedure is essentially correct by construction:
  For any two neighboring nodes $u,v \in V(G)$, since $u$ and $v$ are connected
  by an edge, their respective views $\mu_u$ and $\mu_v$ are connected in
  $\mathcal{V}(\mathcal{D},N)$.
  Thus, since $c$ is proper, $\mathcal{D'}(u) \neq \mathcal{A}(v)$.
\end{proof}

To conclude this section, recall the remark in \cref{sec:def-promise-problems}
we made regarding further restricting the class of graphs $\mathcal{H}$ for the
promise problem $\twocol_\mathcal{H}$, namely that a hiding LCP for
$\twocol_\mathcal{H}$ is not necessarily also hiding for
$\twocol_{\mathcal{H}'}$, where $\mathcal{H}' \subseteq \mathcal{H}$.
\Cref{lem:hiding-char} gives another perspective on this issue:
If we restrict the class of the graphs, then potentially we are removing edges
in $\alvgd$; that is, it might be the case that two views in $\alvgd$ are only
yes-instance-compatible because there is a graph $H \in \mathcal{H} \setminus
\mathcal{H}'$ where they are adjacent.
Once we remove $H$ from the class, the edge between the views in $\alvgd$
disappears, and hence this might disconnect odd cycles in $\alvgd$.

\section{Upper bounds in the anonymous case}
\label{sec:anonymous-ub}

Our goal in this section is to prove:

\restateThmDegOneOrEvenCycle*

We address the cases of $\mathcal{H}_1$ and $\mathcal{H}_2$ separately.

\subsection{Graphs with minimum degree 1}
\label{sec:anonymous-ub-mindeg-1}
We show first the following result:
\begin{lemma}
  Let $\mathcal{H}_1$ be the class of graphs $G$ for which $\delta(G) = 1$.
  There is an anonymous, strong and hiding one round LCP for $\twocol_{\mathcal{H}_1}$ using certificates of constant size.
\end{lemma} \label{lemm:degree1hiding}

Recall that, as discussed in \cref{sec:hiding-def}, it is sufficient to hide the
$2$-coloring at a single node.
The strategy is to hide the $2$-coloring at a node with degree one. The coloring is revealed for all the nodes in the graph with the exception of the degree one node and its unique neighbor. In particular, the degree one node is not able to recover its color. The strong soundness requirement follow from the fact that degree one nodes cannot be part of a cycle (particularly they cannot be part of an odd cycle).

\begin{proof}

Let $G=(V,E)$ be an instance of $2$-coloring. The prover assigns to each node a certificate $\ell:V \mapsto \{0,1,\bot, \top\}$. This certificate can be interpreted as the prover assigning $\bot$ to a degree $1$ node, $\top$ to its unique neighbor and $\{0,1\}$ to the rest of the nodes in the graph, representing a $2$-coloring of that part. Observe that this can be verified easily verified in the LCP model since: 
\begin{itemize}
    \item  A node labeled $\bot$ can verify that it has exactly one neighbor and its unique neighbor is labeled as $\top.$
    \item A node labeled $\top$ can verify that it has a unique neighbor labeled as $\bot.$
    \item A node labeled as $0$ or $1$ can locally check the validity of the coloring.
\end{itemize}

In a yes-instance $G$, since we are assuming $G \in \mathcal{H}$ then, there exists at least one node $u$ such that the degree of $u$ is $1$. Observe that nodes receiving $\{0,1\}$ are actually extracting its coloring from their certificate thus, if the graph is bipartite, there exists one certificate such that any node accepts. For soundness it suffices to observe that if $G$ is a yes-instance and $C$ is a cycle, then, by construction, no node in $C$ can be labeled as $\bot$

We define the decoder $\mathcal{D}$ as follows for a node $v \in V$, a labeling $\ell$ and a port assignment $\ports$:
\begin{enumerate}
    \item If node $v$ receives a certificate $\ell(v) = \bot$,   $\mathcal{D}(v,\ell,\ports) = \textbf{accept}$ if
    \begin{enumerate}
        \item $\delta(v) = 1$ 
        \item Its only neighbor $u$ is labeled by $\ell(u) = \top$.
    \end{enumerate} Otherwise, $\mathcal{D}(v,\ell,\ports)= \textbf{reject.}$
    \item If node $v$ receives a certificate $\ell(v) = \top$, $\mathcal{D}(v,\ell,\ports) = \textbf{accept}$ if:
    \begin{enumerate}
        \item there exists a unique $u \in N(v)$ such that $\ell(u) = \bot$.
        \item For all $w \in N(v) \setminus \{u\}, \ell(w) = \beta \in \{0,1\}$. 
    \end{enumerate}  Otherwise, $\mathcal{D}(v,\ell,\ports)= \textbf{reject.}$
    \item If node $v$  receives certificate $\ell(v) \in \{0,1\}$,  $\mathcal{D}(v,\ell,\ports) = \textbf{accept}$ if:
    \begin{enumerate}
        \item There exists at most one neighbor $u \in N(v)$ such that $\ell(u) = \top$.
        \item For every $w \in N(v) \setminus \{u\}$, we have that $\ell(w) \in \{0,1\}$ and $\ell(v) = \ell(w) +1 \mod 2$. 
    \end{enumerate}
Otherwise, $\mathcal{D}(v)= \textbf{reject.}$
\end{enumerate}

We claim that the decoder $\mathcal{D}$ is a hiding and strong LCP. First we show that $\mathcal{D}$ is an LCP, i.e. it satisfies soundness and completeness.

If $G$ is a yes-instance for $2$-coloring equipped with some port assignment $\ports$  then, there exists a certificate $\ell$ assigned by the prover to each node in $G$ such that:

\begin{enumerate}
    \item $\bot$ is assigned to some node $u$ such that $\delta(u) =1$, i.e. $\ell(u) = \bot$.
    \item $\top$ is assigned to the unique node $v$ such that $v \in N(u)$, i.e. $\ell(v)= \top$.
    \item Some color $\beta \in \{0,1\}$ is assigned to each node $w \in N(v)$ i.e. $\ell(w) = \beta \in \{0,1\}$. 
    \item For the nodes $x \in V\setminus\{u,v\}$, the prover assigns a proper coloring, i.e. $\forall x \in V\setminus\{u,v\} , \forall y \in N(x)\setminus \{v\}$ the labeling $\ell$ satisfies  $\ell(x) = \ell(y)+1\mod 2$.
\end{enumerate}
The existence of this latter labeling follows from the fact that $G$ is a yes-instance and by construction we have that $\forall v \in V, \mathcal{D}(v,\ell,\ports) = \textbf{accept}$. Thus, completeness holds.
\end{proof}
Let us show that the decoder is strong (soundness then holds from the fact that strong soundness implies soundness). By contradiction, let us assume that there exists some instance $G=(V,E)$ some port assignment $\ports$ and a labeling $\ell$ such that $G' = G [\{v\in V\mid \mathcal{D}(v)=1\}]$ contains some odd cycle $C$. Since nodes in $C$ accept, we must have that there must exists some node $x \in V(C)$ such that its certificate does not contain a color, i.e. $\ell(x) \in \{\top,\bot\}$. This node $x$ must also accept, i.e. $\mathcal{D}(x,\ell,\ports) = \textbf{accept.}$ Thus, this node cannot have a $\bot$ labeling since it must have degree $1$ to accept and since $x\in V(C)$ we have that $\delta(x)\geq 2$. The only choice then is that $\ell(x) = \top$. But in that case, since $x$ accepts, then it has exactly one neighbor $u$ such that $\ell(u)=\bot$ and all the rest of its neighbors are labeled by some color $\beta \in \{0,1\}$. Since $C$ is odd, then $u \in V(C)$ which is a contradiction since $u$ must have degree $1$. Strong promise soundness holds.



Finally, to argue that $\mathcal{D}$ is hiding, consider the two instances on
\cref{fig:instancesdeg1}. 
Then $\mathcal{V}(\mathcal{D},4)$ contains the odd cycle depicted on
\cref{fig:oddcycledegree1}. 
We apply \cref{lem:hiding-char}.

\subsection{Cycle}
\label{sec:ub-cycle}

\begin{lemma} \label{lemm:cyclehiding}
  Let $\mathcal{H}_2$ be the class of even cycles.
  There is an anonymous, strong and hiding single-round LCP for $\twocol_{\mathcal{H}_2}$ using certificates of constant size.
\end{lemma}

The strategy in this case is to reveal a $2$-edge-coloring of the cycle.
Unlike the case of $\delta(G) = 1$, this hides the $2$-coloring
\emph{everywhere}.

\begin{proof}

We describe a Hiding LCP $\mathcal{D}(v,\ell, \ports)$ for $2$-coloring restricted to $\mathcal{H}_2$. 

We consider labeling maps of the form $\ell\colon V\mapsto (\binom{\{1,2\}}{2}\times \{0,1\})^2$. Intuitively a valid labeling will encode a $2$-edge coloring by representing each edge by the port assignment of their endpoints. In fact, observe that given an instance $G=(V,E) \in \mathcal{H}_2$ and a port assignment $\ports$ for $G$, for each $e=uv \in E$ there exists a (locally) unique pair of ports $\ports(u,uv)\ports(v,uv)$ which represents the edge $e$.  Then, if $\varphi:E\mapsto\{0,1\}$ is an edge coloring, there exists a labeling $\ell_{\varphi}:V\mapsto (\binom{\{1,2\}}{2}\times \{0,1\})^2$, such that for each node $v$ and its two neighbors $v_1,v_2$ we have that 
\[\ell(v) = \left(
  \left(\ports(v,v_1v)\ports(v_1,v_1v),c_1(v)\right)
  \left(\ports(v,v_2v)\ports(v_2,v_2v),c_2(v)\right)
  \right),\]
where $c_1(v) = \varphi(\ports(v,v_1v)\ports(v_1,v_1v))$ and $c_2(v) = \varphi(\ports(v,v_2v)\ports(v_2,v_2v)).$
Observe that if $G \in \mathcal{H}_2$ then $G$ is $2$-colorable if and only if $G$ is $2$-edge-colorable. The idea is that, the nodes can verify a proper $2$-edge-coloring but they are not able to extract the $2$-coloring from this information.

Let $(G,\ports) \in \mathcal{H}_2$ be some instance with port assignment $\ports.$ We call a labeling of the form $\ell:V\mapsto (\{1,2\}^2 \times \{0,1\})^2,$ such that for every $v\in V(G)$ with unique neighbors $v_1,v_2$  we have:
$$\ell(v) = ((\ports(v,v_1v),\ports(v_1,v_1v)), c_1(v), (\ports(v,v_2v),\ports(v_
2,v_2v), c_2(v)),$$ where $c_i(v) \in \{0,1\}, i=1,2.$ Otherwise we call it \emph{non valid.}

We define the decoder $\mathcal{D}$ as follows for node $v \in V$, a labeling $\ell$ and a port assignment $\ports:$

\begin{enumerate}
    \item Node $v$ verifies if the labeling is valid by comparing the information in the certificate with its own port numbering. If the labeling is non valid then, $\mathcal{D}(v,\ell,\ports)=\textbf{reject}$.
    \item If $\ell$ is valid, node $v$ checks if:
    \begin{itemize}
       \item $c_1(v) \not = c_2(v)$
        \item $\ell(v_1)$ contains edge $p_1(v)p_2(v)$ and color $c_1(v)$
        \item  $\ell(v_2)$ contains edge $p_3(v)p_4(v)$ and color $c_2(v)$
    \end{itemize} 
    Otherwise $\mathcal{D}(v,\ell,\ports)=\textbf{reject.}$
\end{enumerate}

Now we show that $\mathcal{D}$ is a hiding and strong LCP. First, $\mathcal{D}$ satisfies completeness because for any $G=(V,E) \in \mathcal{H}$, we have that $G \in \mathcal{G}(\twocol)$ if and only if there exists a proper $2$-edge-coloring $\varphi:E\mapsto\{0,1\}$. By the previous construction, we have that, for any assignation of ports $\ports$, there exists some (valid) labeling $\ell_{\varphi}$ such that (by construction) for every $v \in V$, $\mathcal{D}(v,\ell,\ports) = \textbf{accept}$. Thus, $\mathcal{D}$ satisfies completeness.

To see that $\mathcal{D}$ is hiding, consider the instance on
\cref{fig:instancescycles} as well as the odd cycle (self-loop) on \cref{fig:oddcyclecycles}
and apply \cref{lem:hiding-char}.
\begin{figure}
  \centering
  \includestandalone{figs/instances_deg1}
  \caption{Labeled instances $I_1$ and $I_2$ for the proof of
  \cref{lemm:degree1hiding}.
  The symbols inside the nodes represent the identifiers.
  The labels (certificates) are written above the nodes. Port numbers are above the edges for odd numbered nodes and bellow for even number nodes (the order is considered from left to right). For example, in $I_1,$ node $u_1$ has port $p_1$ and node $u_2$ has ports $p_1',p'_2.$ In $I_2,$ node $v_2$  has
  The neighborhoods of the node $u_1$ and $u_5$, which are marked in red, are identical in
  the two instances.}
  \label{fig:instancesdeg1}
\end{figure}

\begin{figure}
  \centering
  \includestandalone[width=\textwidth]{figs/prueba}
  \caption{Odd cycle in $\mathcal{V}(\mathcal{D},6)$ for graphs with minimum
  degree $1$. 
  The center node of a view is marked in orange.
  The blue, green, and yellow shades correspond to the areas of the instances
  from \cref{fig:instancesdeg1}.}
  \label{fig:oddcycledegree1}
\end{figure}

\begin{figure}
    \centering
    \includestandalone{figs/instances_cycle}
    \caption{Instances $I_1$ for the proof of
    \cref{lemm:cyclehiding}. In cyan, the center of the neighborhood we are considering for a self-loop in  $\mathcal{V}(\mathcal{D},4)$}
    \label{fig:instancescycles}
\end{figure}

\begin{figure}
    \centering
    \includestandalone[width=\textwidth]{figs/oddcycle_cycle}
  \caption{Self-loop in $\mathcal{V}(\mathcal{D},4)$.
    The single view is part of $I$ in \cref{fig:instancescycles}.}
    \label{fig:oddcyclecycles}
\end{figure}

Now, let us show that $\mathcal{D}$ is strong. Let us consider a graph $G$ and some port assignment $\ports$. Let us call $G'=G [\{v\in V\mid \mathcal{D}(v)=1\}]$ and let us assume that $G'$ has an odd cycle $C$. Observe that since all the nodes in $C$ accept, any node $v \in V(C)$ is such that $\ell(v)$ assigns different colors for its two incident edges. By construction, we have that this cannot be possible since the parity of $C$ is odd. Thus, we must have that some node $v \in V(C)$ is such that $\mathcal{D}(v) = \textbf{reject}$ which is a contradiction. Since $G'$ contains no odd cycle then, $G'$ is bipartite and thus $\mathcal{D}$ is strong. 
\end{proof}

\section{Soundness implies strong soundness on constant protocols}
\label{sec:stgo-postulate}

In this section, we aim to prove \Cref{thm:sound-implies-strong}, that we restate here for convenience.

\restateSoundImpliesStrong*


In the following, we assume that $\mathcal{L}$ is hereditary and that
$\mathcal{L}$ admits a single-round LCP with \emph{constant-size} certificates on graphs of  maximum degree $\Delta$, where $\Delta$ is a fixed constant. 
Formally, we fix a constant $c \in \mathbb{N}$ and a single-round binary decoder
$\mathcal{D}$ such that, for every $n$ and every $n$-node input graph of maximum degree $\Delta$, the
prover uses labelings $\ell \colon V(G) \to \{0,1\}^c$.
We denote the (finite) certificate alphabet by $\Sigma := \{0,1\}^c .$

Given a labeled instance $(G,\ports,\ids,\ell)$ and a node
$v \in V(G)$, let us denote by $\mu = \view(v)$ the view of $v$ and by $\cent(\mu) \in \Sigma$ the certificate of the center, that is, $\cent(\mu) := \ell(v).$ Recall that we call $\aviewsd$ the set of all views that are
accepted by $\mathcal{D}$. Notice that its cardinality depends only on the certificate size $c$ and on the maximum
degree $\Delta$.

\subsection{Authentic accepting views and normalized protocols}

Before proceeding, we notice that protocol may admit accepting views that are not realizable in any
yes-instance: views that are locally accepted in isolation, but whose occurrence in a global instance inevitably forces some node to reject.
Such views are undesirable and play no meaningful role in the behavior of the
protocol.

We therefore \enquote{normalize} the decoder $\mathcal{D}$ so that
all of its accepting views are globally compatible, in the sense that each of
them can occur in a labeled yes-instance in which every node accepts.

More formally, we say that a view $\mu \in \aviewsd$ is
\emph{authentic} if there exists a labeled yes-instance 
$\mathcal{H}_\mu  = (H,\ports,\ids,\ell)$ of size at most $n$ such that $\mathcal{D}(w) = 1$ for every $w \in V(H)$
and $\view(v) = \mu$ for at least one node $v \in V(H)$. In such a case we call $\mathcal{H}_\mu$ a \emph{witness} for $\mu$.

\begin{definition}
  \label{def:normalized-decoder}
  Let $\raviewsd := \bigl\{ \mu \in \aviewsd \mid \mu \text{ is authentic} \bigr\}$
denote the set of authentic accepting views of $\mathcal{D}$ on graphs of size
at most $n$.
We define a decoder $\mathcal{D}^{\mathrm{norm}}$ from $\mathcal{D}$ by modifying
its acceptance condition so that it accepts exactly the views in $\raviewsd$.
Formally, $\textsf{Aviews}(\mathcal{D}^{\mathrm{norm}}, n) = \raviewsd.$

\end{definition}

Note that $\mathcal{D}^{\mathrm{norm}}$ differs from $\mathcal{D}$ only on
accepting views that are not realizable.
In the following lemma, we show that the normalized decoder inherits the key
properties of the original one.

\begin{lemma}
  \label{lem:normalization}
  The normalized decoder $\mathcal{D}^{\mathrm{norm}}$ satisfies:
  \begin{enumerate}
    \item If $\mathcal{D}$ is complete for $\mathcal{L}$, then
          $\mathcal{D}^{\mathrm{norm}}$ is also complete for $\mathcal{L}$.
    \item If $\mathcal{D}$ is sound for $\mathcal{L}$, then
          $\mathcal{D}^{\mathrm{norm}}$ is also sound for $\mathcal{L}$.
    \item If $\mathcal{L}$ is a distributed language of pairs $(G,x)$ with $x: V(G) \rightarrow \{0,1\}^*$ and  $\mathcal{D}$ is hiding for $\mathcal{L}$, then
          $\mathcal{D}^{\mathrm{norm}}$ is also hiding for $\mathcal{L}$.
  \end{enumerate}
\end{lemma}

\begin{proof}
For~(1), fix any yes-instance $G\in\mathcal{L}$.
  Since $\mathcal{D}$ is complete, there exists a labeling $\ell$
  such that $\mathcal{D}(v)=1$ for all $v\in V(G)$.
  For each $v$, the view $\mu(v)$ is therefore an accepting view of
  $\mathcal{D}$.
  As the instance is globally accepting, each such view $\mu(v)$ is
  authentic.
  Hence $  \mu(v)\in\raviewsd 
      = \textsf{Aviews}(\mathcal{D}^{\mathrm{norm}},n ),$
  and therefore $\mathcal{D}^{\mathrm{norm}}(v)=1$ for all $v$.
  This shows that $\mathcal{D}^{\mathrm{norm}}$ is complete.

  For~(2), let $G\notin \mathcal{L}$ and consider any labeling
  $\ell$ of $G$.
  Since $\mathcal{D}$ is sound, there exists $v\in V(G)$ with $\mathcal{D}(v)=0$.
  The definition of $\mathcal{D}^{\mathrm{norm}}$ only modifies the output of
  $\mathcal{D}$ on views that $\mathcal{D}$ originally accepted.
  Hence, if $\mathcal{D}(v)=0$, then
  $\mathcal{D}^{\mathrm{norm}}(v)=0$ as well.
  Thus no no-instance can be fully accepted under $\mathcal{D}^{\mathrm{norm}}$,
  proving soundness.

  For~(3), let $\mathcal{D}$ be hiding.
  Consider any decoder $\mathcal{A}$.
  By the hiding property of $\mathcal{D}$, there exists
  a labeled yes-instance $(G,\ports,\ids,\ell)$ such that:
  \begin{itemize}
    \item $\mathcal{D}(v)=1$ for all $v\in V(G)$, and
    \item the extracted witness $(\mathcal{A}(v))_{v\in V(G)}$ is not valid.
  \end{itemize}
  Since every node accepts under $\mathcal{D}$, every view $\mu(v)$ is an
  accepting view of $\mathcal{D}$ that appears in a labeled yes-instance; hence
  each $\mu(v)$ is authentic, and therefore
  \[
    \mu(v)\in \textsf{AuthenticAviews}(\mathcal{D},n)
      = \textsf{Aviews}(\mathcal{D}^{\mathrm{norm}},n).
  \]
  Thus $\mathcal{D}^{\mathrm{norm}}(v)=1$ for all $v\in V(G)$.
  The decoder $\mathcal{D}^{\mathrm{norm}}$ “reveals” no more information than
  $\mathcal{D}$ (it only rejects more views), so the extraction performed by
  $\mathcal{A}$ on this instance remains invalid.
  Hence $\mathcal{D}^{\mathrm{norm}}$ is also hiding.
\end{proof}

For our purposes, the mere existence of a normalized protocol is not enough.
We need to be able to compute, from an arbitrary protocol, a normalized one.
To this end, we introduce an algebraic characterization of authentic views.

\subsection{Algebraic representation of views}

For each view $\mu \in \aviewsd$ and each
$B \in \Sigma$, we denote by $
  d_{\cent(\mu)\to B}(\mu) \in \mathbb{N}$ 
the number of neighbors of the center of $\mu$ that
carry certificate $B$. We now associate to each view $\mu$ an integer vector,
denoted $x(\mu) \in \mathbb{Z}^{\Sigma\times\Sigma} $ and called the \emph{demand
vector} of $\mu$.  For every ordered pair $(A,B) \in \Sigma^2$ 
we introduce one
coordinate, which we denote by $(AB)$. Whenever $\cent(\mu) = A$, we define for each $B\in \Sigma \setminus \{A\}$,
$  x(\mu)_{AB} := d_{A\to B}(\mu)$
and $x(\mu)_{BA} := -x(\mu)_{AB}$, and all other coordinates are zero. In particular, coordinate $x(\mu)_{AA}$ is~$0$. 

Intuitively, $x(\mu)_{AB}$ records how many neighbors with certificate $B$ the
center of $\mu$, whose certificate is $A$, \enquote{demands} from its neighbors. The antisymmetric definition ensures that each actual edge between certificates $A$ and $B$ contributes $+1$ to the demand vector of one endpoint and $-1$ to that of the other. Thus, when summing the demand vectors  over a collection of views, all \emph{internal} demands, that is, demands satisfied by an edge whose both endpoints appear in the collection, cancel out automatically. What remains in the sum are precisely the \emph{external} demands: the demands that each view imposes toward neighbors that are not part of the collection. This representation excludes the demands of the form $A$-to-$A$, which will be treated explicitly in a different way.

Let $\mathcal{M}$ be a finite multiset of views, and let $m_\mu$ denote the
multiplicity of $\mu$ in $\mathcal{M}$. Its \emph{demand vector} is 
\[
  X(\mathcal{M}) := \sum_{\mu \in \mathcal{M}} m_\mu x(\mu)
  \in \mathbb{Z}^{\Sigma\times\Sigma}.
\]

We say that $\mathcal{M}$ is \emph{closed} if $  X(\mathcal{M}) = 0. $

\begin{remark}
    If $\mathcal{M}$ is obtained from a labeled instance $(H,\ports,\ids,\ell)$  by taking one copy of $\view(v)$ for each $v \in V(H)$, then $\mathcal{M}$ is closed. Indeed, every edge $uv$ with certificates $(A,B)$ contributes $+1$ to the coordinate $(AB)$ through the view at $u$ and $-1$ to the same coordinate
through the view at $v$, and hence all contributions cancel out.
\end{remark}

\paragraph{Representation of local demands by stubs.} A view $\mu$ specifies the certificate of its center $\cent(\mu)$ and, for each certificate $B \in \Sigma$, the number
$d_{\cent(\mu)\to B}(\mu)$ of neighbors of certificate $B$ that the center must
see.  For our algebraic arguments it is convenient to represent these neighbor
requirements as explicit combinatorial objects that we call \emph{stubs}.

A \emph{stub} (or half edge) of type $(A,B)$ should be understood as an abstract
placeholder representing one required neighbor of certificate $B$ for a node
whose certificate is $A$.  
Stubs do not yet correspond to actual neighbors; they only encode the local
demand prescribed by the view.

Let $\mathcal{M}$ be a finite multiset of   $\aviewsd$. For each $\mu \in \mathcal{M}$ with center certificate $A=\cent(\mu)$, we
associate an abstract object which we call the node of type $\mu$.
This object is not part of a graph; it only packages the local neighborhood
requirements of~$\mu$ in symbolic form.

To this abstract node we attach stubs corresponding exactly to the neighbor
requirements appearing in $\mu$:
\begin{itemize}
  \item for each $B\neq A$, we attach $d_{A\to B}(\mu)$ stubs of type $(A,B)$
  \item we attach $d_{A\to A}(\mu)$ stubs of type $(A,A)$, representing neighbors
    of the same certificate
\end{itemize}
Thus each view $\mu$ is accompanied by a collection of stubs that encode
precisely its local certificate neighborhood structure.

We say that a stub of type $(A,B)$ is \emph{compatible} with a stub of type $(B,A)$.
Matching such a pair represents the creation of an edge between a node of
certificate $A$ and a node of certificate $B$.

Once two stubs are matched, the resulting edge is realized by placing it on the
ports of each endpoint, prescribed by the corresponding views. In other words, port numberings do not constrain which stubs may be matched; they only specify how the edge is embedded in the local views after the matching
has been chosen.

Matching stubs therefore amounts to combining local views in a way that is consistent with certificate compatibility, while realizing the resulting edges
on the ports required by the views.

The antisymmetric demand vector $
x(\mu)$ is the algebraic encoding of these stubs. Furthermore, the demand vector $X(\mathcal{M})$ of a multiset $\mathcal{M}$ of views records the net imbalance of stubs between certificates.

\begin{itemize}
  \item If $X(\mathcal{M})_{AB}=0$, then in principle all stubs of type
        $(A,B)$ and $(B,A)$ can be matched internally in a port consistent way
  \item If $X(\mathcal{M})_{AB}\neq 0$, then the imbalance indicates external
        demands, that is, stubs that cannot be matched within $\mathcal{M}$ and
        that must be matched to nodes outside the current multiset
\end{itemize}

\paragraph{Atomic views associated with stubs.}
For certificates $A,B\in\Sigma$ with $A\neq B$, define the \emph{atomic view}
$\alpha_{A\to B}$ to be the radius one view consisting of a center labeled $A$
with exactly one neighbor labeled $B$ and no other neighbors.  
Its demand vector is
$x(\alpha_{A\to B})_{AB}=1,$
  $x(\alpha_{A\to B})_{BA}=-1, $
and all other coordinates are zero.
For $A=B$ we similarly define $\alpha_{A\to A}$ as a center labeled $A$ with one
neighbor labeled $A$.

A stub of type $(A,B)$ is simply an instance of the atomic view $\alpha_{A\to B}$.
Thus every collection $S$ of stubs can be viewed as a multiset of atomic views
and has a well defined demand vector
\[
  X(S) := \sum_{\alpha\in S} x(\alpha).
\]

The following lemma shows that any finite multiset of accepting views can be witnessed by a labeled instance with a bounded replication factor. In
particular, the size of the construction grows only by a multiplicative factor
depending on the maximum degree $\Delta$, and not on the size of the multiset
itself.

\begin{lemma}  \label{lem:global-realizability}
Let $\mathcal{M}$ be a finite multiset of accepting views from
  $\aviewsd$, with multiplicities $(m_\mu)_{\mu\in\mathcal{M}}$.  Then there exist a labeled instance $(H,\ports,\ids,\ell)$, a collection of
  stubs attached to the nodes of $H$, and a positive integer $k = 2^{O(\Delta)}  $ such that:
  \begin{enumerate}
    \item Every node of $H$ has a view in $\mathcal{M}$.
      Moreover, for each $\mu\in\mathcal{M}$, the number of nodes $v\in V(H)$
      satisfying $\view(v)=\mu$ is exactly $k \cdot m_\mu$.

    \item Attach to every node $v\in V(H)$ the stubs prescribed by its view
      $\view(v)$, and for each edge $\{u,v\}\in E(H)$ remove one appropriate
      pair of stubs (one at $u$, one at $v$) corresponding to that edge.
      Let $S_{\mathrm{rem}}$ be the multiset of all remaining, unmatched stubs.

      Then:
      \begin{itemize}
        \item no stub of type $(A,B)$ with $A=B$ remains in $S_{\mathrm{rem}}$,  
              that is, all $(A,A)$ demands are internally matched in $H$,
        \item and the global demand vector of the remaining stubs satisfies
          \[
            X(S_{\mathrm{rem}}) = k\cdot X(\mathcal{M}).
          \]
      \end{itemize}
  \end{enumerate}
  In particular, if $X(\mathcal{M}) = 0$, then the labeled graph $H$ realized from
$\mathcal{M}$ contains no free stubs.
\end{lemma}

\begin{proof}
  We construct $H$ in several steps.
  Throughout the proof we maintain that every node we create is tagged with
  some target view $\mu \in \mathcal{M}$ and carries a certificate and a
  collection of stubs that are consistent with~$\mu$.
  Refer to \cref{fig:stgo-pipeline} for a depiction of the overall process.

  \begin{figure}
    \label{fig:stgo-pipeline}
    \centering
    \includestandalone[width=\textwidth]{figs/soundness_pipeline}
    \caption{Illustration of the pipeline involved in the proof of
    \cref{lem:global-realizability}.
    Starting from a multiset of accepting views, we replace endpoints with stubs
    and determine a demand vector for each view.
    Matching the stubs may result in multiedges, which we remove by a bounded
    blow-up.
    The stubs where both endpoints of the edge have the same certificate are
    dealt with last.}
  \end{figure}

  \paragraph{Step 1: Abstract nodes and stubs.}
  For each $\mu \in \mathcal{M}$ and each copy index
  $i \in [m_\mu]$ we create an abstract node $v_{\mu,i}$, intended to witness
  the view~$\mu$.
  We fix its certificate to be
  $ \ell(v_{\mu,i}) := \cent(\mu).$ At each node $v_{\mu,i}$ we attach a collection of pending stubs that encode
  the neighbor requirements of~$\mu$:
  \begin{itemize}
    \item For every $B \in \Sigma$ with $B \neq \cent(\mu)$ we create
      exactly $d_{\cent(\mu)\to B}(\mu)$ stubs of type $(\cent(\mu),B)$.
      These are intended to be matched to neighbors with center
      certificate~$B$.
    \item We also create $d_{A\to A}(\mu)$ stubs of type
      $(A,A)$ at $v_{\mu,i}$, where $A = \cent(\mu)$.
      These represent neighbors that share the same certificate as the
      center and will be handled separately.
  \end{itemize}

  For distinct certificates $A,B\in\Sigma$ we let $S_{AB}$ be the multiset
  of all stubs of type $(A,B)$ over all nodes $v_{\mu,i}$.

  \paragraph{\boldmath Step 2: Matching cross-certificate stubs $A\neq B$.}
  We now match as many stubs of type $(A,B)$ with $(B,A)$ as possible, for
  every unordered pair $\{A,B\}$ with $A\neq B$. Fix distinct certificates $A,B\in\Sigma$.
  Let $S_{AB}$ and $S_{BA}$ be the corresponding multisets of stubs, and let
  $t_{AB} := \abs{S_{AB}}$ and $t_{BA} := \abs{S_{BA}}$.
  We arbitrarily choose a bijection
  $\varphi_{AB} : S'_{AB} \to S'_{BA},$ where $S'_{AB} \subseteq S_{AB}$ and $S'_{BA} \subseteq S_{BA}$ have size
  $\min(t_{AB},t_{BA})$.
  For every pair of matched stubs $(s,\varphi_{AB}(s))$, one of type $(A,B)$
  and the other of type $(B,A)$, we create an undirected edge between the
  corresponding abstract nodes.
  At both endpoints we record that this edge consumes exactly one stub of the
  appropriate type.
  These edges respect the port constraints, since the degree and port budgets
  at each node are determined by the underlying view and are uniformly bounded
  by~$\Delta$.

  After performing this construction for every unordered pair $\{A,B\}$ with
  $A\neq B$, we obtain a multigraph $H_0$ whose node set is
  $\{v_{\mu,i}\}$.
  At this point:
  \begin{itemize}
    \item Every matched pair of stubs $(A,B)$ and $(B,A)$ has been turned into
      a genuine edge of~$H_0$ and removed from the stub multiset.
    \item For each pair of distinct certificates $A,B$ we may still have
      unmatched stubs of type $(A,B)$ or $(B,A)$, precisely
      $\abs{t_{AB}-t_{BA}}$ such stubs in total.
      These encode the net imbalance between $A$-to-$B$ and $B$-to-$A$
      demands inside the current collection.
    \item All stubs of type $(A,A)$ remain untouched.
  \end{itemize}

  Let $S_{\mathrm{rem}}^{(2)}$ be the multiset of all stubs that remain
  unmatched after Step~2 (both cross-certificate and $(A,A)$ stubs).
  By construction, every time we match one stub of type $(A,B)$ with one stub
  of type $(B,A)$, their contributions cancel in the antisymmetric demand
  vector.
  Therefore the algebraic sum of the contributions of the unmatched stubs is
  still $X(\mathcal{M})$, that is,
  $X\bigl(S_{\mathrm{rem}}^{(2)}\bigr) = X(\mathcal{M}).$

  \paragraph{Step 3: Removing parallel edges by a bounded blow-up.}
  The multigraph $H_0$ may contain parallel edges between two nodes, since
  several cross-certificate demands can be satisfied by the same pair of
  endpoints.
  We now construct a simple graph $H_1$ by a bounded blow-up that preserves
  the local certificate patterns and all remaining stubs.

  For every unordered pair $\{u,v\}$ of vertices in $H_0$, let $m_{uv} \ge 0$
  denote the multiplicity of the edge $\{u,v\}$, that is, the number of
  parallel edges between $u$ and $v$ in~$H_0$.
  All $m_{uv}$ are bounded by $\Delta$.

  Let $ M = \mathrm{lcm}\{m_{uv} \mid \{u,v\}\in E(H_0)\}$
  be the least common multiple of all positive multiplicities.
  Since every $m_{uv} \le \Delta$, we have
  $M \le \mathrm{lcm}(1,2,\dots,\Delta)= 2^{O(\Delta)}$. 
  
  We construct a new labeled multigraph $H_1$ as follows.
  Each vertex $u \in V(H_0)$ is replaced by $M$ copies $u^{(1)},u^{(2)},\dots,u^{(M)},$
  and every copy $u^{(i)}$ receives the same certificate as $u$, that is
  $\ell_{H_1}(u^{(i)}) = \ell_{H_0}(u)$.
  All stubs that were attached to $u$ in $H_0$ are replicated at each copy
  $u^{(i)}$ in $H_1$.

  Now fix a multiedge $\{u,v\}$ with multiplicity $m_{uv} = m \ge 1$.
  We replace the $m$ parallel edges between $u$ and $v$ in $H_0$ by a simple
  $m$ regular bipartite graph between the two clusters $S(u) = \{u^{(1)},\dots,u^{(M)}\}$ and     $S(v) = \{v^{(1)},\dots,v^{(M)}\}.$
  Formally, for each $i \in \{1,\dots,M\}$ and each $j \in \{0,\dots,m-1\}$,
  we add a single edge $u^{(i)} v^{(i+j)}$, where the superscript $i+j$ is
  taken modulo $M$ (with values in $\{1,\dots,M\}$).
  Because $m \le M$, all these edges are pairwise distinct.
  Moreover, each copy $u^{(i)}$ has exactly $m$ neighbors in $S(v)$, and each
  copy $v^{(t)}$ has exactly $m$ neighbors in $S(u)$.
  Thus, between the clusters $S(u)$ and $S(v)$ we obtain a simple bipartite
  graph in which every vertex in $S(u)$ and in $S(v)$ has exactly $m_{uv}$
  neighbors in the other cluster.

  We perform this construction independently for every unordered pair
  $\{u,v\}$ with $m_{uv} \ge 1$.
  The result is a simple graph, which we denote by $H_1$.
  By construction, for every original vertex $u \in V(H_0)$ and every copy
  $u^{(i)}$ we have:
  \begin{itemize}
    \item The multiset of certificates of the neighbors of $u^{(i)}$ in $H_1$
      is exactly the same as the multiset of certificates of the neighbors
      of $u$ in $H_0$.
    \item All stubs attached to $u$ are replicated at each copy $u^{(i)}$,
      so the collection of stubs remains consistent with the target view
      associated with $u$.
  \end{itemize}

  Let $S_{\mathrm{rem}}^{(3)}$ be the multiset of all stubs attached to the
  copies $u^{(i)}$ in $H_1$.
  Since each node of $H_0$ has been replaced by $M$ copies, we obtain
  \[
    X\bigl(S_{\mathrm{rem}}^{(3)}\bigr) = M \cdot X\bigl(S_{\mathrm{rem}}^{(2)}\bigr)
    = M \cdot X(\mathcal{M}).
  \]

  \paragraph{\boldmath Step 4: Satisfying all $(A,A)$ demands.}
  After Step~3, all cross-certificate stubs \((A,B)\) with $A\neq B$ remain
  exactly as in Step~2, just replicated $M$ times.
  Now we intend to eliminate all demands of type
  $(A,A)$.

  For each accepting view $\mu\in\mathcal{M}$ whose center has certificate $A$,
  we define $t(\mu) := d_{A\to A}(\mu),$
  the number of neighbors with certificate $A$ prescribed by~$\mu$.
  Each node in $H_1$ of type~$\mu$ carries exactly $t(\mu)$ unmatched
  $(A,A)$ stubs.

  Now let $
    L := \operatorname{lcm}\{t(\mu) + 1 \mid \mu\in\mathcal{M}\}
    = 2^{O(\Delta)},$   and take $L$ disjoint copies of $H_1$,
   namely $H_1^{(1)}, H_1^{(2)},\dots, H_1^{(L)}.$
  For every vertex $v$ of $H_1$ we write $v^{(j)}$ for its copy in $H_1^{(j)}$.
  Each $v^{(j)}$ inherits its certificate $\ell(v)$, all incident edges to
  neighbors of certificate $B\neq A$, and the same number $t(\mu)$ of
  $(A,A)$-stubs as $v$ had in~$H_1$, whenever $v$ is of type~$\mu$ with
  center certificate~$A$.

  At this point, the $L$ copies are disjoint and simple, and all cross-certificate
  stubs remain untouched.
  The collection of all stubs attached to the nodes of the disjoint union of
  copies has global demand vector
  \[
    X\bigl(S_{\mathrm{rem}}^{(4)}\bigr)
      = L \cdot X\bigl(S_{\mathrm{rem}}^{(3)}\bigr)
      = M L \cdot X(\mathcal{M}).
  \]

  Fix a certificate $A\in\Sigma$ and an accepting view
  $\mu\in\mathcal{M}$ with $\cent(\mu)=A$ and $t(\mu)>0$.
  Let $v$ be any node of type $\mu$ in $H_1$.
  Across the $L$ copies, the nodes $v^{(1)}, v^{(2)},\dots, v^{(L)}$
  each carry exactly $t(\mu)$ stubs of type $(A,A)$.
  Since $L$ is a multiple of $t(\mu)+1$, we can write
  \[
    L = q_\mu\cdot \bigl(t(\mu)+1\bigr)
  \]
  for some positive integer $q_\mu$.

  We now partition the index set $\{1,\dots,L\}$ into
  $q_\mu$ blocks, each of size $t(\mu)+1$:
  \[
    \{1,\dots,L\}
      = B_1 + B_2 + \dots + B_{q_\mu},
    \qquad \abs{B_r}=t(\mu)+1\ \text{for all }r.
  \]
  For each block $B_r$, we add all edges of the clique on the
  vertices $\{v^{(j)} \mid j\in B_r\}$.
  In other words, for every pair of distinct indices $j,j'\in B_r$ we add the
  edge $v^{(j)}v^{(j')}$.
  The block size is $t(\mu)+1$, so in the clique on
  $\{v^{(j)} \mid j\in B_r\}$ each node $v^{(j)}$ gains exactly
  $t(\mu)$ new neighbors of certificate~$A$.
  This consumes all $(A,A)$ stubs attached to the nodes $v^{(j)}$ with
  $j\in B_r$.

  We perform this construction independently for every node type $\mu$ with
  center certificate $A$ and for each certificate $A\in\Sigma$.
  Since different types $\mu$ correspond to different base nodes $v$ in $H_1$,
  the cliques we add for one node $v$ do not interact with those for another
  node $v'$.

  Note that every newly created edge in this step consumes two stubs of type
  $(A,A)$, one at each endpoint.
  Since these stubs do not contribute to the  demand vector
  (we ignore coordinates $(A,A)$), the value of $X(\cdot)$ is unchanged.
  Thus, after matching all $(A,A)$ stubs in this way, the final collection of
  remaining stubs $S_{\mathrm{rem}}$ satisfies
  \[
    X(S_{\mathrm{rem}}) = M L \cdot X(\mathcal{M}).
  \]

  Let $H$ be the resulting simple graph after Step~4.
  Each original node corresponding to a view $\mu$ has been blown up into
  exactly $M L$ copies, so each view $\mu\in\mathcal{M}$ appears as the target
  type of exactly $k\cdot m_\mu$ nodes of $H$, where
  \[
    k := M L = 2^{O(\Delta)}.
  \]
  Moreover, all $(A,A)$ stubs have been matched to internal edges, and the
  remaining unmatched stubs constitute $S_{\mathrm{rem}}$, whose global demand
  vector is $X(S_{\mathrm{rem}}) = k \cdot X(\mathcal{M})$, as required.
\end{proof}

We can now derive the following algebraic characterization of authentic views.

\begin{lemma}
  \label{lem:realizability-vector}
  Let $\mu \in \aviewsd$ be an accepting view. There exists a function $$F(\Delta, c) = 2^{2^{2^{O(c + \log\log \Delta)}}}$$ such that, if $\mu$ is authentic, then $\mu$ admits a witness $\mathcal{H}_\mu = (H,\ports,\ids,\ell)$ with $H$ of size $F(\Delta,c)$, provided that $n\geq F(\Delta,c)$. 
\end{lemma}

\begin{proof}
  Assume that $\mu$ is authentic.
  By definition, there exists a labeled yes-instance
  $(H',\ports',\ids',\ell')$ of size at most $n$ such that all nodes accept and a node
  $v^\ast \in V(H')$ satisfies $\view(v^\ast) = \mu$.
  Let $m_\nu$ be the number of nodes $w \in V(H')$ whose view equals $\nu$.

  By \Cref{lem:global-realizability}, the multiset $\mathcal{M} := \{ \view(w) \mid w \in V(H') \}$
  is closed, that is, $X(\mathcal{M}) = 0$.
  Since $m_\mu \ge 1$, we can write
  $m_\mu = 1 + \lambda_\mu$ and $m_\nu = \lambda_\nu$ for $\nu \neq \mu$, and obtain
  \[
    0 = X(\mathcal{M})
      = \sum_{\nu} m_\nu x(\nu)
      = x(\mu) + \sum_{\nu} \lambda_\nu x(\nu).
  \]
  We deduce that the vector $(-\lambda_\nu)_{\nu \in \mathcal{M}}$ is a solution of the following linear equation over the variables $(\alpha_\nu)_\nu$:
  \begin{equation}\label{eq:realizability-system}
    x(\mu) = \sum_{\nu} \alpha_\nu x(\nu).
  \end{equation}

  We now show that \eqref{eq:realizability-system} admits another solution whose coefficients are uniformly bounded in terms of $c$ and $\Delta$ only.
  Recall that the certificate alphabet is $\Sigma = \{0,1\}^c$ and that the demand vectors $x(\nu)$ live in $\mathbb{Z}^d$, with
  \[
    d := \abs*{\left\{(A,B)\in\Sigma^2 \mid A\neq B\right\}} = \abs*{\Sigma}(\abs*{\Sigma}-1) \le 2^{2c}.
  \]
  Furthermore, for every view $\nu$ we have
  \[
    \sum_{(A,B)\in \Sigma^2} \abs*{x(\nu)_{AB}} \le \Delta.
  \]

  Among all nonnegative integer solutions of \eqref{eq:realizability-system}, we pick one that minimizes the number of non-zero coefficients.
  We use the following consequence of Carath\'eodory’s theorem: whenever a vector in $\mathbb{R}^d$ can be written as a non-negative linear combination of a family of vectors, it can be expressed as a non-negative combination of at most $d$ of them.
  Applying this to $x(\mu)$ and the family $\{x(\nu)\}$, we obtain a solution $(\beta_\nu)_\nu$ of \eqref{eq:realizability-system} such that there exists a set $J$ of at most $d$ views with $\beta_\nu>0$ for all $\nu \in J$, and $\beta_\nu = 0$ for all $\nu \notin J$.

  Restricted to $J$, equation \eqref{eq:realizability-system} becomes a square system of $d$ linear equations with integer coefficients.
  Moreover, by construction, the sum of the absolute values of the entries in any column is at most $\Delta$.

  By Cramer’s rule, each $\beta_\nu$ is a ratio of two determinants of $d \times d$ integer matrices.
  Although this yields non-negative \emph{rational} coefficients, this is sufficient for our purposes.
  Let $K$ be a common denominator of the numbers $\{\beta_\nu \mid \nu \in J\}$.
  Then $(K\beta_\nu)_\nu$ is a non-negative \emph{integer} solution of
  \begin{equation}\label{eq:realizability-system-K}
    Kx(\mu) = \sum_{\nu} \alpha_\nu x(\nu).
  \end{equation}

  By Hadamard’s inequality, the determinant of a $d \times d$ matrix $A$ is bounded by the product of the Euclidean norms of its columns.
  Since each column of $A$ has Euclidean norm at most $\Delta$, we obtain $\abs{\det(A)} \le \Delta^d$.
  Therefore, each denominator of $\beta_\nu$ is at most $\Delta^d$, and we may take $K \le \Delta^d$.
  It follows that $0 \le K\beta_\nu \le \Delta^{2d}$, and in particular
  \[
    \sum_{\nu} K\beta_\nu \le d \Delta^{2d}.
  \]

  Applying Lemma~\ref{lem:global-realizability}, we obtain a labeled yes-instance
  $(H,\ports,\ids,\ell)$ where $H$ has size
  \[
    F(\Delta,c) = d \Delta^{2d} k,
    \qquad\text{with } k = 2^{O(\Delta)}.
  \]
  and where all views belong to $\aviewsd$. Note that
  \[
    F(\Delta,c)
    = 2^{2^c} \cdot \Delta^{2^{2^c+1}} \cdot 2^{O(\Delta)}
    = 2^{2^{2^{O(c + \log \log \Delta)}}}.
  \]
  Assuming that $n \ge F(\Delta,c)$, all nodes accept in $\mathcal{H}_\mu$.
  By soundness of $\mathcal{D}$, $H$ is a yes-instance, and $\mu$ appears as a view in $\mathcal{H}_\mu$.
  This concludes the proof.
\end{proof}

\begin{remark}\label{rem:computing-realizable}
Lemma~\ref{lem:realizability-vector} has an important algorithmic consequence.
It implies that whether an accepting view $\mu$ can be witnessed is decidable by a computable procedure that depends only on $\Delta$ and $c$.
Indeed, if $\mu$ is authentic, then it admits a witness of size at most $F(\Delta,c)$, and conversely, if no such witness exists up to this bound, then $\mu$ is not authentic.
In particular, this decision procedure does not require any knowledge of the global parameter $n$.
\end{remark}

\subsection{Soundness implies strong soundness on normalized protocols}

We now prove that soundness implies strong soundness on normalized protocols. 

\begin{lemma}
  \label{lem:normalized-sound-implies-strong}
  Let $\mathcal{L}$ be a hereditary distributed language, and let
  $\mathcal{D}$ be an anonymous normalized single-round decoder for $\mathcal{L}$ on
  graphs of maximum degree~$\Delta$ with constant-size certificates.
  If $\mathcal{D}$ is sound for $\mathcal{L}$, then $\mathcal{D}$ is also
  strongly sound.
\end{lemma}

\begin{proof}
  Let $(G,\ports,\ids,\ell)$ be an arbitrary labeled instance.
  Define $S := \{ v \in V(G) \mid \mathcal{D}(v)=1 \}$
  to be the set of accepting nodes, and let $H := G[S]$ be the subgraph
  induced by~$S$.
  We aim to prove that $H$ is a yes-instance. 

  \paragraph{Step 1: Finding complements for the views in $S$.} 
  For each $v \in S$ write $\mu(v) := \view(v)$
  for its view, and define the multiset
  $\mathcal{M}_S := \{\{\mu(v) \mid v \in S\}\}$.
  All views in $\mathcal{M}_S$ are accepting views of $\mathcal{D}$.
  Let $X(\mathcal{M}_S)$  be its demand vector.  Under the stub interpretation, if we only look at the nodes of $S$ in
  $(G,\ports,\ids,\ell)$ and cut all edges from $S$ to $V(G)\setminus S$,
  the vector $X(\mathcal{M}_S)$ captures precisely the net imbalance of
  cross-certificate stubs $(A,B)$, $A\neq B$, i.e., the external demands
  that the nodes of $S$ impose toward neighbors outside~$S$.

  Since $\mathcal{D}$ is normalized, every accepting view is authentic.
  In particular, every $\mu \in \mathcal{M}_S$ is authentic.
  By Lemma~\ref{lem:realizability-vector}, for each such $\mu$ there exist
  nonnegative integers $(\lambda_\nu^{(\mu)})_{\nu \in \aviews}$
  such that
  \[
    x(\mu) + \sum_{\nu \in \aviews}
      \lambda_\nu^{(\mu)} x(\nu) = 0.
  \]
  For each $\mu \in \mathcal{M}_S$ define the multiset $
    \mathcal{Y}_\mu := \{\{\nu
      \text{ with multiplicity }\lambda_\nu^{(\mu)}\}\}.$
  By construction,
  \[
    X(\mathcal{Y}_\mu)
      = \sum_{\nu} \lambda_\nu^{(\mu)} x(\nu)
      = - x(\mu).
  \]

  Now consider the multiset $\overline{\mathcal{M}_S}$ obtained as the disjoint union (with multiplicities) of all views in $\mathcal{Y}_\mu$, for each $\mu \in \mathcal{M}_S$. Summing the equalities $X(\mathcal{Y}_\mu) = -x(\mu)$ over all
  $\mu \in \mathcal{M}_S$ yields
  \[
    X(\overline{\mathcal{M}_S})
      = \sum_{\mu\in\mathcal{M}_S} X(\mathcal{Y}_\mu)
      = -\sum_{\mu\in\mathcal{M}_S} x(\mu)
      = -X(\mathcal{M}_S).
  \]
  Intuitively, the views in $\overline{\mathcal{M}_S}$ contribute exactly the opposite
  demands to those arising from the views in~$\mathcal{M}_S$.

  \paragraph{Step 2: Realization of $\overline{\mathcal{M}_S}$.}
  We now apply Lemma~\ref{lem:global-realizability} to the multiset
  $\overline{\mathcal{M}_S}$.
  To be explicit, we obtain a labeled instance $(H',\ports',\ids',\ell')$, a
  multiset of stubs $S_{\mathrm{rem}}$, and a constant $k = 2^{O(\Delta)}$
  such that:
  \begin{itemize}
    \item Every node of $H'$ has an accepting view in $\overline{\mathcal{M}_S}$, and
          for each $\nu \in \overline{\mathcal{M}_S}$ the number of nodes
          $v \in V(H')$ satisfying $\view(v) = \nu$ is exactly
          $k$ times its multiplicity in $\overline{\mathcal{M}_S}$.
    \item If we attach to every $v \in V(H')$ the stubs prescribed by its
          view and remove one appropriate pair of stubs for each edge of
          $H'$, then the set of remaining stubs $S_{\mathrm{rem}}$
          satisfies
          \[
            X(S_{\mathrm{rem}}) = k\cdot X(\overline{\mathcal{M}_S})
              = -k\cdot X(\mathcal{M}_S),
          \]
          and contains no stubs of type $(A,A)$.
  \end{itemize}
  Thus $S_{\mathrm{rem}}$ represents (up to the factor $k$) exactly the
  opposite cross-certificate demands to those coming from the nodes of~$S$.

  \paragraph{Step 3: Attaching copies of $H$ to $H'$.}
  Now consider $k$ disjoint copies of $H = G[S]$,
  \[
    H^{(1)}, H^{(2)}, \dots, H^{(k)},
  \]
  each endowed with the restriction of $(\ports,\ids,\ell)$ to~$S$.
  Interpreting the views of nodes in $S$ in terms of stubs, each copy
  $H^{(j)}$ uses its internal edges to satisfy part of the local demands,
  and the edges from $S$ to $V(G)\setminus S$ are represented as external
  stubs.
  The net demand vector of these stubs in one copy of $H$ is exactly
  $X(\mathcal{M}_S)$, so the total demand of stubs over the $k$ copies is $k\cdot X(\mathcal{M}_S).$

  By construction,
  $X(S_{\mathrm{rem}}) + k\cdot X(\mathcal{M}_S) = 0$.
  That is, when we aggregate all stubs from $S_{\mathrm{rem}}$ and from
  the $k$ copies of $H$, the antisymmetric demand vector on pairs
  $(A,B)$, $A\neq B$, vanishes.
  As in Lemma~\ref{lem:global-realizability}, this allows us to match all
  stubs of type $(A,B)$ with those of type $(B,A)$ in a port-consistent
  way, creating edges that satisfy all cross-certificate demands exactly.

  The stubs of type $(A,A)$ are handled by the same explicit procedure
  used in the last step of the proof of
  Lemma~\ref{lem:global-realizability}: for each certificate $A$ and each
  view type, we perform a bounded blow-up and then form cliques on
  carefully chosen sets of copies, so that every node obtains exactly the
  prescribed number of neighbors with certificate $A$, without creating
  loops and without altering the degree pattern prescribed by its view.

  At the end of this matching and blow-up process we obtain a simple
  labeled graph $(H'',\ports'',\ids'',\ell'')$ in which:
  \begin{itemize}
    \item Each node comes either from one of the copies $H^{(j)}$ or from
          $H'$, and
    \item The view of every node in $H''$ coincides exactly with some
          view in $\mathcal{M}_S$ or $\overline{\mathcal{M}_S}$, all of which are
          belong to $\aviewsd$.
  \end{itemize}
  In particular, every node of $H''$ accepts under~$\mathcal{D}$.

All blow-up factors used in the construction (the coefficients
$\lambda_\nu$ from Lemma~\ref{lem:realizability-vector}, the integers
$M$ and $L$ introduced in the intermediate steps, and the constant
$k$ from Lemma~\ref{lem:global-realizability}) depend only on the fixed
constants $\Delta$ and $c$.  Consequently, there exists a constant $G = G(\Delta,c)$ such that, for every input graph $G$ on $n$ nodes, the graph $H''$ produced by the above construction satisfies
\[
   \abs*{V(H'')} \le G(\Delta,c)\cdot n .
\]

Recall that the certificates are designed for instances of size at most $n$,
while the nodes do not know the exact value of $n$ and only have access to an
upper bound $N$ on the size of the graph.

Since this upper bound $N$ is guaranteed to be polynomial in $n$, it suffices to
assume that $N \ge n^{1+\epsilon}$, for any fixed $\epsilon>0$ and all
sufficiently large $n$, in order to dominate the constant factor $G(\Delta,c)$ and have that $N\geq n\cdot G(\Delta,c)$. Therefore, by the soundness of $\mathcal{D}$, any labeled instance of size at
most $N$ that is accepted by every node and whose views are drawn from $\aviewsd$ must be a
yes-instance.

\paragraph{Step 4: Applying heredity.}
By the previous arguments, we have $H''$ belongs to $\mathcal{L}$.
Moreover, each copy $H^{(j)}$ of $H = G[S]$ appears as an induced subgraph of
$H''$, with the same edge structure and the same labeling.
Since $\mathcal{L}$ is hereditary, every induced subgraph of a graph in
$\mathcal{L}$ also belongs to $\mathcal{L}$.
It follows that $H = G[S]$ is a yes-instance.\\

  Since $(G,\ports,\ids,\ell)$ was arbitrary, this shows that for every
  labeled instance the subgraph induced by the set of accepting nodes is
  always a yes-instance of~$\mathcal{L}$.
  Therefore $\mathcal{D}$ satisfies strong soundness.
\end{proof}

\subsection{Putting everything together}
We are now ready to prove \Cref{thm:sound-implies-strong}.

\begin{proof}[Proof of Theorem~\ref{thm:sound-implies-strong}]
Let $\mathcal{L}$ be a hereditary distributed language, and assume that
$\mathcal{L}$ admits an anonymous single-round LCP $\mathcal{D}$ with
constant-size certificates on graphs of maximum degree~$\Delta$.
We first construct the normalized decoder $\mathcal{D}^{\mathrm{norm}}$
by restricting acceptance to authentic accepting views.

By Lemma~\ref{lem:normalization}, the decoder $\mathcal{D}^{\mathrm{norm}}$
is complete and sound for $\mathcal{L}$.
Moreover, if $\mathcal{L}$ is a language of network pairs $(G,x)$ and
$\mathcal{D}$ is hiding for $\mathcal{L}$, then
$\mathcal{D}^{\mathrm{norm}}$ is also hiding.

By Lemma~\ref{lem:normalized-sound-implies-strong}, any anonymous
normalized single-round decoder with constant-size certificates that is
sound for a hereditary language is strongly sound.
Applying this lemma to $\mathcal{D}^{\mathrm{norm}}$, we conclude that
$\mathcal{D}^{\mathrm{norm}}$ is strongly sound for $\mathcal{L}$.

Thus, $\mathcal{L}$ admits an anonymous single-round LCP with constant-size
certificates that is strongly sound.
Furthermore, if $\mathcal{L}$ is a language of network pairs and the
original decoder $\mathcal{D}$ is hiding, then the resulting strongly
sound decoder $\mathcal{D}^{\mathrm{norm}}$ preserves the hiding
property.
This completes the proof.
\end{proof}

\section{Impossibility in the order-invariant case}
\label{sec:oi-lb}

In this section, our goal is to show our impossibility result in the
order-invariant case (\cref{thm:oi-bip-impossibility}).
Before we can state the proof, we first need to set up some technical notions
regarding views and translating views into graphs (\cref{sec:oi-lb-realization}).  
Using these new tools, we then prove our theorem in \cref{sec:proof-oi-lb}.

\subsection{Realizing subgraphs of the accepting neighborhood graph}
\label{sec:oi-lb-realization}

Fix an $r$-round LCP $\mathcal{D}$.
We are interested in technical results of the following nature:
Given a subgraph $H$ of $\alvgd$ (possibly satisfying some requirements),
construct a graph $G_\bad$ that contains a subgraph $H'$ isomorphic to $H$ and
such that $V(H')$ is a subset of the nodes accepted by $\mathcal{D}$ in
$G_\bad$.
Here the subscript \enquote{bad} foreshadows the later application of these
observations in \cref{sec:proof-oi-lb}.
More specifically, if $H$ is some forbidden structure that $\mathcal{D}$ should
not accept, then the instance $G_\bad$ disproves the strong soundness of
$\mathcal{D}$.

First we introduce some notions concerning views.
Let $\mu_1, \mu_2 \in \mathcal{V}_r$ be two (not necessarily accepting) views
(that we here perceive as graphs).
Then $\mu_2$ is a \emph{subview} of $\mu_1$---in symbols $\mu_2 \subseteq
\mu_1$---if it is an induced subgraph (including identifiers, port numbers, and
labels) of $\mu_1$.
In addition, for a node $u$ that appears in both $\mu_1$ and $\mu_2$ (where we
identify nodes by their identifier assignment), we say that $\mu_1$ is
\emph{compatible with $\mu_2$ with respect to $u$} if there is a view $\mu_u$
such that $\view(\mu_1)(u) \subseteq \mu_u$ and $\view(\mu_2)(u) \subseteq
\mu_u$.
Finally, $\mu_1$ and $\mu_2$ are \emph{compatible} (without any qualification)
if they are compatible with respect to every node that appears in both views.
In particular, two views are compatible if and only if they coexist in some
graph.

The concept is related to the yes-instance-compatibility we defined in
\cref{sec:alvg}, but among other things here we do not require the views to be
accepting or to coexist in a yes-instance.
For an example, see \cref{fig:compatible-view-ex}.

\begin{figure}
  \centering
  \includestandalone[width=.75\textwidth]{figs/compatible_views}
  \caption{Example for compatible views $\mu_u = \view(u)$ and $\mu_v =
    \view(v)$ in a graph $G$ where $r=2$.
    The numbers indicate the nodes' identifiers.
    Port numbers and labels have been omitted for simplicity.
    Note how the edge between nodes $1$ and $4$ is not present in $\mu_1$ since
    it is not visible to $u$.
    Still this does not violate compatibility between the views with respect to,
    say, node $1$ since $\view(\mu_u)(1)$ has radius $2$ and hence contains
    nodes $1$, $9$, $7$, and $5$ but not $4$.}
  \label{fig:compatible-view-ex}
\end{figure}

Notice that compatibility is not transitive in general; that is, we may have
views $\mu_1$, $\mu_2$, and $\mu_3$ where $\mu_1$ and $\mu_2$ as well as $\mu_2$
and $\mu_3$ are compatible and yet there is no single graph where all three
views coexist.
See \cref{fig:realizability-ex} for an example.
More generally, it is \emph{not} the case that a walk $W = (\mu_1,\dots,\mu_m)$
where every pair of views $\mu_i$ and $\mu_{i+1}$ is compatible admits a
\emph{single} graph $G$ containing all of the $\mu_i$.

\begin{figure}
  \centering
  \includestandalone[width=.65\textwidth]{figs/compatibility_transitivity}
  \caption{An example for how compatibility of views fails to be transitive.
    As before, port numbers and labels are omitted for simplicity.
    The views $\mu_u$ and $\mu_v$ as well as $\mu_v$ and $\mu_w$ are compatible
    but $\mu_u$ and $\mu_w$ are not, and hence there is no single graph
    containing all three views.
    One explanation why this occurs is that, since node $4$ disappears from
    $\mu_u$ to $\mu_v$, we have essentially free hand in redefining its
    neighborhood in $\mu_w$ (e.g., we could even have changed its degree and
    compatibility with $\mu_v$ would still hold).
  }
  \label{fig:realizability-ex}
\end{figure}

In order to treat this concept formally, let $H$ be a subgraph of $\alvgd$, and
let $\mathcal{I} \subseteq [N]$ be the set of identifiers that appear in at
least some view of $H$.
For $i \in \mathcal{I}$, let $S(i) = S_H(i)$ denote the subgraph of $H$ induced
by the views of $H$ that contain a node with identifier $i$.
We say that $H$ is \emph{realizable} if it satisfies the following:
\begin{quote}
  For every identifier $i \in \mathcal{I}$, there is a \emph{single} (not
  necessarily accepting) view $\mu_i \in \mathcal{V}_r$ such that, for every
  view $\mu \in S(i)$ where a node $u_i$ with $\ids_\mu(u_i) = i$ appears, $\mu$
  is compatible with $\mu_i$ with respect to $u_i$.
\end{quote}
This definition comes with the following lemma, which motivates its name:

\begin{lemma}
  \label{lem:realizable}
  Let $\mathcal{D}$ be an $r$-round LCP.
  If $H$ is a realizable subgraph of $\alvgd$, then there is an instance
  $G_\bad$ with at most $N$ nodes that contains a subgraph isomorphic to $H$
  whose nodes are all accepted by $\mathcal{D}$.
\end{lemma}

We will not be needing this detail later on, but it is interesting to note that
we do not need the order-invariance of $\mathcal{D}$ for this particular result.

\begin{proof}
  First we make an quick observation regarding the property required in the
  statement.
  Let $\mathcal{C} \subseteq \mathcal{I}$ be the set of identifiers of center
  nodes in the views of $H$.
  Note that, for $i \in \mathcal{C}$, there is a unique view $\mu_i$ that
  satisfies the requirement for realizability.
  In particular, we have that there is a single view in $H$ with a center node
  having $i$ as its identifier and also that $\mathcal{D}(\mu_i) = 1$.

  Let us now describe the construction of $G_\bad$.
  Take a copy of each $\mu_i$ for $i \in \mathcal{I}$ and consider the graph
  $G_\base$ obtained from their (disjoint) union.
  (The identifiers in $G_\base$ may not be unique, but this is immaterial since
  $G_\base$ is simply an intermediate step in the construction.)
  We let $G_\bad$ be the graph obtained from $G_\base$ by identifying nodes
  possessing the same identifier; that is, for each $i \in \mathcal{I}$, the
  vertex with identifier $i$ in $G_\bad$ is the subset of $V(G_\base)$
  containing every $v \in V(G_\base)$ that has identifier $i$.
  To simplify matters, we let nodes in $G_\bad$ be synonymous with their
  identifiers (i.e., the node set of $G_\bad$ becomes simply $\mathcal{I}$).
  For $i,j \in \mathcal{I}$, we draw an edge between $i$ and $j$ in $G_\bad$ if
  and only if there is some view $\mu_k$ (where $k \in \mathcal{I}$) that
  contains nodes $v_i$ and $v_j$ having identifiers $i$ and $j$, respectively,
  that have an edge between them.
  Note that this is well-defined only because $H$ is realizable, that is,
  identified nodes in distinct views have suitable neighbors due to the
  compatibility between views.

  We extend $G_\bad$ with port and label assignments as specified by the
  $\mu_i$.
  For the label assignment $\ell_{G_\bad}$, we simply let $\ell_{G_\bad}(i) =
  \ell_{\mu_i}(i)$ for every $i$.
  Meanwhile, for the port assignments, we do the following: 
  For every $i \in \mathcal{I}$ and every edge $e = ij \in E(G_\bad)$, we let
  $\ports_{G_\bad}(i, e) = \ports_{\mu_i}(v_i, v_iv_j)$, where $v_i$ and $v_j$
  are the vertices with identifiers $i$ and $j$ in $\mu_i$, respectively.

  The identifier, port, and label assignments are well-defined by definition of
  $\mu_i$.
  What is left to argue is that $G_\bad[\mathcal{C}]$ is isomorphic to $H$:
  Let $i,j \in \mathcal{C}$ and consider $\mu_i$ and $\mu_j$ in $\alvgd$.
  If there is an edge between $\mu_i$ and $\mu_j$, then in $\mu_i$ we have a
  node $v_j$ with $\ids_{\mu_i}(v_j)$ and such that $v_iv_j \in E(\mu_i)$, where
  $v_i$ is the center node of $\mu_i$; hence there is an edge between $i$ and
  $j$ in $G_\bad$.
  Conversely, if there is no edge between $\mu_i$ and $\mu_j$ in $\alvgd$, then
  there is no such node $v_j$ in $\mu_i$, implying there is also no edge between
  $i$ and $j$ in $G_\bad$.
\end{proof}

Next we now show how to use the order-invariance of $\mathcal{D}$ to obtain the
conclusion of \cref{lem:realizable} based on a weaker assumption on $H$.
To that end, consider the following weakening of realizability, which we shall
refer to as \emph{component-wise realizability}:
\begin{quote}
  For every $i \in \mathcal{I}$ \emph{and for every connected component $C$ in
  $S(i)$,} there is a (not necessarily accepting) view $\mu_i(C) \in
  \mathcal{V}_r$ such that, for every view $\mu \in V(C)$ where a node $u_i$
  with $\ids_\mu(u_i) = i$ appears, $\mu$ is compatible with $\mu_i$ with
  respect to $u_i$.
\end{quote}

\begin{lemma}
  \label{lem:realizable-oi}
  Let $\mathcal{D}$ be order-invariant, and let $H$ be a component-wise
  realizable subgraph of $\alvgd$.
  In addition, suppose that $\abs{V(H)} \le \sqrt{N / \Delta^r}$.
  Then the conclusion of \cref{lem:realizable} holds; that is, there is an
  instance $G_\bad$ with at most $N$ nodes that contains a subgraph isomorphic
  to $H$ and whose nodes are all accepted by $\mathcal{D}$.
\end{lemma}

We stress that here (unlike the case of \cref{lem:realizable}) the assumption
that $\mathcal{D}$ is order-invariant is essential for the proof to go through.
Albeit the statement is rather technical, the idea for the proof is simple:
If $S(i)$ has more than one component, we simply replace $i$ in one of the
components with a fresh identifier that does not appear yet in $H$.
To do so, we must take care that we are not exceeding our budget of $N$
identifiers and also that we are respecting the local ordering of identifiers.

\begin{proof}
  First note that, if $S(i)$ has a single component for every $i$, then we may
  apply \cref{lem:realizable} directly.
  Hence it is sufficient to show a reduction to that case.
  Without restriction, we assume that $\mathcal{I} = [t]$ for some positive
  integer $t$ (i.e., there are no \enquote{gaps} in the set of identifiers
  appearing in $H$).

  Let thus $i \in \mathcal{I}$ be given such that $S(i)$ has two or more
  components, and pick some component $C$ of $S(i)$.
  For the sake of simplicity, consider first the case where $\mathcal{D}$ is
  anonymous.
  Then, for any view $\mu \in \aviewsd$ that contains some node $u_i$ with
  $\ids_\mu(u_i) = i$, the view $\mu'$ obtained by replacing the identifier of
  $u_i$ with some $i' \neq i$ is also in $\aviewsd$ (as long as $i' \le N$).
  If we perform this modification consistently for all views in the component
  $C$, then the subgraph $H'$ of $\alvgd$ that we obtain is isomorphic to $H$.
  Since there are at most $\Delta^r \abs{V(H)} \le N$ nodes that are visible
  across all the views in $H$, we can guarantee that the replacement $i'$ is
  smaller than $N$.

  For the general case where $\mathcal{D}$ is order-invariant we use the same
  strategy but take care to pick $i'$ adequately.
  More specifically, we associate every $i \in \mathcal{I}$ with the set $I_i =
  [(i-1) \abs{V(H)} + 1, i \abs{V(H)}]$.
  Since every $S(i)$ does not have more than $\abs{V(H)}$ components, this
  allows us to pick a different replacement $I_i$ (as described above) for every
  component of $S(i)$.
  Moreover, we note that, for any $i,j \in \mathcal{I}$, we have $i < j$ if and
  only if $i' < j'$ for every $i' \in I_i$ and $j' \in I_j$.
  Hence this replacement of identifiers preserves the local order of
  identifiers, and hence the behavior of $\mathcal{D}$ in the respective view is
  the same.
  Finally, for the bound on the number of nodes, as before, the largest
  identifier that appears in $H$ is $\Delta^r \abs{V(H)}$, and hence the largest
  identifier that we replace it with is $\Delta^r \abs{V(H)}^2 \le N$.
\end{proof}

Notice that, although we assumed $H$ to be a subgraph of $\alvgd$, it was not
necessary to use that we realize every view at most once.
In particular, we may use the same arguments to obtain the following:

\begin{lemma}
  \label{lem:realizable-oi-walk}
  Let $\mathcal{D}$ be order-invariant, and let $W$ be a component-wise
  realizable closed walk in $\alvgd$.
  In addition, suppose that $\abs{V(W)} \le \sqrt{N / \Delta^r}$.
  Then the conclusion of \cref{lem:realizable} holds; that is, there is an
  instance $G_\bad$ with at most $N$ nodes that contains a closed walk
  isomorphic to $W$ over nodes accepted by $\mathcal{D}$.
\end{lemma}

\subsection{Proving \texorpdfstring{\cref{thm:oi-bip-impossibility}}{Theorem
\ref{thm:oi-bip-impossibility}}}
\label{sec:proof-oi-lb}

We now turn to the proof of \cref{thm:oi-bip-impossibility} proper.
First it is necessary to specify what is meant by an $r$-forgetful graph.

\begin{definition}[$r$-forgetfulness]
  Let $r$ be a positive integer, $G$ a graph, $v \in V(G)$ a node of $G$, and $u
  \in N(v)$ a direct neighbor of $v$.
  A path $P = (v_0 = v, \dots, v_m)$ starting at $v$ is said to be
  \emph{$r$-forgetful to $u$} if, for every $w \in N^r(u)$ at distance $d =
  \dist(w,u) < \dist(w,v)$ from $u$, the distance to $u$ increases monotonically
  as we follow $P$ up to $v_{r-d}$; that is, we have $\dist(v_i,u) = d + i + 1$
  for every $i \le r-d$.
  In particular, note that $u \notin N^r(v_{r-d})$.
  By extension, we say $G$ is \emph{$r$-forgetful} if, for every $v \in V(G)$
  and $u \in N(v)$, there is an $r$-forgetful path to $u$ starting at $v$.
\end{definition}

We now restate and prove the main result of this section:

\restateThmOIBipImpossibility*

We already discussed the main ideas involved in the proof in
\cref{sec:intro-results-impossibility}.
To recall our approach, suppose we have a hiding LCP $\mathcal{D}$ for
$\twocol_\mathcal{G}$.
From the statement, $\mathcal{G}$ consists of $r$-forgetful connected graphs $G$
with $\delta(G) \ge 2$, and containing at least two cycles.
The main idea is to apply \cref{lem:hiding-char} and then construct a
counter-example to $\mathcal{D}$ being a strong LCP:
Fix $n$ such that $\mathcal{G}$ restricted to graphs on $n$ nodes is not empty.
Using \cref{lem:hiding-char}, we have that $\alvgd$ contains an odd cycle $C$.
Nevertheless, $C$ may arrange views in $\alvgd$ in a way that is not realizable
(see \cref{sec:oi-lb-realization}).
We show how to construct an odd \emph{closed walk} in $\alvgd$ that always
avoids this problem.

Before we delve into the technicalities, let us introduce a new notion for
closed walks in $\alvgd$:
We say a walk $W$ in $\alvgd$ is \emph{non-backtracking} if, for every view
$\mu$ in $W$, the successor $\mu_>$ and predecessor $\mu_<$ of $\mu$ in $W$ are
such that $\ids_{\mu_>}(v_>) \neq \ids_{\mu_<}(v_<)$, where $v_>$ and $v_<$ are
the center nodes of $\mu_>$ and $\mu_<$, respectively.
Notice that it is always possible to construct a non-backtracking walk in graphs
with minimum degree at least two simply by repeatedly choosing a successor that
differs from the predecessor of the currently visited node.
In addition, note that paths in $\alvgd$ may not be non-backtracking, but if we
take a path $P$ in a graph $G$ and consider the path induced by $P$ in $\alvgd$,
then that path is non-backtracking (since, by definition, $P$ does not visit
each node in $G$ more than once).

It is evident from the material in \cref{sec:oi-lb-realization} that being
non-backtracking is a necessary condition for the realizability of closed walks.
As we show next, this condition turns out to also be \emph{sufficient} if we are
in the class of graphs $\mathcal{G}$.

\begin{lemma}
  \label{lem:odd-walk-realizability}
  Let $\mathcal{D}$ be an $r$-round LCP for $\twocol_\mathcal{G}$.
  In addition, let $\alvgd$ contain a non-backtracking odd walk $W$.
  Then there is an odd closed walk $W'$ in $\alvgd$ that is component-wise
  realizable.
  Moreover, we have $\abs{V(W')} = O(n) \cdot \abs{V(W)}$.
\end{lemma}

The idea for the proof is to replace every edge of $W$ with a path long enough
so that the view we started with is entirely \enquote{forgotten}.

\begin{proof}
  For every edge $e = \mu_1\mu_2 \in E(W)$, fix a yes-instance $G_e \in
  \mathcal{G}$ on $n$ nodes that asserts that $\mu_1$ and $\mu_2$ are
  yes-instance-compatible.
  Let $v$ and $u$ denote the center nodes of $\mu_1$ and $\mu_2$, respectively.
  In addition, let $P_v = (v_0 = v, \dots, v_r)$ be $r$-forgetful path to $u$
  starting at $v$ (see \cref{sec:def-graphs}) and, similarly, let $P_u = (u_0 =
  u, \dots, u_r)$ be the $r$-forgetful path to $v$ starting at $u$.
  Construct a closed walk $W_e$ in $G_e$ by executing the following procedure:
  \begin{enumerate}
    \item Start at $u$ and take the edge $uv$ to $v$.
    \item Follow $P_v$ up to the node $v_r$.
    \item Since $\delta(G) \ge 2$, there is a path $P_{vu} = (v_r, a, \dots, b,
    u_r)$ where (possibly $a = b$ but) $a \neq v_{r-1}$ and $b \neq u_{r-1}$.
    Follow $P_{vu}$ from $v_r$ to $u_r$.
    \item Finally, follow the path $P_u$ in reverse to $u$, thus closing the
    walk.
  \end{enumerate}
  See \cref{fig:proof-lem-odd-walk-realizability} for an illustration.
  Notice that $W_e$ is a closed walk in the yes-instance $G_e$, which means it
  must be even.
  Since $W_e$ results from composing the edges and paths where we have taken
  care not to use the same edge twice in a row, it is non-backtracking.
  We lift $W_e$ to a closed walk $L_e$ in $\alvgd$ by associating every node
  visited by $W_e$ with its view in $\alvgd$.

  \begin{figure}
    \centering
    \includestandalone[width=.5\textwidth]{figs/odd_walk_realizability}
    \caption{Construction of the closed walk $W_e$ in the proof of
      \cref{lem:odd-walk-realizability}.
      The paths $P_v$, $P_u$, and $P_{uv}$ are marked using distinct colors.
      The arrows indicate the direction of the walk $W_e$.
    }
    \label{fig:proof-lem-odd-walk-realizability}
  \end{figure}

  Now consider the walk $W'$ in $\alvgd$ obtained from $W$ by inserting $L_e$
  before every $e$ in $W$.
  Since the $L_e$ are even closed walks and $W$ is odd, $W'$ is an odd closed
  walk.
  Furthermore, $\abs{L_e} = O(n)$ since every node in $G_e$ is visited by
  $W_e$ at most constantly many times.

  To obtain that $W'$ is non-backtracking, observe that every $L_e$ is
  non-backtracking since $W_e$ itself is non-backtracking by construction (i.e.,
  we take care not to use the same edge in $G_e$ twice in a row). 
  Using that $W$ is non-backtracking as well, the only node where we are
  potentially violating the property at all is the successor $\mu_v$ of the last
  node $\mu_u$ in $L_e = (\mu_u, \mu_v, \dots, \mu_{u_1}, \mu_u)$, where
  $\mu_u$, $\mu_{u_1}$ and $\mu_v$ are the views corresponding to $u$, $u_1$,
  and $v$, respectively.
  However, in this case it is clear that $\mu_{u_1} \neq \mu_v$ since $u_1$ and
  $v$ are distinct nodes in $G_e$.
  Hence $W'$ is non-backtracking.

  It remains to show that $W'$ is component-wise realizable.
  To that end, let $\mathcal{I}$ be the set of identifiers that appear in the
  views of $W'$, and let $i \in \mathcal{I}$ be some identifier.
  Take a component $K$ in $S(i)$.
  Notice that $K$ cannot contain any $L_e$ fully:
  If $L_e = (\mu_u, \mu_v, \dots, \mu_u, \mu_v)$ is fully contained in $K$, then
  $i$ appears in $\mu_u$; however, by construction, there is a view in $L_e$
  that has no node in common with $\mu_u$ (namely the one corresponding to $v_r$
  in the construction of $W_e$), so this view cannot be in $K$.

  Hence we can write $K = L_{e_0}' e_1 L_{e_1}'$ where $e_0$ and $e_1$ are
  consecutive edges in $W$, $L_{e_0}'$ is a suffix of $L_{e_0}$, and $L_{e_1}'$
  is a prefix of $L_{e_1}$.
  Let $G_0$ and $G_1$ be the graphs corresponding to $e_0$ and $e_1$,
  respectively.
  (That is, $G_i$ is the graph where the views that are the endpoints of $e_i$
  are adjacent in.)
  Letting $e_1 = \mu_u \mu_v$, notice that, since we are considering only
  bipartite graphs, one of the two center nodes of these two views must be
  closer to the respective node having identifier $i$ than the other.
  Hence there are two cases to consider:
  \begin{enumerate}
    \item If the node $u_i$ with identifier $i$ is closer to the center node $u$
    of $\mu_u$ than its counterpart in $\mu_v$, then consider the view $\mu_i$
    of $u_i$ in $G_{e_0}$.
    Since we are taking this view from $G_{e_0}$, all of the views in $L_{e_0}'$
    are automatically compatible with it with respect to $u_i$.
    At the same time, using the $r$-forgetfulness property we know that
    $L_{e_1}'$ must be a subpath of $P_v$, that is, $L_{e_1}' = (\mu_v,
    \mu_{v_1}, \dots, \mu_{v_j})$ for some $j \le r$ and also that
    \[
      \view(\mu_{v_j})(u_i) 
      \subseteq \view(\mu_{v_{j-1}})(u_i) 
      \subseteq \cdots 
      \subseteq \view(\mu_v)(u_i) 
      \subseteq \view(\mu_u)(u_i)
    \]
    Since the subview relation is transitive, this means that every view in
    $L_{e_1}'$ is compatible with $\mu_i$ with respect to $u_i$.
    \item The complementary case is the one where the node with identifier $i$
    is closer to the center node $v$ of $\mu_v$.
    This case is similar:
    We take the view $\mu_i$ of the node in $G_{e_1}$, in which case
    compatibility for the views in $L_{e_1}'$ is immediately guaranteed.
    In turn, for the views in $G_{e_0}$ we argue as before using that $P_u$ is
    an $r$-forgetful path and that the subview relation is transitive.
    \qedhere
  \end{enumerate}
\end{proof}

The final ingredient is showing how to find a non-backtracking odd cycle in
$\alvgd$.
Here it will be necessary to use that $\mathcal{G}$ contains only graphs with at
least two cycles, which is the one assumption that we have not used thus far.

\begin{lemma}
  \label{lem:non-backtracking-odd-cycle}
  Let $\mathcal{D}$ be an $r$-round LCP for $\twocol_\mathcal{G}$.
  If $\alvgd$ contains an odd cycle $C$, then it also contains a
  non-backtracking odd walk of length $O(n) \cdot \abs{C}$.
\end{lemma}

\begin{proof}
  Suppose there is a view $\mu \in V(C)$ that prevents $C$ from being
  non-backtracking; that is, $\mu$ has a successor $\mu_>$ and predecessor
  $\mu_<$ in $C$ such that $\ids_{\mu_>}(v_>) = \ids_{\mu_<}(v_<)$, where $v_>$
  and $v_<$ are the center nodes of $\mu_>$ and $\mu_<$, respectively.
  We show it is possible to replace the edge $e = \mu_< \mu$ with a
  non-backtracking odd walk $W = (\mu_0 = \mu_<, \dots, \mu_k = \mu)$ from
  $\mu_<$ to $\mu$ such that $\ids_{\mu_1}(v_1) = \ids_{\mu}(v)$ and
  $\ids_{\mu_{k-1}}(v_{k-1}) \neq \ids_{\mu_>}(v_>)$, where $v$, $v_1$, and
  $v_{k-1}$ are the center nodes of $\mu$, $\mu_1$, and $\mu_{k-1}$,
  respectively.

  By construction of $\alvgd$, there is a graph $G$ where the center nodes $v_>$
  and $v$ of the views $\mu_>$ and $\mu$ are adjacent.
  Let $G'$ denote the graph obtained from $G$ by deleting the edge between $v_>$
  and $v$.
  If $G'$ consists of a single connected component, then the removal of this
  edge decreases the number of cycles in $G'$ by at most one compared to $G$.
  If the removal of the edge disconnects $G'$ into two components, then there is
  still at least one cycle in either component.
  In either case, there is a cycle $C$ in the same component as $v$ in $G'$.
  Let $u$ be a node in $C$ that is at minimal distance from $v$.
  Since $G'$ is obtained from $G$ by deleting a single edge, $C$ is also a cycle
  in $G$ and it avoids $v_<$.
  Moreover, since $G$ is bipartite, it is an even cycle.

  Let $P_{vu}$ denote a path from $v$ to $u$ of minimal length in $G'$.
  Note $P_{vu}$ avoids edges of $C$.
  Consider the walk $W_G = (v_>v) P_{vu} C_u P_{uv}$ where $P_{uv}$ denotes
  $P_{vu}$ in reverse and $C_u$ the closed walk around $C$ starting and ending
  at $u$.
  Since $C$ is even, $W_G$ is an odd walk.
  We lift $W_G$ to a walk $W$ in $\alvgd$.
  Since $W$ originates from $W_G$, $W$ is non-backtracking.
  In addition, we have $\ids_{\mu_1}(v_1) = \ids_{\mu}(v)$ since $v_>v$ is the
  first edge of $W_G$ and $\ids_{\mu_{k-1}}(v_{k-1}) \neq \ids_{\mu_>}(v_>)$
  because $P_{vu}$ avoids $v_<$.
\end{proof}

We now combine all of the above to obtain the proof of
\cref{thm:oi-bip-impossibility}.

\begin{proof}[Proof of \cref{thm:oi-bip-impossibility}]
  Let $\mathcal{D}$ be a hiding order-invariant $r$-round LCP for
  $\twocol_\mathcal{G}$.
  Due to \cref{lem:hiding-char}, $\alvgd$ contains an odd cycle $C$.
  By \cref{lem:non-backtracking-odd-cycle}, we may as well assume that $C$ is
  non-backtracking.
  \Cref{lem:odd-walk-realizability} then gives us an odd closed walk $W$ that is
  component-wise realizable.
  Finally, by \cref{lem:realizable-oi-walk}, we have an instance $G_\bad \notin
  \twocolg$ containing $W$ as a subgraph and, moreover, this entire subgraph is
  accepted by $\mathcal{D}$.
  It follows that $\mathcal{D}$ is not strongly sound.
\end{proof}

\section{Impossibility of constant-size certificates in the non-anonymous case}
\label{sec:ramsey}

In this section, we show that, in general, there is no strong and hiding LCP
with constant-size certificates in the non-anonymous case. 
The argument is based on the fact that the existence of a hiding and strong LCP under these assumptions implies the existence of an equivalent LCP that satisfies the same properties but is also order-invariant. The core of the proof for this result is
inspired in a argument presented in \cite{balliu24_local_podc} which relies on a Ramsey-type theorem (see for example \cite{cholak01_strength_jsl,naor95_what_siamjc} for similar applications). We introduce Ramsey's theorem first and then we state and prove the first result of the section. 

In the next lines, we use the following notation: for fixed $n \in \N$ and some
set $S$ we define the set of all subsets of $S$ of cardinality $n$,  $S^{(n)} =
\{\{s_1,\hdots,s_n\}\mid s_1,\hdots s_n \in S.\}$ 


\begin{lemma}[Ramsey's Theorem] \label{lem:ramsey}
    Let $k,n \in \mathbb{N}$ and let $\mathcal{P}$ be a $k$-coloring of $\N^{(n)}$. There exists $s \in [k]$ and some infinite subset $Y \subseteq \N$ such that for all $I \in Y^{(n)}$ we have that $\mathcal{P}(I)=s$. 
\end{lemma}

The core of the proof will come from the lemma further above.
To prepare the statement, let $\Delta,r>0$ and let us call
$\mathcal{B}(\Delta,r)$ the class of graphs such that:
\begin{enumerate}
    \item  For all $G \in \mathcal{B}(\Delta,r)$ we have that $\Delta(G) \leq \Delta$ 
\item There exists some $G \in \mathcal{B}(\Delta,r) $ such that:
\begin{enumerate}
    \item $G$ is  a connected $r$-forgetful graph.
    \item $G$ is not a cycle
    \item  $\delta(G)\geq2.$
\end{enumerate}
\end{enumerate}
\begin{lemma} \label{lem:ordinvconstdeg}
        Let $\Delta,r>0.$  If there exists an $r$-round hiding and strong LCP $\mathcal{D}$ for $\twocol_{\mathcal{B}(\Delta,r)}$ using constant-size certificates then, there exists an $r$-round hiding, strong \emph{and order-invariant} LCP $\mathcal{D}'$ for $\twocol_{\mathcal{B}(\Delta,r)}$ using constant-size certificates.
\end{lemma}

\begin{proof}
        Let $\mathcal{D}$ be a hiding and strong decoder for $\twocol$ restricted to class $\mathcal{B}(\Delta,r)$.  We proceed following the same construction shown in \cite{balliu24_local_podc} to show that there exists some hiding and strong decoder $\mathcal{D}'$ for $\twocol$ restricted to class $\mathcal{B}(\Delta,r)$ which is order-invariant.
    
    Indeed, observe that, for some instance $G$, an id assignation $\ids$, some port assignation $\ports$ and some labeling $\ell$, the decoder $\mathcal{D}$ is a function which maps values in \[\view(v) = 
  \left(G_v^r, \ports|_{N^r(v)}, \ids|_{N^r(v)}, I|_{N^r}(v)\right)\] to $\{0,1\}$. Thus, this information can be seen as two separated elements: 
  \begin{enumerate}
      \item The identifier assignment $X = \ids|_{N^r(v)}$.
      \item The structure $S=(G_v^r,\ports|_{N^r},I|_{N^r}(v))$ which contains the information about the graph structure $G_v^r$ and the labels of the view contained in $I|_{N^r}(v)$.
  \end{enumerate}
  Thus, $\mathcal{D} = \mathcal{D}(X,S)$. We can see $\mathcal{D}$ as a function that assigns $X$ to some function $F(S)$ depending only on $S$. Formally, $\mathcal{D}(X,S) = \mathcal{D}(X)(S) = F(S)$. We use the same notation as in \cite{balliu24_local_podc} and we call $F(S)$ a \emph{type.} Observe that since the size of the certificates $c$ is constant and $\Delta(G) \leq \Delta$ there is a constant bound on the number of nodes in $\view(v)$ and a constant bound in the number of possible labelings.  More precisely, there are at most $s \leq \Delta^{r+1} $ nodes in $N^r(v)$ and since $\ell:V \mapsto \{0,1\}^c$ there are at most $2^{cs}$ possible labelings for $\view(v)$. Thus, there exists some constant $m$ such there are $m$ different types. We assume that types are numbered from $1$ to $m$. Observe that we can assume that all the identifier assignations have size $s$ (otherwise we can pad $X$ with extra identifiers). Let $\mathcal{X} = \N^{s}$ be the set of all possible id assignations for nodes in $N^r(v)$. 
 
  Observe that $\mathcal{D}$ induces an $m$-coloring $\mathcal{P}$ in $\mathcal{X}$ by assigning to $X \in \mathcal{X}$ some particular type $\mathcal{D}(X)$ between the $m$ possible choices. Observe that we can associate to each color $C \in [m]$ a particular canonical type $F$ such that for all id assignation $X$ satisfying $\mathcal{P}(X)=C$ we have that $\mathcal{D}(X)(S) = F(S)$. Thus, there are $m$ different of these canonical types $\{F_1,\hdots,F_m\}$. By \cref{lem:ramsey} there exists some infinite set of identifier assignations $I^*\subseteq \mathcal{X}^{(s)}$ and some canonical type $F^*$ such that for all $X \in I^{*(s)}$ we have $\mathcal{D}(X)(S)=F^*(S)$.

  It follows that for $\mathcal{D}$ there exists some infinite set of id assignations $B$ such that for any $X \in B$, $\mathcal{D}$ does not depend on the actual numerical value of the identifiers but only on the structure $S$.

  We would like to define a hiding decoder $\mathcal{D}'$ for $\twocol$ restricted to $\mathcal{B}(\Delta,r)$ that is also order-invariant. Given some instance $G \in \mathcal{B}(\Delta,r)$ together with an identifier assignation $\ids$, a port assignation $\ports$, and a labeling $\ell$, for all $v\in V$ decoder $\mathcal{D}'$ will consider the local order induced by $\ids$ in $N^r(v)$ and will assign identifiers to nodes in $N^r(v)$ according to some id assignment $\ids'$ in set $B$ 
  given by the previous application of \cref{lem:ramsey}. This id assignment $\ids'$ is such that the original order in $N^r(v)$ is preserved, i.e. for all $u,w \in N^r(v)$ we have that $\ids(w)<\ids(v) \iff \ids'(w) < \ids'(v).$ Observe that new assignment $\ids'$ exists since $B$ is infinite. Additionally, $\ids'$, depends only on the original (local) order induced by $\ids$, and thus $\mathcal{D}'$ depends solely on this order as well. Furthermore, note that any assignment in $B$ is mapped by $\mathcal{D}$ to some type $F^*(S)$; that is, we have $\mathcal{D}(X)(S) = F^*(S) $ for all $ X \in B $.
  Now, we would like to say that $\mathcal{D'}$ can simulate $\mathcal{D}$ in $G$ with this new id assignation. However, the problem we face now (as same as in \cite{balliu24_local_podc}) is $\mathcal{D}$ works with $\ids$ which maps each node to some value in $\{1,\hdots, \poly(n)\}$ but $B$ may contain assignations with very large numerical values.

We fix this by considering first some new instance $G'$ defined as $G$ together with some independent set of nodes $W$. Formally, we define $G' = G\cup W$ (and its corresponding id assignations and port assignations) where $W$ is a graph composed by  isolated nodes that are not connected to nodes in $V$. We also consider that size of $W$ is large enough so the identifiers for $G'$ are assigned from some set $\{1,\hdots, N^{'} \}$.  This $N^{'}$ is a large number that allow us to choose some $X \in B$ such that $X$ takes values in $\{1,\hdots, N^{'}\}$.  Observe also that $G' \in \mathcal{B}(\Delta,r)$. Now, we can correctly define $\mathcal{D}'$ as a decoder that first re-writes the original id assignation $\ids$ in $G$ and then, assigns ids according to some assignation $X \in B$ and then runs decoder $\mathcal{D}$ in $G'$. More precisely, we have that $\mathcal{D}'(v) = \mathcal{D}(v)$ for all $v \in V(G')$. Observe that $\mathcal{D}'(v)$ is order-invariant in the component $G$ of $G'$ by construction. We claim that $\mathcal{D}'$ is hiding. Indeed, observe that $\mathcal{D}'$ cannot be hiding in $W$ since any map $c:W\mapsto \{0,1\}$ is a proper coloring for $W$ and thus it is always possible to extract a valid witness for $W$. Since $\mathcal{D}$ is hiding then, $\mathcal{D'}$ must be hiding in $G$. For the strong part, we observe that since nodes in $W$ are always $2$-colorable and $\mathcal{D}$ is strong, then $\mathcal{D}'$ is strong too. We conclude that $\mathcal{D'}$ is hiding, strong and order invariant. The lemma holds.
\end{proof}

Finally, we get the desired result by using \cref{thm:oi-bip-impossibility}.
 We state the main result of the section:

\begin{theorem}[\cref{thm:full-bip-impossibility}, restated]
    Let $\Delta,r>0$ and $\mathcal{D}$ an $r$-round LCP for $\twocol_{\mathcal{B}(\Delta,r)}$ using certificates of constant size. If $\mathcal{D}$ is hiding, then $\mathcal{D}$ is not strong. 
\end{theorem}
\begin{proof}
Let $\mathcal{D}$ an $r$-round LCP for $\twocol$ restricted to the class $\mathcal{B}(\Delta,r)$ using certificates of constant size. Let us assume that $\mathcal{D}$ is hiding and strong. By \cref{lem:ordinvconstdeg} there exists an $r$-round hiding, strong and order-invariant LCP for $\twocol$ restricted to the class $\mathcal{B}(\Delta,r)$ using certificates of constant size which contradicts \cref{thm:oi-bip-impossibility}. 
\end{proof}

\section{Upper bounds in the non-anonymous case}
\label{sec:non-anon-ubs}

In this section, we prove our upper bounds in the non-anonymous setting. 
As already discussed in the introduction (\cref{sec:results-variable-cert}), our
results are twofold:
\begin{enumerate}
  \item For the standard $\twocol$ problem (i.e., the set of yes-instances is
  the entire class of bipartite graphs), we present a hiding and strongly sound
  single-round LCP $P_1$ with $O(\log n)$ certificate size.
  \item Since $P_1$ relies on graphs with diameter at least four in order for it
  to guarantee the hiding property, we also consider the promise problem
  $\twocol_{\clbipdiamle{3}}$ where the yes-instances are the class
  $\clbipdiamle{3}$ of bipartite graphs with diameter at most three.
  In this case we show that a hiding and strongly sound single-round LCP $P_2$
  still exists, though the certificate size becomes $O(n \log n)$.
\end{enumerate}
These two results are proven in the
\cref{sec:ub-variable-size,sec:ub-diam-three} that follow, respectively.

\subsection{General bipartite graphs}
\label{sec:ub-variable-size}

We restate the relevant theorem for the reader's convenience.

\restateThmUpperBoundVariableSize*

\begin{proof}
Let $G$ be a bipartite graph.
We define a single-round decoder $\mathcal{D}$ that expects a labeling $\ell$ of
$G$.
The labeling $\ell$ itself is one of two possible labelings:
\begin{enumerate}
  \item $\ell_1$, which simply reveals a $2$-coloring of $G$ and has certificate
  complexity $O(1)$.
  \item $\ell_2$ (described in more detail further below), which is valid only
  when $G$ has diameter four and has certificate complexity $O(\log n)$.
\end{enumerate}
We let $\ell_1$ and $\ell_2$ use distinct sets of labels so nodes can discern
which $\ell_i$ they see labels from.
The labeling is valid only if labels from a \emph{single} $\ell_i$ are used.
Nodes reject immediately if they see labels from both $\ell_i$ at the same time.

\begin{description}
  \item[Description of $\ell_2$.]
Assuming $\diam(G) \ge 4$, the labeling $\ell_2$ assigns labels to nodes so as
to partition $G$ into three components:
\begin{enumerate}
  \item A singleton component with a node $v \in V(G)$ that has eccentricity
  at least $4$.
  (Note we do not need to certify the eccentricity of $v$.)
  \item The neighborhood $N(v)$ of $v$.
  \item All nodes at distance at least $2$ from $v$.
\end{enumerate}
The identifier of $v$ is given to every node in $G$, which dominates the
complexity of the certificate.
In addition, having fixed an arbitrary $2$-coloring $c\colon V(G) \to \binalph$
of $G$, for every node $u \in V(G)$ of the third type (i.e., such that
$\dist(u,v) \ge 2$), we also reveal its own color $c(u)$ and the color $c(N(v))$
of the neighbors of $v$.
(Note it is crucial that we have the liberty of choosing an \emph{arbitrary} and
not a fixed $2$-coloring here.)

The verification procedure for a node $u \in V(G)$ is as follows:
\begin{enumerate}
  \item If $u = v$, then $u$ checks that it does not see any colors.
  \item If $u \in N(v)$, then either $u$ has degree one or $u$ has a neighbor
  $u' \neq v$.
  If we are in the latter case, then $u'$ must have a color and contain
  $c(N(v))$ in its certificate (otherwise $u$ can reject).
  The node $u$ then checks that the value of $c(N(v))$ is the same in all of
  $N(u) \setminus \{ v \}$ and also that it agrees with the colors assigned to
  the nodes in $N(u) \setminus \{ v \}$.
  \item Every other node $u$ checks that what the value it received for
  $c(N(v))$ is consistent with what its neighbors have and also that its own
  color $c(u)$ is consistent with that of its (colored) neighbors.
\end{enumerate}
In addition, every node also verifies that it has received the same value for
$\id(v)$ as its neighbors.
\end{description}
Finally, we analyze the properties of our decoder:
\begin{description}
  \item[Completeness.] 
  For any bipartite $G$, $\ell_1$ trivially certifies that $G$ is bipartite.
  \item[Strong soundness.]
  Let $G=(V,E)$ be given along with some labeling that induces a set of
  accepting nodes $A \subseteq V$, and let $C$ be a cycle in $G[A]$.
  If $G[A]$ is labeled with $\ell_1$, then the claim is trivial; hence we assume
  that $G[A]$ is labeled using $\ell_2$.
  Observe that every node $u \in V(C) \setminus N[v]$ receives a color $c(u)$
  that is consistent with its neighborhood.
  We consider two cases:
  \begin{enumerate}
    \item If $v$ is not a node of $C$, then every $u \in V(C) \cap N(v)$ is such
    that its two neighbors $u_1$ and $u_2$ in $C$ are unequal $v$. 
    Since $u$ is accepting, $u_1$ and $u_2$ agree on their colors $c(u_1) =
    c(u_2)$ as well as on their value for $c(N(v)) = 3 - c(u_1)$.
    Hence we may take the proper $2$-coloring $c$ of $V(C) \setminus N[v]$ and
    extend it in a consistent manner (i.e., so that it remains a proper
    $2$-coloring) to also include the set $N(v)$. 
    This yields a $2$-coloring on $V(C) = (V(C) \setminus N[v]) \cup N(v)$. 
    Therefore $C$ admits a proper $2$-coloring, which implies it is an even
    cycle.
    \item Otherwise, $v$ is a node of $C$.
    By a similar reasoning as in the previous case, every node $u \in N(v)$ that
    is not a predecessor or successor of $v$ in $C$ must be a neighbor of two
    nodes on $C$ that agree both on their colors and their value for $c(N(v))$. 
    Hence we can extend the $2$-coloring $c$ consistently to a $2$-coloring of
    $V(C) \setminus \{ v, u_1, u_2 \}$, where $u_1, u_2 \in N(v)$ are the
    predecessor and successor of $v$ in $C$, respectively.
    Moreover, it follows that the value for $c(N(v))$ is the same for all nodes
    in $V(C) \setminus N[v]$; otherwise one of the aforementioned nodes in
    $(N(v) \cap V(C)) \setminus \{ v,u_1,u_2 \}$ would be rejecting.
    Hence we can color $c(u_1) = c(u_2) = c(N(v))$ in a manner consistent with
    the neighborhoods of $u_1$ and $u_2$ and thus also $c(v) = 3 - c(N(v))$.
    As before, this gives a proper $2$-coloring of $C$ and hence that it is
    even.
  \end{enumerate}
  \item[Hiding.] 
  Given \cref{lem:hiding-char}, we only need to exhibit an odd cycle in the
  accepting local view graph. This odd cycle is  \Cref{fig:oddcycledegree1} using as $\id$'s the values inside the nodes. Note that the two instances of \Cref{fig:instancesdeg1} have diameter at least $4$.
  \qedhere
\end{description}
\end{proof}

\subsection{Bipartite graphs with diameter at most three}
\label{sec:ub-diam-three}

As mentioned in the previous section, the protocol from
\cref{thm:upper-bound-variable-size} depends on an instance with diameter four
in order to guarantee the hiding property.
In this section, we prove it is also possible to obtain a protocol when the
yes-instances are restricted to having diameter at most three---at the cost of
blowing up the certificate complexity:

\restateThmUpperBoundDiamThree*

In particular, to ensure the hiding property, we are interested in what we dub
\emph{$2$-wise graphs}.
We say a graph $G=(V,E)$ is said to be \emph{$2$-wise} if there exist two
\emph{wise nodes} $a,b\in V$ such that $G$ can be partitioned into $N(a)$ and
$N(b)$, that is, we have $ab\in E$, $V=N(a) + N(b)$, and $xy \notin E$
whenever $x$ and $y$ are both in $N(a)$ or both in $N(b)$.
See \Cref{fig:2wise} for a graphical example.

\begin{figure}
  \centering
  \includestandalone{figs/2-wise}
  \caption{Illustration of a $2$-wise graph $G$ with wise nodes $a$ and $b$.
  The sets $N(a)$ and $N(b)$ form a partition of $G$.}
  \label{fig:2wise}
\end{figure}

As shown in the following lemma, $2$-wise graphs form a class in-between
(bipartite) graphs with diameter two (i.e., complete bipartite graphs) and
graphs with diameter three.

\begin{lemma}\label{lem:relations-wise-graphs}
  Let $\clwise$ be the class of $2$-wise graphs, $\clbipdiameq{2}$ the
  class of bipartite graphs with diameter $2$, and $\cldiameq{i}$ the class of
  graphs with diameter $i$. Then the following holds:
  \begin{enumerate}
    \item $\clbipdiameq{2} \subset \clwise$.
    \item $\left(\cldiameq{2} \setminus \clbipdiameq{2}\right)\cap
    \clwise = \emptyset.$
    \item $\clwise \subset \cldiameq{3}$.
  \end{enumerate}
\end{lemma}
\begin{proof}
For the first item, let $G=(V,E)$ be a bipartite graph of diameter $2$, and let
$A\dot{\cup} B =V$ be a partition of the nodes. 
Since $G$ has diameter $2$, this means that $G$ is the complete bipartite graph
$K_{A,B}$ (since otherwise there are nodes $a \in A$ and $b \in B$ that are not
connected by an edge and hence have distance $3$).
Therefore, by picking any two nodes $a\in A$ and $b\in B$, we can partition the
vertex set as $V =  N(a) + N(b)$ and $ab\in E$.

For the second item, suppose that there exists a non-bipartite graph $G=(V,E)$ of diameter $2$ such that $G\in \clwise$. Then there exist two nodes $a,b\in V$ such that $ab\in E$ and $V=N(a)+ N(b)$, but then $G$ is bipartite with sides $N(a)$ and $N(b)$, a contradiction.

Finally, for the last item, let $G$ be a $2$-wise graph and $a,b\in V$ wise nodes. Given any two nodes $u,v\in V$, if $u,v\in N(a)$ or $u,v\in N(b)$ then they are at distance at most $2$. If not, then $u\in N(a)$ and $v\in N(b)$, but then $(u,a,b,v)$ is a valid path between $u$ and $v$ and therefore they are at distance at most $3$. 
\end{proof}

We now turn to the proof of \cref{thm:upper-bound-diam-three}.
As in \cref{thm:upper-bound-variable-size}, we focus on designing a
certification scheme that is hiding on a subset of yes-instances, namely
$2$-wise graphs.
The scheme takes a partitioning of the nodes of $G$ into $A$ and $B$, places
special symbols to indicate wise nodes, and reveals to the remaining nodes the
full list of identifiers to the nodes that are located on the same partition as
itself (i.e., if $u \in A$, then $u$ receives a full list of the identifiers in
$A$).
The hiding property stems from the fact that this does not give wise nodes any
information beyond what they already learn from the view.
Meanwhile, the strong soundness follows from a blend of an argument similar to
the one presented in \cref{thm:upper-bound-variable-size} along with structural
properties of $2$-wise graphs.

\begin{proof}
Let $G$ be bipartite with $\diam(G) \le 3$.
As in the proof of \cref{thm:upper-bound-variable-size}, we define a
single-round decoder $\mathcal{D}$ that expects a labeling $\ell$ of $G$ that
itself is one of two possible labelings:
\begin{enumerate}
  \item $\ell_1$, which simply reveals a $2$-coloring of $G$ and has
  certificate complexity $O(1)$.
  \item $\ell_2$, which is valid only when $G$ is a $2$-wise graph and whose
  certificate complexity is $O(n \log n)$.
\end{enumerate}
Again, we use distinct sets of labels for either $\ell_i$ and which are
disallowed to coexist in the same labeled instance.

\begin{description}
\item[Description of $\ell_2$.] 
Assuming $G$ is a $2$-wise graph, each $u \in V(G)$ is labeled using (only)
one of the two following possibilities:
\begin{enumerate}
  \item A special symbol $\bot$, indicating that node $u$ is a \emph{wise}
  node (and nothing else).
  \item A (sorted) list $L(u)$ of the identifiers of all nodes in its own
  side of a (supposedly) valid partitioning of the nodes into two disjoint
  sets $A$ and $B$. 
\end{enumerate}
The certificate clearly has size $O(n\log n)$. 
The verification procedure at $u \in V(G)$ is as follows:
\begin{enumerate}
\item Exactly one node in $N[u]$ (i.e., possibly $u$ itself) was selected as a
wise node.
\item All neighbors $v\in N(u)$ receive the same list $L(v)$ and, in case $u$
is not a wise node and received a list $L(u)$, then $L(u)$ and $L(v)$ have no
element in common.
\item If $u$ is a wise node, then it verifies that, for each $v \in N(u)$, the
list $L(v)$ that $v$ received correctly lists all the identifiers in $N(u)$
(and nothing else).
\item If $u$ is not a wise node, then it verifies that, for each $v \in N(u)$,
the list $L(v)$ that $v$ received contains all identifiers it sees in $N(u)$.
\end{enumerate}
\end{description}

Now we analyze the properties of our decoder.

\begin{description}
\item[Completeness.] 
For $G \in \clbipdiamle{3}$, any $2$-coloring of $G$ trivially yields a valid
labeling under $\ell_1$.

\item[Strong soundness.] Let $G=(V,E)$ be a (labeled) graph, $A\subseteq V$
the set of accepting nodes of $G$, and $H=G[A]$ the subgraph induced by $A$.
If $G[A]$ is labeled using $\ell_1$, then it is trivially bipartite; hence we
assume that it is labeled with $\ell_2$.
We prove that every cycle $C = (v_1, \dots, v_k, v_1)$ of $H$ has even length.
We say that the node $v_i$ is in position $i$.

Let us first deal with the simple cases.
If $C$ does not contain a wise node, then, since $L(v_{i-1})$ and $L(v_{i+1})$
must be identical and disjoint from $L(v_i)$ in order for $v_i$ to accept, we
have that $L(v_1) = L(v_3) = \dots$ and $L(v_2) = L(v_4) = \dots$ are disjoint
and hence induce a $2$-coloring on $C$.
Similarly, if $C$ contains a single wise node $v_i$, then this argument induces
a $2$-coloring of the path obtained by excluding $v_i$ from $C$ and, since $v_i$
is a single node, we can extend the $2$-coloring to $v_i$ appropriately (in a
similar fashion to the proof of \cref{thm:upper-bound-variable-size}).

Hence the interesting case is the one where $C$ has at least two nodes marked as
wise. 
Let $W_C$ as the set of wise nodes in $C$.
If we remove $W_C$ from $C$, then we obtain a collection $\mathcal{P}_C$ of
node-disjoint paths.
Then we have
\begin{equation}\label{eq:keven}
  k = \abs{W_C} + \sum_{P\in \mathcal{P}_C} \abs{P}.
\end{equation}
In addition, observe that $\abs{P}$ is even for every $P \in \mathcal{P}_C$.
This is because, if $P = (u_1,\dots,u_j)$, then we can argue as before that
$L(u_1) = L(u_3) = \dots$ and $L(u_2) = L(u_4) = \dots$ induce a $2$-coloring of
$P$.
This means that, if $j$ is odd, then $L(u_j) = L(u_1)$ and hence there can only
be a single wise node in $C$ (which is a case we already handled above).

Thus it is sufficient to prove that $\abs{W_C}$ is even.
In fact, we can show that $\abs{W_C} = 2$.
To that end, consider the case where we have three successive nodes $w_1$,
$w_2$, and $w_3$ in $W_C$.
Since $w_2$ is accepting, it cannot be a neighbor of both $w_1$ and
$w_3$.
Hence there is, say, a path $P = (u_1,\dots,u_j) \in \mathcal{P}_C$ between
$w_1$ and $w_2$.
(The other case is similar.)
Arguing as before, we have $L(u_1) = L(u_3) = \dots$ and $L(u_2) = L(u_4) =
\dots$ and, since $u_1$ is accepting and $j$ is even, the identifier of
$w_1$ appears in $L(u_2) = L(u_j)$.
Letting $v$ be the node that comes after $w_2$ in $C$, we then have two
cases:
\begin{enumerate}
  \item If $v$ is a wise node, then its identifier must be in $L(u_j)$.
  Hence $v = w_1$ (as otherwise $w_2$ would see two wise nodes and not
  be accepting), and $\ell = 2$ follows.
  \item Otherwise we have $L(v) = L(u_j)$.
  Let $P' = (v_1,\dots,v_s) \in \mathcal{P}_C$ be the path coming after
  $w_2$ in $C$.
  Then again $L(v_1) = L(v_{s-1})$ must contain the identifier of $w_1$.
  It follows that $v_s$ must be a neighbor of $w_1$, in which case
  $w_3 = w_1$, and thus $\ell = 2$ follows.
\end{enumerate}

\item[Hiding.] Given \Cref{lem:hiding-char}, we only need exhibit an odd cycle in the local view graph. Such an odd cycle is shown in \Cref{fig:c15} derived from the instances shown in \Cref{fig:wiseinstances}.
\qedhere
\begin{figure}[p]
  \centering
  \begin{subfigure}{\textwidth}
    \includestandalone[width=\textwidth]{figs/c15hidingnew}
    \caption{Odd cycle $C_{15}$. The identifiers are above the nodes, and the label of each node is written in red, and the blue node $v$ in each view $\mu$ corresponds to the node such that $\view(v) = \mu$. }
    \label{fig:c15}
  \end{subfigure}

  \vspace*{1em}

  \begin{subfigure}{\textwidth}
    \centering
    \includestandalone[width=\textwidth]{figs/wise_instances}
    \caption{The accepting instances from where \Cref{fig:c15} is derived. Labels are written with red and $\id$'s with black above the nodes.}
    \label{fig:wiseinstances}
  \end{subfigure}
  \caption{Proof of the hiding property for the protocol from
  \cref{thm:upper-bound-diam-three}.}
\end{figure}

\end{description}
\end{proof}

\ifanon\else
\section*{Acknowledgments}
We thank Jukka Suomela, Massimo Equi, and Sebastian Brandt for very fruitful
discussions.
We would also like to thank the remaining participants of the RW-DIST 2024
workshop that were involved in coming up with the upper bound construction for
the case of a single cycle (\cref{sec:ub-cycle}), which formed the cornerstone
case for this project.

Parts of this work were done while Pedro Montealegre and Martín Ríos-Wilson were
visiting Aalto University.
We acknowledge the support of the Research Council of Finland, Grant 363558.

Augusto Modanese was partly supported by the Research Council of Finland, Grant
359104.
Most of this work was done while he was affiliated with Aalto University.
\fi 

\section*{Disclosure}
Large language models were used to generate the base code for some of the TikZ
illustrations.
No other AI tools were employed in the writing of this paper.

\printbibliography

\end{document}

%% file: figs/forgetful_torus.tex
\begin{tikzpicture}[every node/.style={circle,draw,inner sep=2pt, minimum size=6mm}]

  \draw[draw=none,fill=stgogreen!50!white] 
    (0.5,-2.5) -- ++(0,1) -- ++(1,0) -- ++(0,1) -- ++(1,0) -- ++(0,1)
      -- ++(-1,0) -- ++(0,1) -- ++(-1,0) -- ++(0,1) -- ++(-2,0)
      -- ++(0,-1) -- ++(-1,0) -- ++(0,-1) -- ++(-1,0) -- ++(0,-1)
      -- ++(1,0) -- ++(0,-1) -- ++(1,0) -- ++(0,-1);

  \def\xmin{-4}
  \def\xmax{3}
  \def\ymin{-3}
  \def\ymax{3}

  \foreach \x in {\xmin,...,\xmax} {
    \foreach \y in {\ymin,...,\ymax} {
      \node (n\x\y) at (\x,\y) {};
    }
  }

  \foreach \x in {\xmin,...,\xmax} {
    \foreach \y in {\ymin,...,\ymax} {
      \pgfmathtruncatemacro{\xp}{\x+1} 
      \ifnum \xp>\xmax
      \else
        \draw (n\x\y) -- (n\xp\y);
      \fi
    }
  }
  
  \foreach \x in {\xmin,...,\xmax} {
    \foreach \y in {\ymin,...,\ymax} {
      \pgfmathtruncatemacro{\yp}{\y+1}
      \ifnum \yp>\ymax
      \else
        \draw (n\x\y) -- (n\x\yp);
      \fi
    }
  }

  \foreach \y in {\ymin,...,\ymax} {
    \draw[dash pattern=on 0.5pt off 2pt, line cap=round] (n\xmin\y) -- ++(-0.8,0);
  }
  \foreach \y in {\ymin,...,\ymax} {
    \draw[dash pattern=on 0.5pt off 2pt, line cap=round] (n\xmax\y) -- ++(0.8,0);
  }
  \foreach \x in {\xmin,...,\xmax} {
    \draw[dash pattern=on 0.5pt off 2pt, line cap=round] (n\x\ymin) -- ++(0,-0.8);
  }
  \foreach \x in {\xmin,...,\xmax} {
    \draw[dash pattern=on 0.5pt off 2pt, line cap=round] (n\x\ymax) -- ++(0,0.8);
  }

  \node[circle,fill=stgoblue,font=\bfseries] (u) at (-1,0) {$u$};
  \node[circle,fill=stgogreen,font=\bfseries] (v) at (0,0) {$v$};

  \node[circle,fill=stgoorange] at (-1,-1) {};
  \node[circle,fill=stgoorange] at (-1,-2) {};
  \node[circle,fill=stgoorange] at (-2,-1) {};
  \node[circle,fill=stgoorange] at (-2,0) {};
  \node[circle,fill=stgoorange] at (-3,0) {};
  \node[circle,fill=stgoorange] at (-1,1) {};
  \node[circle,fill=stgoorange] at (-1,2) {};
  \node[circle,fill=stgoorange] at (-2,1) {};

  \node[circle,fill=stgogreen,font=\bfseries] (v1) at (1,0) {$v_1$};
  \node[circle,fill=stgogreen,font=\bfseries] (v2) at (2,0) {$v_2$};
  \node[circle,fill=stgogreen,font=\bfseries] (v3) at (3,0) {$v_3$};

\end{tikzpicture}

%% file: figs/basic_view.tex
\begin{tikzpicture}[node distance=1cm, auto]
  \tikzstyle{node_style}=[circle, draw, inner sep=2pt, minimum size=6mm]
  \tikzstyle{edge_style}=[draw, thick, -]
  \tikzstyle{center_style}=[fill=stgoblue!20, very thick]
  \tikzstyle{layer_one}=[fill=stgogreen!25]
  \tikzstyle{layer_two}=[fill=stgoorange!20]
  \tikzstyle{outside_style}=[fill=gray!15]

  \begin{scope}[shift={(0,0)}]
    \node[anchor=south] at (-2,-1) {\LARGE $G$};

    \node[node_style, center_style] (v) at (0,0) {$v$};
    \node[node_style, layer_one, above left=of v] (a) {};
    \node[node_style, layer_one, above right=of v] (b) {};
    \node[node_style, below=of v] (c) {};

    \node[node_style, layer_two, above left=of a] (x) {$x$};
    \node[node_style, layer_two, above right=of b] (z) {};
    \node[node_style, layer_two] (y) at ($(x)!0.5!(z) + (0,0.2)$) {$y$};

    \node[node_style, outside_style, above left=of x] (p) {};
    \node[node_style, outside_style, above right=of z] (q) {};

    \draw[edge_style] (v) -- (a);
    \draw[edge_style] (v) -- (b);
    \draw[edge_style] (v) -- (c);
    \draw[edge_style] (a) -- (x);
    \draw[edge_style] (a) -- (y);
    \draw[edge_style] (b) -- (y);
    \draw[edge_style] (b) -- (z);
    \draw[edge_style] (x) -- (y);
    \draw[edge_style] (x) -- (p);
    \draw[edge_style] (z) -- (q);

    \draw[dashed, gray] (v) circle (4cm);
  \end{scope}

  \begin{scope}[shift={(9,0)}]
    \node[anchor=south] at (-2,-1) {\LARGE $G_v^r$};

    \node[node_style, center_style] (v2) at (0,0) {$v$};
    \node[node_style, layer_one, above left=of v2] (a2) {};
    \node[node_style, layer_one, above right=of v2] (b2) {};
    \node[node_style, below=of v2] (c2) {};

    \node[node_style, layer_two, above left=of a2] (x2) {$x$};
    \node[node_style, layer_two, above right=of b2] (z2) {};
    \node[node_style, layer_two] (y2) at ($(x2)!0.5!(z2) + (0,0.2)$) {$y$};

    \draw[edge_style] (v2) -- (a2);
    \draw[edge_style] (v2) -- (b2);
    \draw[edge_style] (v2) -- (c2);
    \draw[edge_style] (a2) -- (x2);
    \draw[edge_style] (a2) -- (y2);
    \draw[edge_style] (b2) -- (y2);
    \draw[edge_style] (b2) -- (z2);

    \draw[edge_style, dashed, stgored] (x2) -- (y2);

    \draw[dashed, gray] (v2) circle (4cm);
  \end{scope}
\end{tikzpicture}

%% file: figs/instances_deg1.tex
\begin{tikzpicture}
    \colorlet{colorLabel}{stgored}
    \tikzstyle{nodeStyle}=[circle, draw, fill=white, inner sep=2pt,
      minimum size=5ex]
    \tikzstyle{labelStyle}=[draw=none, minimum size=0ex, color=colorLabel]

    \draw[fill=stgogreen!10] (-1.25,2.25) rectangle (10.8,-2);
    \node at (-0.75,1.75) {\Large $I_1$};

    \draw[fill=stgoorange!10] (-0.75,-1.25) rectangle (3-0.5,1.25);

    \draw[fill=stgoorange!10] (-0.75+6.2,-1.25) rectangle (3+7.5,1.25);
    
    \node[nodeStyle, text=red] (1) at (0,0) {$u_1$};
    \node[text=red] at (0,0.7) {$\bot$};
    \node[nodeStyle] (2) at (2,0) {$u_2$};
    \node[] at (2,0.7) {$\top$};

    \node[nodeStyle] (3) at (4,0) {$u_3$};
    \node[] at (4,0.7) {$0$};
    \node[nodeStyle] (4) at (6,0) {$u_4$};
    \node[] at (6,0.7) {$1$};
    \node[nodeStyle,text=red] (5) at (8,0) {$u_5$};
    \node[] at (8,0.7) {$0$};
    \node[nodeStyle] (6) at (10,0) {$u_6$};
    \node[] at (10,0.7) {$1$};

    \draw (1) -- node[above]{$p_1$} (2);
        \draw (1) -- node[below]{$p'_1$} (2);
    \draw (2) -- node[above]{$p_2$} (3);
        \draw (2) -- node[below]{$p'_2$} (3);

    \draw (3) -- node[above]{$p_3$} (4);
        \draw (3) -- node[below]{$p'_3$} (4);

    \draw (4) -- node[above]{$p_4$} (5);
        \draw (4) -- node[below]{$p'_4$} (5);

    \draw (5) -- node[above]{$p_5$}  (6);
        \draw (5) -- node[below]{$p'_5$}  (6);

    \begin{scope}[shift={(0,-5)}]
      \draw[fill=stgoblue!10] (-1.25,2.25) rectangle (10.8,-2);
      \node at (-0.75,1.75) {\Large $I_2$};

      \draw[fill=stgoorange!10] (-0.75,-1.25) rectangle (3-0.5,1.25);
          \draw[fill=stgoorange!10] (-0.75+4.2,-1.25) rectangle (3+5.5,1.25);
    \node[nodeStyle, text=red] (1) at (0,0) {$u_1$};
     \node[text=red] at (0,0.7) {$\bot$};
    \node[nodeStyle] (2) at (2,0) {$u_2$};
        \node[] at (2,0.7) {$\top$};
    \node[nodeStyle] (3) at (4,0) {$u_4$};
        \node[] at (4,0.7) {$1$};
    \node[nodeStyle, text=red] (4) at (6,0) {$u_5$};
            \node[] at (6,0.7) {$0$};
    \node[nodeStyle] (5) at (8,0) {$u_6$};
            \node[] at (8,0.7) {$1$};

    \node[nodeStyle] (6) at (10,0) {$v_7$};
            \node[] at (10,0.7) {$0$};

    \draw (1) -- node[above]{$p_1$} (2);
    \draw (1) -- node[below]{$q_1$} (2);
    \draw (2) -- node[above]{$q'_2$} (3);
    \draw (2) -- node[below]{$q_2$} (3);
    \draw (3) -- node[above]{$p'_4$} (4);
    \draw (3) -- node[below]{$p_4$} (4);
    \draw (4) -- node[above]{$p'_5$} (5);
        \draw (4) -- node[below]{$p_5$} (5);

    \draw (5) -- node[above]{$q'_5$}  (6);
        \draw (5) -- node[below]{$q_5$}  (6);

    \end{scope}
\end{tikzpicture}

%% file: figs/prueba.tex
\begin{tikzpicture}
\colorlet{colorCenter}{stgoorange}
\colorlet{colorLabel}{stgored}

\def\radiusA{10cm}
\def\radiusB{5cm}
\def\rotacion{0}

\def\distNeighb{10mm}        
\def\InnerCircleR{10mm}      
\def\NeighbPad{1.5mm}       
\def\LabelYOffset{7mm}      

\def\BBx{18mm}
\def\BBy{10mm}

\pgfmathsetlengthmacro{\HalfDist}{0.9*\distNeighb}

\tikzset{
  nodeStyle/.style={
    circle, draw, fill=white,
    inner sep=1.2pt,
    font=\Large
  },
  neighbNodeStyle/.style={
    circle, draw, fill=white,
    inner sep=\NeighbPad
  },
  redLabel/.style={
    text=colorLabel,
    font=\Large,
    overlay
  }
}

\foreach \name/\angle in {
  u0/190,
  u1/140,
  u2/100,
  u3/55,
  u4/-20,
  u5/-65,
  u6/-105
} {
  \coordinate (\name) at ({\radiusA*cos(\angle + \rotacion - 25)},
                          {\radiusB*sin(\angle + \rotacion - 25)});
}

\def\radiuss{5.8cm and 2.2cm}
\def\radiusl{11.2cm and 6.2cm}

\draw[draw=none,fill=stgoorange!10] (185:\radiuss) arc (185:-200:\radiuss)
      -- (-200:\radiusl) arc (-200:185:\radiusl) -- cycle;

\draw[draw=none,fill=stgoblue!10] (145:\radiuss) arc (145:-15:\radiuss)
      -- (-15:\radiusl) arc (-15:145:\radiusl) -- cycle;

\draw[draw=none,fill=stgoorange!10] (-15:\radiuss) arc (-15:-65:\radiuss)
      -- (-65:\radiusl) arc (-65:-15:\radiusl) -- cycle;

\draw[draw=none,fill=stgogreen!10] (-65:\radiuss) arc (-65:-165:\radiuss)
      -- (-165:\radiusl) arc (-165:-65:\radiusl) -- cycle;

\draw (0,0) ellipse ({\radiusA} and \radiusB);


\node[neighbNodeStyle] at (u0) {%
  \begin{tikzpicture}[baseline]
    \path[use as bounding box] (-\BBx,-\BBy) rectangle (\BBx,\BBy);

    \node[nodeStyle,fill=colorCenter] (x) at (-\HalfDist,0) {$u_1$};
    \node[redLabel] at ($(x)+(0,\LabelYOffset)$) {$\bot$};

    \node[nodeStyle] (y) at (\HalfDist,0) {$u_2$};
    \node[redLabel] at ($(y)+(0,\LabelYOffset)$) {$\top$};

    \draw (x) -- node[above, yshift=-2.5mm]{\Large $p_1$} (y);
    \draw[draw=none] (0,0) circle (\InnerCircleR);
  \end{tikzpicture}%
};

\node[neighbNodeStyle] at (u1) {%
  \begin{tikzpicture}[baseline]
    \path[use as bounding box] (-\BBx,-\BBy) rectangle (\BBx,\BBy);

    \node[nodeStyle,fill=colorCenter] (x) at (0,0) {$u_2$};
    \node[redLabel] at ($(x)+(0,\LabelYOffset)$) {$\top$};

    \node[nodeStyle] (y1) [left=\distNeighb of x] {$u_1$};
    \node[redLabel] at ($(y1)+(0,\LabelYOffset)$) {$\bot$};

    \node[nodeStyle] (y2) [right=\distNeighb of x] {$u_3$};
    \node[redLabel] at ($(y2)+(0,\LabelYOffset)$) {$0$};

    \draw (x) -- node[above, yshift=-2.5mm]{\Large  $p'_1$} (y1);
    \draw (x) -- node[above, yshift=-2.5mm]{\Large  $p'_2$} (y2);
    \draw[draw=none] (x) circle (\InnerCircleR);
  \end{tikzpicture}%
};

\node[neighbNodeStyle] at (u2) {%
  \begin{tikzpicture}[baseline]
    \path[use as bounding box] (-\BBx,-\BBy) rectangle (\BBx,\BBy);

    \node[nodeStyle,fill=colorCenter] (x) at (0,0) {$u_3$};
    \node[redLabel] at ($(x)+(0,\LabelYOffset)$) {$0$};

    \node[nodeStyle] (y1) [left=\distNeighb of x] {$u_2$};
    \node[redLabel] at ($(y1)+(0,\LabelYOffset)$) {$\top$};

    \node[nodeStyle] (y2) [right=\distNeighb of x] {$u_4$};
    \node[redLabel] at ($(y2)+(0,\LabelYOffset)$) {$1$};

    \draw (x) -- node[above, yshift=-2.5mm]{\Large  $p_2$} (y1);
    \draw (x) -- node[above, yshift=-2.5mm]{\Large  $p_3$} (y2);
    \draw[draw=none] (x) circle (\InnerCircleR);
  \end{tikzpicture}%
};

\node[neighbNodeStyle] at (u3) {%
  \begin{tikzpicture}[baseline]
    \path[use as bounding box] (-\BBx,-\BBy) rectangle (\BBx,\BBy);

    \node[nodeStyle,fill=colorCenter] (x) at (0,0) {$u_4$};
    \node[redLabel] at ($(x)+(0,\LabelYOffset)$) {$1$};

    \node[nodeStyle] (y1) [left=\distNeighb of x] {$u_3$};
    \node[redLabel] at ($(y1)+(0,\LabelYOffset)$) {$0$};

    \node[nodeStyle] (y2) [right=\distNeighb of x] {$u_5$};
    \node[redLabel] at ($(y2)+(0,\LabelYOffset)$) {$0$};

    \draw (x) -- node[above, yshift=-2.5mm]{\Large  $p'_3$} (y1);
    \draw (x) -- node[above, yshift=-2.5mm]{\Large $p'_4$} (y2);
    \draw[draw=none] (x) circle (\InnerCircleR);
  \end{tikzpicture}%
};

\node[neighbNodeStyle] at (u4) {%
  \begin{tikzpicture}[baseline]
    \path[use as bounding box] (-\BBx,-\BBy) rectangle (\BBx,\BBy);

    \node[nodeStyle,fill=colorCenter] (x) at (0,0) {$u_5$};
    \node[redLabel] at ($(x)+(0,\LabelYOffset)$) {$0$};

    \node[nodeStyle] (y1) [left=\distNeighb of x] {$u_4$};
    \node[redLabel] at ($(y1)+(0,\LabelYOffset)$) {$1$};

    \node[nodeStyle] (y2) [right=\distNeighb of x] {$u_6$};
    \node[redLabel] at ($(y2)+(0,\LabelYOffset)$) {$1$};

    \draw (x) -- node[above, yshift=-2.5mm]{\Large  $p_4$} (y1);
    \draw (x) -- node[above, yshift=-2.5mm]{\Large  $p_5$} (y2);
    \draw[draw=none] (x) circle (\InnerCircleR);
  \end{tikzpicture}%
};

\node[neighbNodeStyle] at (u5) {%
  \begin{tikzpicture}[baseline]
    \path[use as bounding box] (-\BBx,-\BBy) rectangle (\BBx,\BBy);

    \node[nodeStyle,fill=colorCenter] (x) at (0,0) {$u_4$};
    \node[redLabel] at ($(x)+(0,\LabelYOffset)$) {$1$};

    \node[nodeStyle] (y1) [left=\distNeighb of x] {$u_2$};
    \node[redLabel] at ($(y1)+(0,\LabelYOffset)$) {$1$};

    \node[nodeStyle] (y2) [right=\distNeighb of x] {$u_5$};
    \node[redLabel] at ($(y2)+(0,\LabelYOffset)$) {$1$};

    \draw (x) -- node[above, yshift=-2.5mm]{\Large $q'_2$} (y1);
    \draw (x) -- node[above, yshift=-2.5mm]{\Large  $p'_4$} (y2);
    \draw[draw=none] (x) circle (\InnerCircleR);
  \end{tikzpicture}%
};

\node[neighbNodeStyle] at (u6) {%
  \begin{tikzpicture}[baseline]
    \path[use as bounding box] (-\BBx,-\BBy) rectangle (\BBx,\BBy);

    \node[nodeStyle,fill=colorCenter] (x) at (0,0) {$u_2$};
    \node[redLabel] at ($(x)+(0,\LabelYOffset)$) {$\top$};

    \node[nodeStyle] (y1) [left=\distNeighb of x] {$u_1$};
    \node[redLabel] at ($(y1)+(0,\LabelYOffset)$) {$\bot$};

    \node[nodeStyle] (y2) [right=\distNeighb of x] {$u_4$};
    \node[redLabel] at ($(y2)+(0,\LabelYOffset)$) {$1$};

    \draw (x) -- node[above, yshift=-2.5mm]{\Large  $q_1$} (y1);
    \draw (x) -- node[above, yshift=-2.5mm]{\Large  $q_2$} (y2);
    \draw[draw=none] (x) circle (\InnerCircleR);
  \end{tikzpicture}%
};

\end{tikzpicture}

%% file: figs/instances_cycle.tex
\begin{tikzpicture}
    \tikzstyle{nodeStyle}=[circle, draw, fill=white, inner sep=2pt,
      minimum size=5ex]
    \colorlet{colorLabel}{stgored}
    \tikzstyle{labelStyle}=[draw=none, minimum size=0ex, color=colorLabel]
    \tikzstyle{portStyle}=[labelStyle, color=black]

    \draw[fill=stgogreen!10] (-1.25,2.75) rectangle (4.25,-5.75);
    \node at (-0.75,2.25) {\Large $I_1$};

    
    \node[nodeStyle] (3) at (0,0) {$3$};
    \node[nodeStyle] (2) at (3,0) {$2$};
    \node[nodeStyle] (4) at (3,-3) {$4$};
    \node[nodeStyle] (1) at (0,-3) {$1$};

    \draw (3) edge 
      node [pos=0.1,above,portStyle] {$1$}
      node [below,labelStyle] {$1$}
      node [pos=0.9,above,portStyle] {$2$}
    (2);
    \draw (2) edge 
      node [pos=0.1,right,portStyle] {$1$}
      node [left,labelStyle] {$0$}
      node [pos=0.9,right,portStyle] {$2$}
    (4);
    \draw (4) edge 
      node [pos=0.1,below,portStyle] {$1$}
      node [above,labelStyle] {$1$}
      node [pos=0.9,below,portStyle] {$1$}
    (1);
    \draw (1) edge 
      node [pos=0.1,left,portStyle] {$2$}
      node [right,labelStyle] {$0$}
      node [pos=0.9,left,portStyle] {$2$}
    (3);

    \begin{scope}[shift={(6,0)}]
      \draw[fill=stgoblue!10] (-1.25,2.75) rectangle (4.25,-5.75);
      \node at (-0.75,2.25) {\Large $I_2$};

    
      \begin{scope}[shift={(0,0.5)}]
        \node[nodeStyle] (4) at (3,-5) {$4$};
        \node[nodeStyle] (1) at (0,-5) {$1$};
        \node[nodeStyle] (2) at (0,-2) {$2$};
        \node[nodeStyle] (3) at (0,1) {$3$};
        \node[nodeStyle] (5) at (3,1) {$5$};
        \node[nodeStyle] (6) at (3,-2) {$6$};
      \end{scope}

      \draw (4) edge 
        node [pos=0.1,below,portStyle] {$1$}
        node [above,labelStyle] {$1$}
        node [pos=0.9,below,portStyle] {$1$}
      (1);
      \draw (1) edge 
        node [pos=0.1,left,portStyle] {$2$}
        node [right,labelStyle] {$0$}
        node [pos=0.9,left,portStyle] {$1$}
      (2);
      \draw (2) edge 
        node [pos=0.1,left,portStyle] {$2$}
        node [right,labelStyle] {$1$}
        node [pos=0.9,left,portStyle] {$1$}
      (3);
      \draw (3) edge 
        node [pos=0.1,above,portStyle] {$2$}
        node [below,labelStyle] {$0$}
        node [pos=0.9,above,portStyle] {$1$}
      (5);
      \draw (5) edge 
        node [pos=0.1,right,portStyle] {$2$}
        node [left,labelStyle] {$1$}
        node [pos=0.9,right,portStyle] {$1$}
      (6);
      \draw (6) edge 
        node [pos=0.1,right,portStyle] {$2$}
        node [left,labelStyle] {$0$}
        node [pos=0.9,right,portStyle] {$2$}
      (4);
    \end{scope}
\end{tikzpicture}

%% file: figs/oddcycle_cycle.tex
\begin{tikzpicture}
    \colorlet{colorCenter}{stgoorange}
    \colorlet{colorLabel}{stgored}
    \tikzstyle{nodeStyle}=[circle, draw, fill=white, inner sep=0pt,
      minimum size=5ex]
    \tikzstyle{neighbNodeStyle}=[circle, draw, fill=white, inner sep=1pt,
      minimum size=3cm]
    \tikzstyle{labelStyle}=[draw=none, minimum size=0ex, color=colorLabel]
    \tikzstyle{portStyle}=[labelStyle, color=black]
  
    \def\radiusA{8cm}
    \def\radiusB{4cm}
    \def\radius{{\radiusA} and \radiusB}

    \coordinate (u) at (180:\radius);
    \coordinate (v1) at (110:\radius);
    \coordinate (w1) at (55:\radius);
    \coordinate (v2) at (-110:\radius);
    \coordinate (w2) at (-55:\radius);

    \def\radiuss{5.5cm and 2cm}
    \def\radiusl{11cm and 6cm}
    \draw[draw=none,fill=stgoblue!10] (210:\radiuss) arc (210:0:\radiuss)
      -- (0:\radiusl) arc (0:225:\radiusl) -- cycle;
    \draw[draw=none,fill=stgogreen!10] (-150:\radiuss) arc (-150:0:\radiuss)
      -- (0:\radiusl) arc (0:-135:\radiusl) -- cycle;

    \draw (0,0) ellipse ({\radiusA} and \radiusB);

    \def\distNeighb{7mm}
    \node[neighbNodeStyle] at (u) {\begin{tikzpicture}
      \node[nodeStyle,fill=colorCenter] (x) at (0,0) {$1$};
      \node[nodeStyle] (y1) [left = \distNeighb of x] {$4$};
      \node[nodeStyle] (y2) [right = \distNeighb of x] {$2$};
      \draw (x) edge 
        node [pos=0.2,above,portStyle] {$1$}
        node [below,labelStyle] {$1$}
        node [pos=0.8,above,portStyle] {$1$}
      (y1);
      \draw (x) edge 
        node [pos=0.2,above,portStyle] {$2$}
        node [below,labelStyle] {$0$}
        node [pos=0.8,above,portStyle] {$1$}
      (y2);
    \end{tikzpicture}};
    \node[neighbNodeStyle] at (v1) {\begin{tikzpicture}
      \node[nodeStyle,fill=colorCenter] (x) at (0,0) {$2$};
      \node[nodeStyle] (y1) [left = \distNeighb of x] {$1$};
      \node[nodeStyle] (y2) [right = \distNeighb of x] {$3$};
      \draw (x) edge 
        node [pos=0.2,above,portStyle] {$1$}
        node [below,labelStyle] {$0$}
        node [pos=0.8,above,portStyle] {$2$}
      (y1);
      \draw (x) edge 
        node [pos=0.2,above,portStyle] {$2$}
        node [below,labelStyle] {$1$}
        node [pos=0.8,above,portStyle] {$1$}
      (y2);
    \end{tikzpicture}};
    \node[neighbNodeStyle] at (v2) {\begin{tikzpicture}
      \node[nodeStyle,fill=colorCenter] (x) at (0,0) {$4$};
      \node[nodeStyle] (y1) [left = \distNeighb of x] {$2$};
      \node[nodeStyle] (y2) [right = \distNeighb of x] {$1$};
      \draw (x) edge 
        node [pos=0.2,above,portStyle] {$2$}
        node [below,labelStyle] {$0$}
        node [pos=0.8,above,portStyle] {$1$}
      (y1);
      \draw (x) edge 
        node [pos=0.2,above,portStyle] {$1$}
        node [below,labelStyle] {$1$}
        node [pos=0.8,above,portStyle] {$1$}
      (y2);
    \end{tikzpicture}};
    \node[neighbNodeStyle] at (w1) {\begin{tikzpicture}
      \node[nodeStyle,fill=colorCenter] (x) at (0,0) {$3$};
      \node[nodeStyle] (y1) [left = \distNeighb of x] {$2$};
      \node[nodeStyle] (y2) [right = \distNeighb of x] {$5$};
      \draw (x) edge 
        node [pos=0.2,above,portStyle] {$1$}
        node [below,labelStyle] {$1$}
        node [pos=0.8,above,portStyle] {$2$}
      (y1);
      \draw (x) edge 
        node [pos=0.2,above,portStyle] {$2$}
        node [below,labelStyle] {$0$}
        node [pos=0.8,above,portStyle] {$1$}
      (y2);
    \end{tikzpicture}};
    \node[neighbNodeStyle] at (w2) {\begin{tikzpicture}
      \node[nodeStyle,fill=colorCenter] (x) at (0,0) {$2$};
      \node[nodeStyle] (y1) [left = \distNeighb of x] {$3$};
      \node[nodeStyle] (y2) [right = \distNeighb of x] {$4$};
      \draw (x) edge 
        node [pos=0.2,above,portStyle] {$2$}
        node [below,labelStyle] {$1$}
        node [pos=0.8,above,portStyle] {$1$}
      (y1);
      \draw (x) edge 
        node [pos=0.2,above,portStyle] {$1$}
        node [below,labelStyle] {$0$}
        node [pos=0.8,above,portStyle] {$2$}
      (y2);
    \end{tikzpicture}};
\end{tikzpicture}

%% file: figs/soundness_pipeline.tex
\begin{tikzpicture}[node distance=1cm, auto, >=Latex]
  \tikzstyle{node_style}=[circle, draw, inner sep=2pt, minimum size=6mm]
  \tikzstyle{a_style}=[fill=stgoblue!20]
  \tikzstyle{center_style}=[very thick]
  \tikzstyle{b_style}=[fill=stgogreen!20]
  \tikzstyle{c_style}=[fill=stgoorange!20]
  \tikzstyle{panel_label}=[align=center]
  \tikzstyle{arrow_style}=[->, thick]

  \newlength{\panelgap}
  \setlength{\panelgap}{1.5cm}
  \newlength{\panelcaptiongap}
  \setlength{\panelcaptiongap}{0.4cm}
  \newlength{\panelw}
  \setlength{\panelw}{8.0cm}
  \newlength{\panelh}
  \setlength{\panelh}{5.0cm}

\node[minimum width=\panelw, minimum height=\panelh, inner sep=0, anchor=west] (panel1) at (0,0) {
    \begin{tikzpicture}[node distance=.3cm, baseline]
      \node[node_style, a_style, center_style] (v1a) at (-1,0) {$A$};
      \node[node_style, a_style, above left=of v1a] (v1b) {$A$};
      \node[node_style, b_style, below left=of v1a] (v1c) {$B$};
      \node[node_style, b_style, left=of v1a] (v1ac) {$B$};
      \draw (v1a) -- (v1b);
      \draw (v1a) -- (v1c);
      \draw (v1a) -- (v1ac);

      \node[node_style, a_style, center_style] at (0,0.7) (v1d) {$A$};
      \node[node_style, a_style, above left=of v1d] (v1e) {$A$};
      \node[node_style, b_style, above right=of v1d] (v1f) {$B$};
      \node[node_style, b_style, above=of v1d] (v1dc) {$B$};
      \draw (v1d) -- (v1e);
      \draw (v1d) -- (v1f);
      \draw (v1d) -- (v1dc);

      \node[node_style, a_style, center_style] at (1,0) (v1g) {$A$};
      \node[node_style, a_style, above right=of v1g] (v1h) {$A$};
      \node[node_style, b_style, below right=of v1g] (v1i) {$B$};
      \node[node_style, b_style, right=of v1g] (v1gc) {$B$};
      \draw (v1g) -- (v1h);
      \draw (v1g) -- (v1i);
      \draw (v1g) -- (v1gc);

    \end{tikzpicture}
  };
  \node[panel_label] at ($(panel1.south)+(0,\panelcaptiongap)$)
    {\small Multiset of accepting views};

\node[minimum width=\panelw, minimum height=\panelh, inner sep=0, anchor=west] (panel2) at ($(panel1.east)+(\panelgap,0)$) {
    \begin{tikzpicture}[baseline]
      \node[node_style, a_style, center_style] (v2a) at (6,-0.5) {$A$};
      \draw[thick] (v2a) -- ++(-1,0.5)
        node[pos=1, anchor=east, minimum width=0, minimum height=0] {\scriptsize $(A,A)$};
      \draw[thick] (v2a) -- ++(-1,0.0)
        node[pos=1, anchor=east, minimum width=0, minimum height=0] {\scriptsize $(A,B)$};
      \draw[thick] (v2a) -- ++(-1,-0.5)
        node[pos=1, anchor=east, minimum width=0, minimum height=0] {\scriptsize $(A,B)$};
      \node at (1.5,-2) {\scriptsize $x(\mu)
        = \left(
        \begin{array}{c}
          x_{AA}\\
          x_{AB}\\
          x_{BA}\\
          x_{BB}
        \end{array}
        \right)
        = \left(
        \begin{array}{c}
          0\\
          2\\
          -2\\
          0
        \end{array}
        \right)$};
    \end{tikzpicture}
  };
  \node[panel_label] at ($(panel2.south)+(0,\panelcaptiongap)$)
    {\small Stubs / demand vector};

\node[minimum width=\panelw, minimum height=\panelh, inner sep=0,
anchor=west] (panel3) at ($(panel2.south west)+(0,-2\panelgap)$) {
    \begin{tikzpicture}[baseline]
      \node[node_style, a_style] (a3a) at (-1.8,0.8) {$A$};
      \node[node_style, a_style] (a3b) at (-1.8,0) {$A$};
      \node[node_style, a_style] (a3c) at (-1.8,-0.8) {$A$};
      \node[node_style, b_style] (b3) at (-0.2,0) {$B$};

      \draw (a3a) -- (b3);
      \draw (a3c) -- (b3);
      \draw (a3b) to[bend left=20] (b3);
      \draw (a3b) to[bend right=20] (b3);

      \node at (-3.1,0) {\small $\Rightarrow$};

      \node[node_style, a_style] (a3a2) at (1.3,0.8) {$A$};
      \node[node_style, a_style] (a3b2) at (1.3,0) {$A$};
      \node[node_style, a_style] (a3c2) at (1.3,-0.8) {$A$};
      \node[node_style, b_style] (b3a) at (2.6,0.5) {$B$};
      \node[node_style, b_style] (b3b) at (2.6,-0.5) {$B$};

      \draw (a3a2) -- (b3a);
      \draw (a3c2) -- (b3b);
      \draw (a3b2) -- (b3a);
      \draw (a3b2) -- (b3b);

    \end{tikzpicture}
  };
  \node[panel_label] at ($(panel3.south)+(0,\panelcaptiongap)$)
    {\small Blow-up, remove multiedges};

\node[minimum width=\panelw, minimum height=\panelh, inner sep=0,
anchor=west] (panel4) at ($(panel1.south west)+(0,-2\panelgap)$) {
    \begin{tikzpicture}[baseline]
      \node[node_style, a_style] (c1) at (-0.6,0.6) {$A$};
      \node[node_style, a_style] (c2) at (0.6,0.6) {$A$};
      \node[node_style, a_style] (c3) at (0, -0.4) {$A$};

      \draw (c1) -- (c2);
      \draw (c1) -- (c3);
      \draw (c2) -- (c3);

    \end{tikzpicture}
  };
  \node[panel_label] at ($(panel4.south)+(0,\panelcaptiongap)$)
    {\small Satisfy $(A,A)$ demands};

  \draw[arrow_style] (panel1.east) -- (panel2.west);
  \draw[arrow_style] ($(panel2.south)-(0,0.2)$) -- ($(panel3.north)-(0,0.6)$);
  \draw[arrow_style] (panel3.west) -- (panel4.east);
\end{tikzpicture}

%% file: figs/compatible_views.tex
\begin{tikzpicture}
    \tikzstyle{nodeStyle}=[circle, draw, fill=white, inner sep=2pt]

    \colorlet{uColor}{stgoblue}
    \colorlet{vColor}{stgored}

    \node[nodeStyle, fill=uColor!20, very thick] (u) at (0,0) {$7$};
    \node[nodeStyle, fill=vColor!20, very thick] (v) at (7,0) {$9$};

    \draw[thick, dashed, uColor] (u) circle (3cm);

    \draw[thick, dashed, vColor] (v) circle (3cm);

    \node[uColor] (mu1) at (-2.5,2.5) {\Large $\mu_u$};
    \node (ulabel) at (0.4,0.4) {\large $u$};
    \node[nodeStyle, fill=uColor!20] (u1) at (120:1.5) {$9$};
    \node[nodeStyle, fill=uColor!20] (u2) at (300:1.5) {$2$};
    \node[nodeStyle, fill=uColor!20] (u3) at (200:1.5) {$5$};
    \node[nodeStyle, fill=uColor!20] (u4) at (250:2) {$4$};
    \node[nodeStyle, fill=uColor!20] (u5) at (70:2) {$1$};

    \foreach \i in {1,...,3} {
        \draw[uColor] (u) -- (u\i);
    }
    \draw[uColor] (u2) -- (u4);
    \draw[uColor] (u1) -- (u3);
    \draw[uColor] (u1) -- (u5);

    \begin{scope}[shift={(7,0)}]
      \node[vColor] (mu2) at (-2.5,2.5) {\Large $\mu_v$};
      \node (vlabel) at (-0.4,-0.4) {\large $v$};
      \node[nodeStyle, fill=vColor!20] (v1) at (45:1.5) {$7$};
      \node[nodeStyle, fill=vColor!20] (v2) at (150:1.5) {$5$};
      \node[nodeStyle, fill=vColor!20] (v3) at (320:1.5) {$1$};
      \node[nodeStyle, fill=vColor!20] (v4) at (90:2) {$2$};
      \node[nodeStyle, fill=vColor!20] (v5) at (250:2) {$3$};
      \node[nodeStyle, fill=vColor!20] (v6) at (290:2.5) {$4$};

      \foreach \i in {1,...,3} {
          \draw[vColor] (v) -- (v\i);
      }
      \draw[vColor] (v1) -- (v2);
      \draw[vColor] (v1) -- (v4);
      \draw[vColor] (v3) -- (v5);
      \draw[vColor] (v3) -- (v6);
    \end{scope}

    \begin{scope}[shift={(2.5,4.5)}]
      \draw[fill=stgogreen!10] (-3,2.75) rectangle (5,-0.75);
      \node (G) at (-2.5,2.25) {\Large $G$};
      
      \node[nodeStyle,fill=uColor!20] (7) at (0,0) {$7$};
      \node at (0.4,0.4) {\large $u$};
      \node[nodeStyle] (5) at (-2,0) {$5$};
      \node[nodeStyle,fill=vColor!20] (9) at (0,2) {$9$};
      \node at (0.4,2.4) {\large $v$};
      \node[nodeStyle] (1) at (2,2) {$1$};
      \node[nodeStyle] (2) at (2,0) {$2$};
      \node[nodeStyle] (4) at (4,0) {$4$};
      \node[nodeStyle] (3) at (4,2) {$3$};

      \draw (7) -- (5);
      \draw (7) -- (9);
      \draw (7) -- (2);
      \draw (5) -- (9);
      \draw (9) -- (1);
      \draw (4) -- (2);
      \draw (4) -- (1);
      \draw (1) -- (3);
    \end{scope}
\end{tikzpicture}

%% file: figs/compatibility_transitivity.tex
\begin{tikzpicture}
    \tikzstyle{nodeStyle}=[circle, draw, fill=white, inner sep=2pt]

    \colorlet{uColor}{stgoblue}
    \colorlet{vColor}{stgored}
    \colorlet{wColor}{stgogreen}

    \node[nodeStyle, fill=uColor!20, very thick] (u) at (0,0) {$1$};
    \node[nodeStyle, fill=vColor!20, very thick] (v) at (6,0) {$3$};
    \node[nodeStyle, fill=wColor!20, very thick] (w) at (3,-5) {$6$};

    \draw[thick, dashed, uColor] (u) circle (2.5cm);
    \draw[thick, dashed, vColor] (v) circle (2.5cm);
    \draw[thick, dashed, wColor] (w) circle (2.5cm);

    \node[uColor] (mu1) at (-2.5,2) {\Large $\mu_u$};
    \node (ulabel) at (-0.4,0.4) {\large $u$};
    \node[nodeStyle, fill=uColor!20] (u1) at (60:1) {$4$};
    \node[nodeStyle, fill=uColor!20] (u2) at (240:1) {$2$};
    \node[nodeStyle, fill=uColor!20] (u3) at (240:2) {$3$};

    \draw[uColor] (u) -- (u1);
    \draw[uColor] (u) -- (u2);
    \draw[uColor] (u2) -- (u3);

    \begin{scope}[shift={(6,0)}]
      \node[vColor] (mu2) at (-2.5,2) {\Large $\mu_v$};
      \node (vlabel) at (-0.4,0.4) {\large $v$};
      \node[nodeStyle, fill=vColor!20] (v1) at (60:1) {$2$};
      \node[nodeStyle, fill=vColor!20] (v2) at (60:2) {$1$};
      \node[nodeStyle, fill=vColor!20] (v3) at (240:1) {$5$};
      \node[nodeStyle, fill=vColor!20] (v4) at (240:2) {$6$};

      \draw[vColor] (v) -- (v1);
      \draw[vColor] (v1) -- (v2);
      \draw[vColor] (v) -- (v3);
      \draw[vColor] (v3) -- (v4);
    \end{scope}

    \begin{scope}[shift={(3,-5)}]
      \node[wColor] (mu3) at (-2.5,2) {\Large $\mu_w$};
      \node (wlabel) at (-0.4,0.4) {\large $w$};
      \node[nodeStyle, fill=wColor!20] (w1) at (60:1) {$5$};
      \node[nodeStyle, fill=wColor!20] (w2) at (60:2) {$3$};
      \node[nodeStyle, fill=wColor!20] (w3) at (240:1) {$4$};

      \draw[wColor] (w) -- (w1);
      \draw[wColor] (w1) -- (w2);
      \draw[wColor] (w) -- (w3);
    \end{scope}
\end{tikzpicture}

%% file: figs/odd_walk_realizability.tex
\begin{tikzpicture}[every node/.style={circle, draw, inner sep=1pt, fill=black}]
    \colorlet{uColor}{stgoblue!80}
    \colorlet{vColor}{stgored!80}
    \colorlet{uvColor}{stgogreen!80}

    \node[uColor,label=150:$u$] (u) at (0,0) {};
    \node[vColor,label=210:$v$] (v) at (0,-1) {};

    \node[vColor,label=240:$v_1$] (v1) at (1,-1.5) {};
    \node[vColor,label=270:$v_2$] (v2) at (2,-1.75) {};
    \node[vColor,draw=none,fill=none,rotate=-7.5] (vdots) at (3,-1.9) {$\cdots$};
    \node[vColor,label=280:$v_{r-1}$] (vr-1) at (4,-2) {};
    \node[vColor,label=300:$v_r$] (vr) at (5,-2) {};

    \node[uColor,label=120:$u_1$] (u1) at (1,0.5) {};
    \node[uColor,label=90:$u_2$] (u2) at (2,0.75) {};
    \node[uColor,draw=none,fill=none,rotate=7.5] (udots) at (3,0.9) {$\cdots$};
    \node[uColor,label=80:$u_{r-1}$] (ur-1) at (4,1) {};
    \node[uColor,label=60:$u_r$] (ur) at (5,1) {};

    \node[uvColor,label=330:$a$] (a) at (5.5,-1.3) {};
    \node[uvColor,draw=none,fill=none,rotate=90] (abdots) at (5.5,-0.5) {$\cdots$};
    \node[uvColor,label=30:$b$] (b) at (5.5,0.2) {};

    \draw[->] (u) -- (v);
    \draw[->,vColor] (v) -- (v1);
    \draw[->,vColor] (v1) -- (v2);
    \draw[->,vColor] (v2) -- (vdots);
    \draw[->,vColor] (vdots) -- (vr-1);
    \draw[->,vColor] (vr-1) -- (vr);
    \draw[->,uColor] (u1) -- (u);
    \draw[->,uColor] (u2) -- (u1);
    \draw[->,uColor] (udots) -- (u2);
    \draw[->,uColor] (ur-1) -- (udots);
    \draw[->,uColor] (ur) -- (ur-1);
    \draw[->,uvColor] (vr) -- (a);
    \draw[->,uvColor] (a) -- (abdots);
    \draw[->,uvColor] (abdots) -- (b);
    \draw[->,uvColor] (b) -- (ur);

    \node[vColor,draw=none,fill=none] at (2.5,-1.25) {\Large $P_v$};
    \node[uColor,draw=none,fill=none] at (2.5,0.25) {\Large $P_u$};
    \node[uvColor,draw=none,fill=none] at (4.75,-0.5) {\Large $P_{uv}$};
\end{tikzpicture}

%% file: figs/2-wise.tex
\begin{tikzpicture}[
  vertex/.style={circle,draw,inner sep=1.4pt},
  wise/.style={circle,draw,thick,inner sep=1.6pt,fill=stgored!50}
]

  \draw[rounded corners, fill=stgoblue!50] (-2,-1.25) rectangle (-0.5,2);
  \node at (-1.25,2.3) {$N(a)$};

  \draw[rounded corners, fill=stgogreen!50] (0.5,-1.25) rectangle (2,2);
  \node at (1.25,2.3) {$N(b)$};

  \node[wise,label=left:$b$] (b)  at (-1.25,-0.75) {};
  \node[vertex]              (x1) at (-1.25,0.75) {};
  \node[vertex]              (x2) at (-1.25,0) {};
  \node[vertex]              (x3) at (-1.25,1.5) {};

  \node[wise,label=right:$a$] (a) at (1.25,1.5) {};
  \node[vertex]              (y1) at (1.25,0.75) {};
  \node[vertex]              (y2) at (1.25,0) {};
  \node[vertex]              (y3) at (1.25,-0.75) {};

  \draw (a) -- (b);         

  \foreach \v in {x1,x2,x3}
    \draw (a) -- (\v);

  \foreach \v in {y1,y2,y3}
    \draw (b) -- (\v);

\end{tikzpicture}

%% file: figs/c15hidingnew.tex
\tikzset{every picture/.style={line width=0.75pt}} 

\begin{tikzpicture}[x=0.75pt,y=0.75pt,yscale=-1,xscale=1]

\draw  [fill={rgb, 255:red, 238; green, 225; blue, 108 }  ,fill opacity=1 ][dash pattern={on 0.84pt off 2.51pt}] (31,59.8) .. controls (31,41.13) and (46.13,26) .. (64.8,26) -- (211.2,26) .. controls (229.87,26) and (245,41.13) .. (245,59.8) -- (245,161.2) .. controls (245,179.87) and (229.87,195) .. (211.2,195) -- (64.8,195) .. controls (46.13,195) and (31,179.87) .. (31,161.2) -- cycle ;
\draw  [fill={rgb, 255:red, 238; green, 225; blue, 108 }  ,fill opacity=1 ][dash pattern={on 0.84pt off 2.51pt}] (296,54.8) .. controls (296,36.13) and (311.13,21) .. (329.8,21) -- (476.2,21) .. controls (494.87,21) and (510,36.13) .. (510,54.8) -- (510,156.2) .. controls (510,174.87) and (494.87,190) .. (476.2,190) -- (329.8,190) .. controls (311.13,190) and (296,174.87) .. (296,156.2) -- cycle ;
\draw  [fill={rgb, 255:red, 238; green, 225; blue, 108 }  ,fill opacity=1 ][dash pattern={on 0.84pt off 2.51pt}] (583,56.8) .. controls (583,38.13) and (598.13,23) .. (616.8,23) -- (789.2,23) .. controls (807.87,23) and (823,38.13) .. (823,56.8) -- (823,158.2) .. controls (823,176.87) and (807.87,192) .. (789.2,192) -- (616.8,192) .. controls (598.13,192) and (583,176.87) .. (583,158.2) -- cycle ;
\draw  [fill={rgb, 255:red, 238; green, 225; blue, 108 }  ,fill opacity=1 ][dash pattern={on 0.84pt off 2.51pt}] (873.67,54.8) .. controls (873.67,36.13) and (888.8,21) .. (907.47,21) -- (1092.87,21) .. controls (1111.53,21) and (1126.67,36.13) .. (1126.67,54.8) -- (1126.67,156.2) .. controls (1126.67,174.87) and (1111.53,190) .. (1092.87,190) -- (907.47,190) .. controls (888.8,190) and (873.67,174.87) .. (873.67,156.2) -- cycle ;
\draw  [fill={rgb, 255:red, 238; green, 225; blue, 108 }  ,fill opacity=1 ][dash pattern={on 0.84pt off 2.51pt}] (1188,51.8) .. controls (1188,33.13) and (1203.13,18) .. (1221.8,18) -- (1409.2,18) .. controls (1427.87,18) and (1443,33.13) .. (1443,51.8) -- (1443,153.2) .. controls (1443,171.87) and (1427.87,187) .. (1409.2,187) -- (1221.8,187) .. controls (1203.13,187) and (1188,171.87) .. (1188,153.2) -- cycle ;
\draw  [fill={rgb, 255:red, 238; green, 225; blue, 108 }  ,fill opacity=1 ][dash pattern={on 0.84pt off 2.51pt}] (1192,264.47) .. controls (1192,245.8) and (1207.13,230.67) .. (1225.8,230.67) -- (1410.2,230.67) .. controls (1428.87,230.67) and (1444,245.8) .. (1444,264.47) -- (1444,365.87) .. controls (1444,384.53) and (1428.87,399.67) .. (1410.2,399.67) -- (1225.8,399.67) .. controls (1207.13,399.67) and (1192,384.53) .. (1192,365.87) -- cycle ;
\draw  [fill={rgb, 255:red, 238; green, 225; blue, 108 }  ,fill opacity=1 ][dash pattern={on 0.84pt off 2.51pt}] (1184,473.47) .. controls (1184,454.8) and (1199.13,439.67) .. (1217.8,439.67) -- (1419.2,439.67) .. controls (1437.87,439.67) and (1453,454.8) .. (1453,473.47) -- (1453,574.87) .. controls (1453,593.53) and (1437.87,608.67) .. (1419.2,608.67) -- (1217.8,608.67) .. controls (1199.13,608.67) and (1184,593.53) .. (1184,574.87) -- cycle ;
\draw  [fill={rgb, 255:red, 238; green, 225; blue, 108 }  ,fill opacity=1 ][dash pattern={on 0.84pt off 2.51pt}] (1179,685.47) .. controls (1179,666.8) and (1194.13,651.67) .. (1212.8,651.67) -- (1425.2,651.67) .. controls (1443.87,651.67) and (1459,666.8) .. (1459,685.47) -- (1459,786.87) .. controls (1459,805.53) and (1443.87,820.67) .. (1425.2,820.67) -- (1212.8,820.67) .. controls (1194.13,820.67) and (1179,805.53) .. (1179,786.87) -- cycle ;
\draw    (1321.67,188.67) -- (1321.67,226.67) ;
\draw    (1316.67,401.67) -- (1316.67,439.67) ;
\draw    (1318.67,610.67) -- (1318.67,648.67) ;
\draw    (1179,738) -- (1031,690) ;
\draw  [fill={rgb, 255:red, 238; green, 225; blue, 108 }  ,fill opacity=1 ][dash pattern={on 0.84pt off 2.51pt}] (390,554.47) .. controls (390,535.8) and (405.13,520.67) .. (423.8,520.67) -- (637.53,520.67) .. controls (656.2,520.67) and (671.33,535.8) .. (671.33,554.47) -- (671.33,655.87) .. controls (671.33,674.53) and (656.2,689.67) .. (637.53,689.67) -- (423.8,689.67) .. controls (405.13,689.67) and (390,674.53) .. (390,655.87) -- cycle ;
\draw    (515.33,689.67) -- (515.33,727.67) ;
\draw    (284.33,730.33) -- (339,771.67) ;
\draw  [fill={rgb, 255:red, 238; green, 225; blue, 108 }  ,fill opacity=1 ][dash pattern={on 0.84pt off 2.51pt}] (8,670.13) .. controls (8,651.47) and (23.13,636.33) .. (41.8,636.33) -- (250.2,636.33) .. controls (268.87,636.33) and (284,651.47) .. (284,670.13) -- (284,771.53) .. controls (284,790.2) and (268.87,805.33) .. (250.2,805.33) -- (41.8,805.33) .. controls (23.13,805.33) and (8,790.2) .. (8,771.53) -- cycle ;
\draw    (129,596.33) -- (129,634.33) ;
\draw  [fill={rgb, 255:red, 238; green, 225; blue, 108 }  ,fill opacity=1 ][dash pattern={on 0.84pt off 2.51pt}] (6.33,461.13) .. controls (6.33,442.47) and (21.47,427.33) .. (40.13,427.33) -- (248.53,427.33) .. controls (267.2,427.33) and (282.33,442.47) .. (282.33,461.13) -- (282.33,562.53) .. controls (282.33,581.2) and (267.2,596.33) .. (248.53,596.33) -- (40.13,596.33) .. controls (21.47,596.33) and (6.33,581.2) .. (6.33,562.53) -- cycle ;
\draw    (127.33,388.33) -- (127.33,426.33) ;
\draw  [fill={rgb, 255:red, 238; green, 225; blue, 108 }  ,fill opacity=1 ][dash pattern={on 0.84pt off 2.51pt}] (14.67,265.4) .. controls (14.67,246.95) and (29.62,232) .. (48.07,232) -- (241.6,232) .. controls (260.05,232) and (275,246.95) .. (275,265.4) -- (275,365.6) .. controls (275,384.05) and (260.05,399) .. (241.6,399) -- (48.07,399) .. controls (29.62,399) and (14.67,384.05) .. (14.67,365.6) -- cycle ;
\draw    (672,667) -- (754.33,729.33) ;
\draw    (905,688) -- (905,721) ;
\draw    (1127,109) -- (1189,110) ;
\draw    (824.33,114.33) -- (872.33,114.33) ;
\draw    (511,113) -- (583,114.33) ;
\draw    (245.67,110.33) -- (293.67,110.33) ;
\draw    (126,199) -- (126,237) ;
\draw  [fill={rgb, 255:red, 238; green, 225; blue, 108 }  ,fill opacity=1 ][dash pattern={on 0.84pt off 2.51pt}] (741,757.47) .. controls (741,738.8) and (756.13,723.67) .. (774.8,723.67) -- (987.2,723.67) .. controls (1005.87,723.67) and (1021,738.8) .. (1021,757.47) -- (1021,858.87) .. controls (1021,877.53) and (1005.87,892.67) .. (987.2,892.67) -- (774.8,892.67) .. controls (756.13,892.67) and (741,877.53) .. (741,858.87) -- cycle ;
\draw  [fill={rgb, 255:red, 238; green, 225; blue, 108 }  ,fill opacity=1 ][dash pattern={on 0.84pt off 2.51pt}] (339,757.47) .. controls (339,738.8) and (354.13,723.67) .. (372.8,723.67) -- (585.2,723.67) .. controls (603.87,723.67) and (619,738.8) .. (619,757.47) -- (619,858.87) .. controls (619,877.53) and (603.87,892.67) .. (585.2,892.67) -- (372.8,892.67) .. controls (354.13,892.67) and (339,877.53) .. (339,858.87) -- cycle ;
\draw  [fill={rgb, 255:red, 238; green, 225; blue, 108 }  ,fill opacity=1 ][dash pattern={on 0.84pt off 2.51pt}] (806,552.47) .. controls (806,533.8) and (821.13,518.67) .. (839.8,518.67) -- (1052.2,518.67) .. controls (1070.87,518.67) and (1086,533.8) .. (1086,552.47) -- (1086,653.87) .. controls (1086,672.53) and (1070.87,687.67) .. (1052.2,687.67) -- (839.8,687.67) .. controls (821.13,687.67) and (806,672.53) .. (806,653.87) -- cycle ;

\draw (258,77) node [anchor=north west][inner sep=0.75pt]   [align=left] {$\displaystyle G_{1}$};
\draw (536.33,81.33) node [anchor=north west][inner sep=0.75pt]   [align=left] {$\displaystyle G_{2}$};
\draw (842.67,90) node [anchor=north west][inner sep=0.75pt]   [align=left] {$\displaystyle G_{3}$};
\draw (1158.67,84) node [anchor=north west][inner sep=0.75pt]   [align=left] {$\displaystyle G_{4}$};
\draw (1332.67,197.67) node [anchor=north west][inner sep=0.75pt]   [align=left] {$\displaystyle G_{5}$};
\draw (1327.67,410.67) node [anchor=north west][inner sep=0.75pt]   [align=left] {$\displaystyle G_{6}$};
\draw (1329.67,619.67) node [anchor=north west][inner sep=0.75pt]   [align=left] {$\displaystyle G_{7}$};
\draw (1101.67,722.67) node [anchor=north west][inner sep=0.75pt]   [align=left] {$\displaystyle G_{8}$};
\draw (911.67,694.67) node [anchor=north west][inner sep=0.75pt]   [align=left] {$\displaystyle G_{9}$};
\draw (683.67,697) node [anchor=north west][inner sep=0.75pt]   [align=left] {$\displaystyle G_{10}$};
\draw (526.67,699.67) node [anchor=north west][inner sep=0.75pt]   [align=left] {$\displaystyle G_{11}$};
\draw (294.67,766) node [anchor=north west][inner sep=0.75pt]   [align=left] {$\displaystyle G_{11}$};
\draw (99,604.33) node [anchor=north west][inner sep=0.75pt]   [align=left] {$\displaystyle G_{12}$};
\draw (97.33,396.33) node [anchor=north west][inner sep=0.75pt]   [align=left] {$\displaystyle G_{12}$};
\draw (100.67,203) node [anchor=north west][inner sep=0.75pt]   [align=left] {$\displaystyle G_{1}$};
\draw (89.43,35.91) node [anchor=north west][inner sep=0.75pt]    {$1$};
\draw  [color={rgb, 255:red, 0; green, 0; blue, 0 }  ,draw opacity=1 ][line width=1.5]   (164.08, 63.39) circle [x radius= 10.06, y radius= 10.06]   ;
\draw (158.11,60.83) node [anchor=north west][inner sep=0.75pt]  [font=\tiny,rotate=-359.69,xslant=0.02]  {$$};
\draw  [color={rgb, 255:red, 0; green, 0; blue, 0 }  ,draw opacity=1 ][fill={rgb, 255:red, 74; green, 144; blue, 226 }  ,fill opacity=1 ][line width=1.5]   (94.68, 64.26) circle [x radius= 10.06, y radius= 10.06]   ;
\draw (88.71,61.69) node [anchor=north west][inner sep=0.75pt]  [font=\tiny,rotate=-359.69,xslant=0.02]  {$$};
\draw  [color={rgb, 255:red, 0; green, 0; blue, 0 }  ,draw opacity=1 ][line width=1.5]   (164.08, 110.94) circle [x radius= 10.06, y radius= 10.06]   ;
\draw (158.11,108.37) node [anchor=north west][inner sep=0.75pt]  [font=\tiny,rotate=-359.69,xslant=0.02]  {$$};
\draw  [color={rgb, 255:red, 0; green, 0; blue, 0 }  ,draw opacity=1 ][line width=1.5]   (164.08, 160.22) circle [x radius= 10.06, y radius= 10.06]   ;
\draw (158.11,157.65) node [anchor=north west][inner sep=0.75pt]  [font=\tiny,rotate=-359.69,xslant=0.02]  {$$};
\draw (158.08,35.05) node [anchor=north west][inner sep=0.75pt]    {$4$};
\draw (158.08,82.59) node [anchor=north west][inner sep=0.75pt]    {$5$};
\draw (158.08,131.87) node [anchor=north west][inner sep=0.75pt]    {$6$};
\draw (56.41,50.13) node [anchor=north west][inner sep=0.75pt]  [font=\Large,color={rgb, 255:red, 208; green, 2; blue, 27 }  ,opacity=1 ]  {$\bot $};
\draw (169.35,48.4) node [anchor=north west][inner sep=0.75pt]  [font=\Large,color={rgb, 255:red, 208; green, 2; blue, 27 }  ,opacity=1 ]  {$\bot $};
\draw (174.63,97.88) node [anchor=north west][inner sep=0.75pt]  [font=\large,color={rgb, 255:red, 208; green, 2; blue, 27 }  ,opacity=1 ]  {$( 4,5,6)$};
\draw (174.38,147.16) node [anchor=north west][inner sep=0.75pt]  [font=\large,color={rgb, 255:red, 208; green, 2; blue, 27 }  ,opacity=1 ]  {$( 4,5,6)$};
\draw (372.61,38.77) node [anchor=north west][inner sep=0.75pt]    {$1$};
\draw (373.41,88.16) node [anchor=north west][inner sep=0.75pt]    {$2$};
\draw (374.2,133.3) node [anchor=north west][inner sep=0.75pt]    {$3$};
\draw  [color={rgb, 255:red, 0; green, 0; blue, 0 }  ,draw opacity=1 ][line width=1.5]   (379.41, 160.48) circle [x radius= 10.06, y radius= 10.06]   ;
\draw (373.44,157.92) node [anchor=north west][inner sep=0.75pt]  [font=\tiny,rotate=-359.69,xslant=0.02]  {$$};
\draw  [color={rgb, 255:red, 0; green, 0; blue, 0 }  ,draw opacity=1 ][fill={rgb, 255:red, 74; green, 144; blue, 226 }  ,fill opacity=1 ][line width=1.5]   (451.8, 65.95) circle [x radius= 10.06, y radius= 10.06]   ;
\draw (445.82,63.39) node [anchor=north west][inner sep=0.75pt]  [font=\tiny,rotate=-359.69,xslant=0.02]  {$$};
\draw  [color={rgb, 255:red, 0; green, 0; blue, 0 }  ,draw opacity=1 ][line width=1.5]   (377.82, 66.81) circle [x radius= 10.06, y radius= 10.06]   ;
\draw (371.85,64.24) node [anchor=north west][inner sep=0.75pt]  [font=\tiny,rotate=-359.69,xslant=0.02]  {$$};
\draw (445.8,37.92) node [anchor=north west][inner sep=0.75pt]    {$4$};
\draw  [color={rgb, 255:red, 0; green, 0; blue, 0 }  ,draw opacity=1 ][line width=1.5]   (379.41, 115.35) circle [x radius= 10.06, y radius= 10.06]   ;
\draw (373.44,112.78) node [anchor=north west][inner sep=0.75pt]  [font=\tiny,rotate=-359.69,xslant=0.02]  {$$};
\draw (339.24,51.73) node [anchor=north west][inner sep=0.75pt]  [font=\Large,color={rgb, 255:red, 208; green, 2; blue, 27 }  ,opacity=1 ]  {$\bot $};
\draw (455.38,50.02) node [anchor=north west][inner sep=0.75pt]  [font=\Large,color={rgb, 255:red, 208; green, 2; blue, 27 }  ,opacity=1 ]  {$\bot $};
\draw (303.64,103.9) node [anchor=north west][inner sep=0.75pt]  [font=\large,color={rgb, 255:red, 208; green, 2; blue, 27 }  ,opacity=1 ]  {$( 1,2,3)$};
\draw (302.25,149.18) node [anchor=north west][inner sep=0.75pt]  [font=\large,color={rgb, 255:red, 208; green, 2; blue, 27 }  ,opacity=1 ]  {$( 1,2,3)$};
\draw (657.32,88.09) node [anchor=north west][inner sep=0.75pt]    {$2$};
\draw  [color={rgb, 255:red, 0; green, 0; blue, 0 }  ,draw opacity=1 ][line width=1.5]   (739.84, 64.23) circle [x radius= 10.06, y radius= 10.06]   ;
\draw (733.87,61.66) node [anchor=north west][inner sep=0.75pt]  [font=\tiny,rotate=-359.69,xslant=0.02]  {$$};
\draw  [color={rgb, 255:red, 0; green, 0; blue, 0 }  ,draw opacity=1 ][line width=1.5]   (739.84, 113.67) circle [x radius= 10.06, y radius= 10.06]   ;
\draw (733.87,111.1) node [anchor=north west][inner sep=0.75pt]  [font=\tiny,rotate=-359.69,xslant=0.02]  {$$};
\draw  [color={rgb, 255:red, 0; green, 0; blue, 0 }  ,draw opacity=1 ][line width=1.5]   (739.84, 164.91) circle [x radius= 10.06, y radius= 10.06]   ;
\draw (733.87,162.34) node [anchor=north west][inner sep=0.75pt]  [font=\tiny,rotate=-359.69,xslant=0.02]  {$$};
\draw (733.84,35.06) node [anchor=north west][inner sep=0.75pt]    {$4$};
\draw (733.84,84.5) node [anchor=north west][inner sep=0.75pt]    {$7$};
\draw  [color={rgb, 255:red, 0; green, 0; blue, 0 }  ,draw opacity=1 ][fill={rgb, 255:red, 74; green, 144; blue, 226 }  ,fill opacity=1 ][line width=1.5]   (663.32, 116.37) circle [x radius= 10.06, y radius= 10.06]   ;
\draw (657.35,113.8) node [anchor=north west][inner sep=0.75pt]  [font=\tiny,rotate=-359.69,xslant=0.02]  {$$};
\draw (733.84,135.74) node [anchor=north west][inner sep=0.75pt]    {$8$};
\draw (744.63,48.08) node [anchor=north west][inner sep=0.75pt]  [font=\Large,color={rgb, 255:red, 208; green, 2; blue, 27 }  ,opacity=1 ]  {$\bot $};
\draw (587.27,104.87) node [anchor=north west][inner sep=0.75pt]  [font=\large,color={rgb, 255:red, 208; green, 2; blue, 27 }  ,opacity=1 ]  {$( 1,2,3)$};
\draw (752.41,100.48) node [anchor=north west][inner sep=0.75pt]  [font=\large,color={rgb, 255:red, 208; green, 2; blue, 27 }  ,opacity=1 ]  {$( 4,7,8)$};
\draw (753.25,151.71) node [anchor=north west][inner sep=0.75pt]  [font=\large,color={rgb, 255:red, 208; green, 2; blue, 27 }  ,opacity=1 ]  {$( 4,7,8)$};
\draw (956.08,34.82) node [anchor=north west][inner sep=0.75pt]    {$1$};
\draw (956.99,87.98) node [anchor=north west][inner sep=0.75pt]    {$2$};
\draw (957.91,136.57) node [anchor=north west][inner sep=0.75pt]    {$3$};
\draw  [color={rgb, 255:red, 0; green, 0; blue, 0 }  ,draw opacity=1 ][line width=1.5]   (962.08, 166.17) circle [x radius= 10.06, y radius= 10.06]   ;
\draw (956.11,163.6) node [anchor=north west][inner sep=0.75pt]  [font=\tiny,rotate=-359.69,xslant=0.02]  {$$};
\draw  [color={rgb, 255:red, 0; green, 0; blue, 0 }  ,draw opacity=1 ][line width=1.5]   (961.17, 64.42) circle [x radius= 10.06, y radius= 10.06]   ;
\draw (955.2,61.85) node [anchor=north west][inner sep=0.75pt]  [font=\tiny,rotate=-359.69,xslant=0.02]  {$$};
\draw  [color={rgb, 255:red, 0; green, 0; blue, 0 }  ,draw opacity=1 ][fill={rgb, 255:red, 74; green, 144; blue, 226 }  ,fill opacity=1 ][line width=1.5]   (1045.99, 113.92) circle [x radius= 10.06, y radius= 10.06]   ;
\draw (1040.02,111.35) node [anchor=north west][inner sep=0.75pt]  [font=\tiny,rotate=-359.69,xslant=0.02]  {$$};
\draw (1039.99,84.32) node [anchor=north west][inner sep=0.75pt]    {$7$};
\draw  [color={rgb, 255:red, 0; green, 0; blue, 0 }  ,draw opacity=1 ][line width=1.5]   (962.99, 116.67) circle [x radius= 10.06, y radius= 10.06]   ;
\draw (957.02,114.1) node [anchor=north west][inner sep=0.75pt]  [font=\tiny,rotate=-359.69,xslant=0.02]  {$$};
\draw (908.56,153.61) node [anchor=north west][inner sep=0.75pt]  [font=\Large,color={rgb, 255:red, 208; green, 2; blue, 27 }  ,opacity=1 ]  {$\bot $};
\draw (878.1,52.98) node [anchor=north west][inner sep=0.75pt]  [font=\large,color={rgb, 255:red, 208; green, 2; blue, 27 }  ,opacity=1 ]  {$( 1,2,3)$};
\draw (879.01,104.32) node [anchor=north west][inner sep=0.75pt]  [font=\large,color={rgb, 255:red, 208; green, 2; blue, 27 }  ,opacity=1 ]  {$( 1,2,3)$};
\draw (1062.32,100.65) node [anchor=north west][inner sep=0.75pt]  [font=\large,color={rgb, 255:red, 208; green, 2; blue, 27 }  ,opacity=1 ]  {$( 4,7,8)$};
\draw (1274.22,135.62) node [anchor=north west][inner sep=0.75pt]    {$3$};
\draw  [color={rgb, 255:red, 0; green, 0; blue, 0 }  ,draw opacity=1 ][fill={rgb, 255:red, 74; green, 144; blue, 226 }  ,fill opacity=1 ][line width=1.5]   (1278.41, 165.82) circle [x radius= 10.06, y radius= 10.06]   ;
\draw (1272.43,163.25) node [anchor=north west][inner sep=0.75pt]  [font=\tiny,rotate=-359.69,xslant=0.02]  {$$};
\draw  [color={rgb, 255:red, 0; green, 0; blue, 0 }  ,draw opacity=1 ][line width=1.5]   (1361.93, 60.35) circle [x radius= 10.06, y radius= 10.06]   ;
\draw (1355.95,57.78) node [anchor=north west][inner sep=0.75pt]  [font=\tiny,rotate=-359.69,xslant=0.02]  {$$};
\draw  [color={rgb, 255:red, 0; green, 0; blue, 0 }  ,draw opacity=1 ][line width=1.5]   (1361.93, 112.14) circle [x radius= 10.06, y radius= 10.06]   ;
\draw (1355.95,109.57) node [anchor=north west][inner sep=0.75pt]  [font=\tiny,rotate=-359.69,xslant=0.02]  {$$};
\draw  [color={rgb, 255:red, 0; green, 0; blue, 0 }  ,draw opacity=1 ][line width=1.5]   (1361.93, 165.82) circle [x radius= 10.06, y radius= 10.06]   ;
\draw (1355.95,163.25) node [anchor=north west][inner sep=0.75pt]  [font=\tiny,rotate=-359.69,xslant=0.02]  {$$};
\draw (1355.93,30.15) node [anchor=north west][inner sep=0.75pt]    {$4$};
\draw (1355.93,81.94) node [anchor=north west][inner sep=0.75pt]    {$7$};
\draw (1355.93,135.62) node [anchor=north west][inner sep=0.75pt]    {$8$};
\draw (1234.13,152.3) node [anchor=north west][inner sep=0.75pt]  [font=\Large,color={rgb, 255:red, 208; green, 2; blue, 27 }  ,opacity=1 ]  {$\bot $};
\draw (1368.48,153.25) node [anchor=north west][inner sep=0.75pt]  [font=\Large,color={rgb, 255:red, 208; green, 2; blue, 27 }  ,opacity=1 ]  {$\bot $};
\draw (1377.13,47.92) node [anchor=north west][inner sep=0.75pt]  [font=\large,color={rgb, 255:red, 208; green, 2; blue, 27 }  ,opacity=1 ]  {$( 4,7,8)$};
\draw (1378.04,99.72) node [anchor=north west][inner sep=0.75pt]  [font=\large,color={rgb, 255:red, 208; green, 2; blue, 27 }  ,opacity=1 ]  {$( 4,7,8)$};
\draw (1277.83,234.02) node [anchor=north west][inner sep=0.75pt]    {$9$};
\draw (1278.33,291.78) node [anchor=north west][inner sep=0.75pt]    {$10$};
\draw (1279.61,344.56) node [anchor=north west][inner sep=0.75pt]    {$3$};
\draw  [color={rgb, 255:red, 0; green, 0; blue, 0 }  ,draw opacity=1 ][line width=1.5]   (1283.83, 376.06) circle [x radius= 10.06, y radius= 10.06]   ;
\draw (1277.86,373.49) node [anchor=north west][inner sep=0.75pt]  [font=\tiny,rotate=-359.69,xslant=0.02]  {$$};
\draw  [color={rgb, 255:red, 0; green, 0; blue, 0 }  ,draw opacity=1 ][line width=1.5]   (1282.95, 265.52) circle [x radius= 10.06, y radius= 10.06]   ;
\draw (1276.98,262.95) node [anchor=north west][inner sep=0.75pt]  [font=\tiny,rotate=-359.69,xslant=0.02]  {$$};
\draw  [color={rgb, 255:red, 0; green, 0; blue, 0 }  ,draw opacity=1 ][fill={rgb, 255:red, 74; green, 144; blue, 226 }  ,fill opacity=1 ][line width=1.5]   (1365.48, 376.06) circle [x radius= 10.06, y radius= 10.06]   ;
\draw (1359.5,373.49) node [anchor=north west][inner sep=0.75pt]  [font=\tiny,rotate=-359.69,xslant=0.02]  {$$};
\draw  [color={rgb, 255:red, 0; green, 0; blue, 0 }  ,draw opacity=1 ][line width=1.5]   (1284.72, 322.28) circle [x radius= 10.06, y radius= 10.06]   ;
\draw (1278.75,319.72) node [anchor=north west][inner sep=0.75pt]  [font=\tiny,rotate=-359.69,xslant=0.02]  {$$};
\draw (1359.48,344.56) node [anchor=north west][inner sep=0.75pt]    {$8$};
\draw (1240.16,362.46) node [anchor=north west][inner sep=0.75pt]  [font=\Large,color={rgb, 255:red, 208; green, 2; blue, 27 }  ,opacity=1 ]  {$\bot $};
\draw (1371.49,363.46) node [anchor=north west][inner sep=0.75pt]  [font=\Large,color={rgb, 255:red, 208; green, 2; blue, 27 }  ,opacity=1 ]  {$\bot $};
\draw (1198.86,253.93) node [anchor=north west][inner sep=0.75pt]  [font=\large,color={rgb, 255:red, 208; green, 2; blue, 27 }  ,opacity=1 ]  {$( 3,9,10)$};
\draw (1200.63,308.7) node [anchor=north west][inner sep=0.75pt]  [font=\large,color={rgb, 255:red, 208; green, 2; blue, 27 }  ,opacity=1 ]  {$( 3,9,10)$};
\draw (1271.33,505.73) node [anchor=north west][inner sep=0.75pt]    {$10$};
\draw  [color={rgb, 255:red, 0; green, 0; blue, 0 }  ,draw opacity=1 ][line width=1.5]   (1368.33, 478.33) circle [x radius= 10.06, y radius= 10.06]   ;
\draw (1362.36,475.77) node [anchor=north west][inner sep=0.75pt]  [font=\tiny,rotate=-359.69,xslant=0.02]  {$$};
\draw  [color={rgb, 255:red, 0; green, 0; blue, 0 }  ,draw opacity=1 ][line width=1.5]   (1368.33, 533.33) circle [x radius= 10.06, y radius= 10.06]   ;
\draw (1362.36,530.77) node [anchor=north west][inner sep=0.75pt]  [font=\tiny,rotate=-359.69,xslant=0.02]  {$$};
\draw  [color={rgb, 255:red, 0; green, 0; blue, 0 }  ,draw opacity=1 ][line width=1.5]   (1368.33, 590.33) circle [x radius= 10.06, y radius= 10.06]   ;
\draw (1362.36,587.77) node [anchor=north west][inner sep=0.75pt]  [font=\tiny,rotate=-359.69,xslant=0.02]  {$$};
\draw (1362.33,446.73) node [anchor=north west][inner sep=0.75pt]    {$1$};
\draw (1362.33,501.73) node [anchor=north west][inner sep=0.75pt]    {$2$};
\draw  [color={rgb, 255:red, 0; green, 0; blue, 0 }  ,draw opacity=1 ][fill={rgb, 255:red, 74; green, 144; blue, 226 }  ,fill opacity=1 ][line width=1.5]   (1277.33, 536.33) circle [x radius= 10.06, y radius= 10.06]   ;
\draw (1271.36,533.77) node [anchor=north west][inner sep=0.75pt]  [font=\tiny,rotate=-359.69,xslant=0.02]  {$$};
\draw (1362.33,558.73) node [anchor=north west][inner sep=0.75pt]    {$8$};
\draw (1377.33,577.73) node [anchor=north west][inner sep=0.75pt]  [font=\Large,color={rgb, 255:red, 208; green, 2; blue, 27 }  ,opacity=1 ]  {$\bot $};
\draw (1187.33,522.73) node [anchor=north west][inner sep=0.75pt]  [font=\large,color={rgb, 255:red, 208; green, 2; blue, 27 }  ,opacity=1 ]  {$( 3,9,10)$};
\draw (1388.33,465.73) node [anchor=north west][inner sep=0.75pt]  [font=\large,color={rgb, 255:red, 208; green, 2; blue, 27 }  ,opacity=1 ]  {$( 1,2,8)$};
\draw (1389.33,520.73) node [anchor=north west][inner sep=0.75pt]  [font=\large,color={rgb, 255:red, 208; green, 2; blue, 27 }  ,opacity=1 ]  {$( 1,2,8)$};
\draw (1271.67,652.4) node [anchor=north west][inner sep=0.75pt]    {$9$};
\draw (1272.67,710.4) node [anchor=north west][inner sep=0.75pt]    {$10$};
\draw (1273.67,763.4) node [anchor=north west][inner sep=0.75pt]    {$3$};
\draw  [color={rgb, 255:red, 0; green, 0; blue, 0 }  ,draw opacity=1 ][line width=1.5]   (1277.67, 795) circle [x radius= 10.06, y radius= 10.06]   ;
\draw (1271.69,792.43) node [anchor=north west][inner sep=0.75pt]  [font=\tiny,rotate=-359.69,xslant=0.02]  {$$};
\draw  [color={rgb, 255:red, 0; green, 0; blue, 0 }  ,draw opacity=1 ][line width=1.5]   (1276.67, 684) circle [x radius= 10.06, y radius= 10.06]   ;
\draw (1270.69,681.43) node [anchor=north west][inner sep=0.75pt]  [font=\tiny,rotate=-359.69,xslant=0.02]  {$$};
\draw  [color={rgb, 255:red, 0; green, 0; blue, 0 }  ,draw opacity=1 ][fill={rgb, 255:red, 74; green, 144; blue, 226 }  ,fill opacity=1 ][line width=1.5]   (1369.67, 738) circle [x radius= 10.06, y radius= 10.06]   ;
\draw (1363.69,735.43) node [anchor=north west][inner sep=0.75pt]  [font=\tiny,rotate=-359.69,xslant=0.02]  {$$};
\draw (1363.67,706.4) node [anchor=north west][inner sep=0.75pt]    {$2$};
\draw  [color={rgb, 255:red, 0; green, 0; blue, 0 }  ,draw opacity=1 ][line width=1.5]   (1278.67, 741) circle [x radius= 10.06, y radius= 10.06]   ;
\draw (1272.69,738.43) node [anchor=north west][inner sep=0.75pt]  [font=\tiny,rotate=-359.69,xslant=0.02]  {$$};
\draw (1229.67,669.4) node [anchor=north west][inner sep=0.75pt]  [font=\Large,color={rgb, 255:red, 208; green, 2; blue, 27 }  ,opacity=1 ]  {$\bot $};
\draw (1185.67,728.4) node [anchor=north west][inner sep=0.75pt]  [font=\large,color={rgb, 255:red, 208; green, 2; blue, 27 }  ,opacity=1 ]  {$( 3,9,10)$};
\draw (1187.67,782.4) node [anchor=north west][inner sep=0.75pt]  [font=\large,color={rgb, 255:red, 208; green, 2; blue, 27 }  ,opacity=1 ]  {$( 3,9,10)$};
\draw (1390.67,725.4) node [anchor=north west][inner sep=0.75pt]  [font=\large,color={rgb, 255:red, 208; green, 2; blue, 27 }  ,opacity=1 ]  {$( 1,2,8)$};
\draw (872.33,527.73) node [anchor=north west][inner sep=0.75pt]    {$9$};
\draw  [color={rgb, 255:red, 0; green, 0; blue, 0 }  ,draw opacity=1 ][line width=1.5]   (970.33, 558.33) circle [x radius= 10.06, y radius= 10.06]   ;
\draw (964.36,555.77) node [anchor=north west][inner sep=0.75pt]  [font=\tiny,rotate=-359.69,xslant=0.02]  {$$};
\draw  [color={rgb, 255:red, 0; green, 0; blue, 0 }  ,draw opacity=1 ][fill={rgb, 255:red, 74; green, 144; blue, 226 }  ,fill opacity=1 ][line width=1.5]   (877.33, 559.33) circle [x radius= 10.06, y radius= 10.06]   ;
\draw (871.36,556.77) node [anchor=north west][inner sep=0.75pt]  [font=\tiny,rotate=-359.69,xslant=0.02]  {$$};
\draw  [color={rgb, 255:red, 0; green, 0; blue, 0 }  ,draw opacity=1 ][line width=1.5]   (970.33, 613.33) circle [x radius= 10.06, y radius= 10.06]   ;
\draw (964.36,610.77) node [anchor=north west][inner sep=0.75pt]  [font=\tiny,rotate=-359.69,xslant=0.02]  {$$};
\draw  [color={rgb, 255:red, 0; green, 0; blue, 0 }  ,draw opacity=1 ][line width=1.5]   (970.33, 670.33) circle [x radius= 10.06, y radius= 10.06]   ;
\draw (964.36,667.77) node [anchor=north west][inner sep=0.75pt]  [font=\tiny,rotate=-359.69,xslant=0.02]  {$$};
\draw (964.33,526.73) node [anchor=north west][inner sep=0.75pt]    {$1$};
\draw (964.33,581.73) node [anchor=north west][inner sep=0.75pt]    {$2$};
\draw (964.33,638.73) node [anchor=north west][inner sep=0.75pt]    {$8$};
\draw (830.33,544.73) node [anchor=north west][inner sep=0.75pt]  [font=\Large,color={rgb, 255:red, 208; green, 2; blue, 27 }  ,opacity=1 ]  {$\bot $};
\draw (980.33,546.73) node [anchor=north west][inner sep=0.75pt]  [font=\Large,color={rgb, 255:red, 208; green, 2; blue, 27 }  ,opacity=1 ]  {$\bot $};
\draw (986.33,599.73) node [anchor=north west][inner sep=0.75pt]  [font=\large,color={rgb, 255:red, 208; green, 2; blue, 27 }  ,opacity=1 ]  {$( 1,2,8)$};
\draw (984.33,658.73) node [anchor=north west][inner sep=0.75pt]  [font=\large,color={rgb, 255:red, 208; green, 2; blue, 27 }  ,opacity=1 ]  {$( 1,2,8)$};
\draw (844,728.73) node [anchor=north west][inner sep=0.75pt]    {$9$};
\draw (845,786.73) node [anchor=north west][inner sep=0.75pt]    {$4$};
\draw (846,839.73) node [anchor=north west][inner sep=0.75pt]    {$5$};
\draw  [color={rgb, 255:red, 0; green, 0; blue, 0 }  ,draw opacity=1 ][line width=1.5]   (850, 871.33) circle [x radius= 10.06, y radius= 10.06]   ;
\draw (844.03,868.77) node [anchor=north west][inner sep=0.75pt]  [font=\tiny,rotate=-359.69,xslant=0.02]  {$$};
\draw  [color={rgb, 255:red, 0; green, 0; blue, 0 }  ,draw opacity=1 ][fill={rgb, 255:red, 74; green, 144; blue, 226 }  ,fill opacity=1 ][line width=1.5]   (942, 759.33) circle [x radius= 10.06, y radius= 10.06]   ;
\draw (936.03,756.77) node [anchor=north west][inner sep=0.75pt]  [font=\tiny,rotate=-359.69,xslant=0.02]  {$$};
\draw  [color={rgb, 255:red, 0; green, 0; blue, 0 }  ,draw opacity=1 ][line width=1.5]   (849, 760.33) circle [x radius= 10.06, y radius= 10.06]   ;
\draw (843.03,757.77) node [anchor=north west][inner sep=0.75pt]  [font=\tiny,rotate=-359.69,xslant=0.02]  {$$};
\draw (936,727.73) node [anchor=north west][inner sep=0.75pt]    {$1$};
\draw  [color={rgb, 255:red, 0; green, 0; blue, 0 }  ,draw opacity=1 ][line width=1.5]   (851, 817.33) circle [x radius= 10.06, y radius= 10.06]   ;
\draw (845.03,814.77) node [anchor=north west][inner sep=0.75pt]  [font=\tiny,rotate=-359.69,xslant=0.02]  {$$};
\draw (802,745.73) node [anchor=north west][inner sep=0.75pt]  [font=\Large,color={rgb, 255:red, 208; green, 2; blue, 27 }  ,opacity=1 ]  {$\bot $};
\draw (952,747.73) node [anchor=north west][inner sep=0.75pt]  [font=\Large,color={rgb, 255:red, 208; green, 2; blue, 27 }  ,opacity=1 ]  {$\bot $};
\draw (758,804.73) node [anchor=north west][inner sep=0.75pt]  [font=\large,color={rgb, 255:red, 208; green, 2; blue, 27 }  ,opacity=1 ]  {$( 4,5,9)$};
\draw (760,858.73) node [anchor=north west][inner sep=0.75pt]  [font=\large,color={rgb, 255:red, 208; green, 2; blue, 27 }  ,opacity=1 ]  {$( 4,5,9)$};
\draw (487.33,587.02) node [anchor=north west][inner sep=0.75pt]    {$4$};
\draw  [color={rgb, 255:red, 0; green, 0; blue, 0 }  ,draw opacity=1 ][line width=1.5]   (584.33, 559.62) circle [x radius= 10.06, y radius= 10.06]   ;
\draw (578.36,557.06) node [anchor=north west][inner sep=0.75pt]  [font=\tiny,rotate=-359.69,xslant=0.02]  {$$};
\draw  [color={rgb, 255:red, 0; green, 0; blue, 0 }  ,draw opacity=1 ][line width=1.5]   (584.33, 614.62) circle [x radius= 10.06, y radius= 10.06]   ;
\draw (578.36,612.06) node [anchor=north west][inner sep=0.75pt]  [font=\tiny,rotate=-359.69,xslant=0.02]  {$$};
\draw  [color={rgb, 255:red, 0; green, 0; blue, 0 }  ,draw opacity=1 ][line width=1.5]   (584.33, 671.62) circle [x radius= 10.06, y radius= 10.06]   ;
\draw (578.36,669.06) node [anchor=north west][inner sep=0.75pt]  [font=\tiny,rotate=-359.69,xslant=0.02]  {$$};
\draw (578.33,528.02) node [anchor=north west][inner sep=0.75pt]    {$1$};
\draw (578.33,583.02) node [anchor=north west][inner sep=0.75pt]    {$2$};
\draw  [color={rgb, 255:red, 0; green, 0; blue, 0 }  ,draw opacity=1 ][fill={rgb, 255:red, 74; green, 144; blue, 226 }  ,fill opacity=1 ][line width=1.5]   (493.33, 617.62) circle [x radius= 10.06, y radius= 10.06]   ;
\draw (487.36,615.06) node [anchor=north west][inner sep=0.75pt]  [font=\tiny,rotate=-359.69,xslant=0.02]  {$$};
\draw (578.33,640.02) node [anchor=north west][inner sep=0.75pt]    {$3$};
\draw (594.33,548.02) node [anchor=north west][inner sep=0.75pt]  [font=\Large,color={rgb, 255:red, 208; green, 2; blue, 27 }  ,opacity=1 ]  {$\bot $};
\draw (400.33,605.02) node [anchor=north west][inner sep=0.75pt]  [font=\large,color={rgb, 255:red, 208; green, 2; blue, 27 }  ,opacity=1 ]  {$( 4,5,9)$};
\draw (600.33,601.02) node [anchor=north west][inner sep=0.75pt]  [font=\large,color={rgb, 255:red, 208; green, 2; blue, 27 }  ,opacity=1 ]  {$( 1,2,3)$};
\draw (598.33,660.02) node [anchor=north west][inner sep=0.75pt]  [font=\large,color={rgb, 255:red, 208; green, 2; blue, 27 }  ,opacity=1 ]  {$( 1,2,3)$};
\draw (426.67,725.02) node [anchor=north west][inner sep=0.75pt]    {$9$};
\draw (427.67,783.02) node [anchor=north west][inner sep=0.75pt]    {$4$};
\draw (428.67,836.02) node [anchor=north west][inner sep=0.75pt]    {$5$};
\draw  [color={rgb, 255:red, 0; green, 0; blue, 0 }  ,draw opacity=1 ][line width=1.5]   (432.67, 867.62) circle [x radius= 10.06, y radius= 10.06]   ;
\draw (426.69,865.06) node [anchor=north west][inner sep=0.75pt]  [font=\tiny,rotate=-359.69,xslant=0.02]  {$$};
\draw  [color={rgb, 255:red, 0; green, 0; blue, 0 }  ,draw opacity=1 ][line width=1.5]   (431.67, 756.62) circle [x radius= 10.06, y radius= 10.06]   ;
\draw (425.69,754.06) node [anchor=north west][inner sep=0.75pt]  [font=\tiny,rotate=-359.69,xslant=0.02]  {$$};
\draw  [color={rgb, 255:red, 0; green, 0; blue, 0 }  ,draw opacity=1 ][fill={rgb, 255:red, 74; green, 144; blue, 226 }  ,fill opacity=1 ][line width=1.5]   (524.67, 867.62) circle [x radius= 10.06, y radius= 10.06]   ;
\draw (518.69,865.06) node [anchor=north west][inner sep=0.75pt]  [font=\tiny,rotate=-359.69,xslant=0.02]  {$$};
\draw  [color={rgb, 255:red, 0; green, 0; blue, 0 }  ,draw opacity=1 ][line width=1.5]   (433.67, 813.62) circle [x radius= 10.06, y radius= 10.06]   ;
\draw (427.69,811.06) node [anchor=north west][inner sep=0.75pt]  [font=\tiny,rotate=-359.69,xslant=0.02]  {$$};
\draw (518.67,836.02) node [anchor=north west][inner sep=0.75pt]    {$3$};
\draw (388.67,853.02) node [anchor=north west][inner sep=0.75pt]  [font=\Large,color={rgb, 255:red, 208; green, 2; blue, 27 }  ,opacity=1 ]  {$\bot $};
\draw (349.67,744.02) node [anchor=north west][inner sep=0.75pt]  [font=\large,color={rgb, 255:red, 208; green, 2; blue, 27 }  ,opacity=1 ]  {$( 4,5,9)$};
\draw (352.67,800.02) node [anchor=north west][inner sep=0.75pt]  [font=\large,color={rgb, 255:red, 208; green, 2; blue, 27 }  ,opacity=1 ]  {$( 4,5,9)$};
\draw (538.67,856.02) node [anchor=north west][inner sep=0.75pt]  [font=\large,color={rgb, 255:red, 208; green, 2; blue, 27 }  ,opacity=1 ]  {$( 1,2,3)$};
\draw (92.67,752.02) node [anchor=north west][inner sep=0.75pt]    {$5$};
\draw  [color={rgb, 255:red, 0; green, 0; blue, 0 }  ,draw opacity=1 ][fill={rgb, 255:red, 74; green, 144; blue, 226 }  ,fill opacity=1 ][line width=1.5]   (96.67, 783.62) circle [x radius= 10.06, y radius= 10.06]   ;
\draw (90.69,781.06) node [anchor=north west][inner sep=0.75pt]  [font=\tiny,rotate=-359.69,xslant=0.02]  {$$};
\draw  [color={rgb, 255:red, 0; green, 0; blue, 0 }  ,draw opacity=1 ][line width=1.5]   (188.67, 671.62) circle [x radius= 10.06, y radius= 10.06]   ;
\draw (182.69,669.06) node [anchor=north west][inner sep=0.75pt]  [font=\tiny,rotate=-359.69,xslant=0.02]  {$$};
\draw  [color={rgb, 255:red, 0; green, 0; blue, 0 }  ,draw opacity=1 ][line width=1.5]   (188.67, 726.62) circle [x radius= 10.06, y radius= 10.06]   ;
\draw (182.69,724.06) node [anchor=north west][inner sep=0.75pt]  [font=\tiny,rotate=-359.69,xslant=0.02]  {$$};
\draw  [color={rgb, 255:red, 0; green, 0; blue, 0 }  ,draw opacity=1 ][line width=1.5]   (188.67, 783.62) circle [x radius= 10.06, y radius= 10.06]   ;
\draw (182.69,781.06) node [anchor=north west][inner sep=0.75pt]  [font=\tiny,rotate=-359.69,xslant=0.02]  {$$};
\draw (182.67,640.02) node [anchor=north west][inner sep=0.75pt]    {$1$};
\draw (182.67,695.02) node [anchor=north west][inner sep=0.75pt]    {$2$};
\draw (182.67,752.02) node [anchor=north west][inner sep=0.75pt]    {$3$};
\draw (52.67,769.02) node [anchor=north west][inner sep=0.75pt]  [font=\Large,color={rgb, 255:red, 208; green, 2; blue, 27 }  ,opacity=1 ]  {$\bot $};
\draw (198.67,660.02) node [anchor=north west][inner sep=0.75pt]  [font=\Large,color={rgb, 255:red, 208; green, 2; blue, 27 }  ,opacity=1 ]  {$\bot $};
\draw (204.67,713.02) node [anchor=north west][inner sep=0.75pt]  [font=\large,color={rgb, 255:red, 208; green, 2; blue, 27 }  ,opacity=1 ]  {$( 1,2,3)$};
\draw (202.67,772.02) node [anchor=north west][inner sep=0.75pt]  [font=\large,color={rgb, 255:red, 208; green, 2; blue, 27 }  ,opacity=1 ]  {$( 1,2,3)$};
\draw (91.67,430.36) node [anchor=north west][inner sep=0.75pt]    {$6$};
\draw (92.67,488.36) node [anchor=north west][inner sep=0.75pt]    {$4$};
\draw (93.67,541.36) node [anchor=north west][inner sep=0.75pt]    {$5$};
\draw  [color={rgb, 255:red, 0; green, 0; blue, 0 }  ,draw opacity=1 ][line width=1.5]   (97.67, 572.96) circle [x radius= 10.06, y radius= 10.06]   ;
\draw (91.69,570.39) node [anchor=north west][inner sep=0.75pt]  [font=\tiny,rotate=-359.69,xslant=0.02]  {$$};
\draw  [color={rgb, 255:red, 0; green, 0; blue, 0 }  ,draw opacity=1 ][line width=1.5]   (96.67, 461.96) circle [x radius= 10.06, y radius= 10.06]   ;
\draw (90.69,459.39) node [anchor=north west][inner sep=0.75pt]  [font=\tiny,rotate=-359.69,xslant=0.02]  {$$};
\draw  [color={rgb, 255:red, 0; green, 0; blue, 0 }  ,draw opacity=1 ][fill={rgb, 255:red, 74; green, 144; blue, 226 }  ,fill opacity=1 ][line width=1.5]   (189.67, 515.96) circle [x radius= 10.06, y radius= 10.06]   ;
\draw (183.69,513.39) node [anchor=north west][inner sep=0.75pt]  [font=\tiny,rotate=-359.69,xslant=0.02]  {$$};
\draw (183.67,484.36) node [anchor=north west][inner sep=0.75pt]    {$2$};
\draw  [color={rgb, 255:red, 0; green, 0; blue, 0 }  ,draw opacity=1 ][line width=1.5]   (98.67, 518.96) circle [x radius= 10.06, y radius= 10.06]   ;
\draw (92.69,516.39) node [anchor=north west][inner sep=0.75pt]  [font=\tiny,rotate=-359.69,xslant=0.02]  {$$};
\draw (53.67,558.36) node [anchor=north west][inner sep=0.75pt]  [font=\Large,color={rgb, 255:red, 208; green, 2; blue, 27 }  ,opacity=1 ]  {$\bot $};
\draw (14.67,449.36) node [anchor=north west][inner sep=0.75pt]  [font=\large,color={rgb, 255:red, 208; green, 2; blue, 27 }  ,opacity=1 ]  {$( 4,5,6)$};
\draw (17.67,505.36) node [anchor=north west][inner sep=0.75pt]  [font=\large,color={rgb, 255:red, 208; green, 2; blue, 27 }  ,opacity=1 ]  {$( 4,5,6)$};
\draw (205.67,502.36) node [anchor=north west][inner sep=0.75pt]  [font=\large,color={rgb, 255:red, 208; green, 2; blue, 27 }  ,opacity=1 ]  {$( 1,2,3)$};
\draw (92.67,234.36) node [anchor=north west][inner sep=0.75pt]    {$6$};
\draw  [color={rgb, 255:red, 0; green, 0; blue, 0 }  ,draw opacity=1 ][line width=1.5]   (190.67, 264.96) circle [x radius= 10.06, y radius= 10.06]   ;
\draw (184.69,262.39) node [anchor=north west][inner sep=0.75pt]  [font=\tiny,rotate=-359.69,xslant=0.02]  {$$};
\draw  [color={rgb, 255:red, 0; green, 0; blue, 0 }  ,draw opacity=1 ][fill={rgb, 255:red, 74; green, 144; blue, 226 }  ,fill opacity=1 ][line width=1.5]   (97.67, 265.96) circle [x radius= 10.06, y radius= 10.06]   ;
\draw (91.69,263.39) node [anchor=north west][inner sep=0.75pt]  [font=\tiny,rotate=-359.69,xslant=0.02]  {$$};
\draw  [color={rgb, 255:red, 0; green, 0; blue, 0 }  ,draw opacity=1 ][line width=1.5]   (190.67, 319.96) circle [x radius= 10.06, y radius= 10.06]   ;
\draw (184.69,317.39) node [anchor=north west][inner sep=0.75pt]  [font=\tiny,rotate=-359.69,xslant=0.02]  {$$};
\draw  [color={rgb, 255:red, 0; green, 0; blue, 0 }  ,draw opacity=1 ][line width=1.5]   (190.67, 376.96) circle [x radius= 10.06, y radius= 10.06]   ;
\draw (184.69,374.39) node [anchor=north west][inner sep=0.75pt]  [font=\tiny,rotate=-359.69,xslant=0.02]  {$$};
\draw (184.67,233.36) node [anchor=north west][inner sep=0.75pt]    {$1$};
\draw (184.67,288.36) node [anchor=north west][inner sep=0.75pt]    {$2$};
\draw (184.67,345.36) node [anchor=north west][inner sep=0.75pt]    {$3$};
\draw (200.67,253.36) node [anchor=north west][inner sep=0.75pt]  [font=\Large,color={rgb, 255:red, 208; green, 2; blue, 27 }  ,opacity=1 ]  {$\bot $};
\draw (15.67,253.36) node [anchor=north west][inner sep=0.75pt]  [font=\large,color={rgb, 255:red, 208; green, 2; blue, 27 }  ,opacity=1 ]  {$( 4,5,6)$};
\draw (206.67,306.36) node [anchor=north west][inner sep=0.75pt]  [font=\large,color={rgb, 255:red, 208; green, 2; blue, 27 }  ,opacity=1 ]  {$( 1,2,3)$};
\draw (204.67,365.36) node [anchor=north west][inner sep=0.75pt]  [font=\large,color={rgb, 255:red, 208; green, 2; blue, 27 }  ,opacity=1 ]  {$( 1,2,3)$};
\draw    (104.75,64.13) -- (154.02,63.52) ;
\draw    (103.03,69.88) -- (155.73,105.32) ;
\draw    (100.58,72.41) -- (158.18,152.06) ;
\draw    (387.88,66.69) -- (441.73,66.07) ;
\draw    (385.53,152.49) -- (445.68,73.95) ;
\draw    (387.72,109.67) -- (443.48,71.63) ;
\draw    (673.38,116.01) -- (729.78,114.03) ;
\draw    (671.82,121.76) -- (731.34,159.52) ;
\draw    (671.64,110.7) -- (731.52,69.9) ;
\draw    (969.86,69.49) -- (1037.29,108.84) ;
\draw    (973.05,116.33) -- (1035.93,114.25) ;
\draw    (970.63,160.85) -- (1037.44,119.24) ;
\draw    (1284.66,157.93) -- (1355.68,68.24) ;
\draw    (1286.87,160.37) -- (1353.46,117.58) ;
\draw    (1288.47,165.82) -- (1351.86,165.82) ;
\draw    (1288.97,273.59) -- (1359.45,367.99) ;
\draw    (1293.1,327.86) -- (1357.1,370.48) ;
\draw    (1293.9,376.06) -- (1355.41,376.06) ;
\draw    (1287.39,536) -- (1358.27,533.66) ;
\draw    (1285.99,541.47) -- (1359.68,585.2) ;
\draw    (1285.82,530.92) -- (1359.84,483.74) ;
\draw    (1285.37,689.05) -- (1360.96,732.95) ;
\draw    (1288.73,740.67) -- (1359.61,738.33) ;
\draw    (1286.22,789.7) -- (1361.11,743.3) ;
\draw    (887.4,559.23) -- (960.27,558.44) ;
\draw    (886.04,564.39) -- (961.63,608.28) ;
\draw    (883.8,567.05) -- (963.87,662.62) ;
\draw    (859.06,760.23) -- (931.94,759.44) ;
\draw    (856.39,863.56) -- (935.61,767.11) ;
\draw    (859.49,811.92) -- (933.51,764.74) ;
\draw    (503.39,617.29) -- (574.27,614.96) ;
\draw    (501.99,622.76) -- (575.68,666.49) ;
\draw    (501.82,612.21) -- (575.84,565.03) ;
\draw    (438.13,764.34) -- (518.2,859.91) ;
\draw    (442.32,818.76) -- (516.01,862.49) ;
\draw    (442.73,867.62) -- (514.6,867.62) ;
\draw    (103.06,775.85) -- (182.28,679.4) ;
\draw    (105.22,778.32) -- (180.11,731.93) ;
\draw    (106.73,783.62) -- (178.6,783.62) ;
\draw    (105.37,467.01) -- (180.96,510.9) ;
\draw    (108.73,518.63) -- (179.61,516.29) ;
\draw    (106.22,567.66) -- (181.11,521.26) ;
\draw    (107.73,265.85) -- (180.6,265.07) ;
\draw    (106.37,271.01) -- (181.96,314.9) ;
\draw    (104.13,273.67) -- (184.2,369.24) ;

\end{tikzpicture}

%% file: figs/wise_instances.tex
\begin{tikzpicture}[x=0.75pt,y=0.75pt,yscale=-1,xscale=1,every node/.style={scale=1.25}]

\draw    (-2.33,272) -- (1383,266.67) ;
\draw    (0.67,546.67) -- (1383,540) ;
\draw    (321.67,40) -- (320.33,796) ;
\draw    (676.33,49.33) -- (676.33,798.67) ;
\draw    (1027,45.33) -- (1027,804) ;

\draw (95,65.07) node [anchor=north west][inner sep=0.75pt]    {$1$};
\draw (96,123.07) node [anchor=north west][inner sep=0.75pt]    {$2$};
\draw (97,176.07) node [anchor=north west][inner sep=0.75pt]    {$3$};
\draw  [color={rgb, 255:red, 0; green, 0; blue, 0 }  ,draw opacity=1 ][line width=1.5]   (102, 206.67) circle [x radius= 10.06, y radius= 10.06]   ;
\draw (96.03,204.1) node [anchor=north west][inner sep=0.75pt]  [font=\tiny,rotate=-359.69,xslant=0.02]  {$$};
\draw  [color={rgb, 255:red, 0; green, 0; blue, 0 }  ,draw opacity=1 ][line width=1.5]   (193, 95.67) circle [x radius= 10.06, y radius= 10.06]   ;
\draw (187.03,93.1) node [anchor=north west][inner sep=0.75pt]  [font=\tiny,rotate=-359.69,xslant=0.02]  {$$};
\draw  [color={rgb, 255:red, 0; green, 0; blue, 0 }  ,draw opacity=1 ][line width=1.5]   (100, 96.67) circle [x radius= 10.06, y radius= 10.06]   ;
\draw (94.03,94.1) node [anchor=north west][inner sep=0.75pt]  [font=\tiny,rotate=-359.69,xslant=0.02]  {$$};
\draw  [color={rgb, 255:red, 0; green, 0; blue, 0 }  ,draw opacity=1 ][line width=1.5]   (193, 150.67) circle [x radius= 10.06, y radius= 10.06]   ;
\draw (187.03,148.1) node [anchor=north west][inner sep=0.75pt]  [font=\tiny,rotate=-359.69,xslant=0.02]  {$$};
\draw  [color={rgb, 255:red, 0; green, 0; blue, 0 }  ,draw opacity=1 ][line width=1.5]   (193, 207.67) circle [x radius= 10.06, y radius= 10.06]   ;
\draw (187.03,205.1) node [anchor=north west][inner sep=0.75pt]  [font=\tiny,rotate=-359.69,xslant=0.02]  {$$};
\draw (187,64.07) node [anchor=north west][inner sep=0.75pt]    {$4$};
\draw (187,119.07) node [anchor=north west][inner sep=0.75pt]    {$5$};
\draw  [color={rgb, 255:red, 0; green, 0; blue, 0 }  ,draw opacity=1 ][line width=1.5]   (102, 153.67) circle [x radius= 10.06, y radius= 10.06]   ;
\draw (96.03,151.1) node [anchor=north west][inner sep=0.75pt]  [font=\tiny,rotate=-359.69,xslant=0.02]  {$$};
\draw (187,176.07) node [anchor=north west][inner sep=0.75pt]    {$6$};
\draw (56,81.07) node [anchor=north west][inner sep=0.75pt]  [font=\Large,color={rgb, 255:red, 208; green, 2; blue, 27 }  ,opacity=1 ]  {$\bot $};
\draw (202,79.07) node [anchor=north west][inner sep=0.75pt]  [font=\Large,color={rgb, 255:red, 208; green, 2; blue, 27 }  ,opacity=1 ]  {$\bot $};
\draw (20,143.07) node [anchor=north west][inner sep=0.75pt]  [font=\large,color={rgb, 255:red, 208; green, 2; blue, 27 }  ,opacity=1 ]  {$( 1,2,3)$};
\draw (17,195.07) node [anchor=north west][inner sep=0.75pt]  [font=\large,color={rgb, 255:red, 208; green, 2; blue, 27 }  ,opacity=1 ]  {$( 1,2,3)$};
\draw (214,137.07) node [anchor=north west][inner sep=0.75pt]  [font=\large,color={rgb, 255:red, 208; green, 2; blue, 27 }  ,opacity=1 ]  {$( 4,5,6)$};
\draw (215,194.07) node [anchor=north west][inner sep=0.75pt]  [font=\large,color={rgb, 255:red, 208; green, 2; blue, 27 }  ,opacity=1 ]  {$( 4,5,6)$};
\draw (445,68.07) node [anchor=north west][inner sep=0.75pt]    {$1$};
\draw (446,126.07) node [anchor=north west][inner sep=0.75pt]    {$2$};
\draw (447,179.07) node [anchor=north west][inner sep=0.75pt]    {$3$};
\draw  [color={rgb, 255:red, 0; green, 0; blue, 0 }  ,draw opacity=1 ][line width=1.5]   (451, 210.67) circle [x radius= 10.06, y radius= 10.06]   ;
\draw (445.03,208.1) node [anchor=north west][inner sep=0.75pt]  [font=\tiny,rotate=-359.69,xslant=0.02]  {$$};
\draw  [color={rgb, 255:red, 0; green, 0; blue, 0 }  ,draw opacity=1 ][line width=1.5]   (543, 98.67) circle [x radius= 10.06, y radius= 10.06]   ;
\draw (537.03,96.1) node [anchor=north west][inner sep=0.75pt]  [font=\tiny,rotate=-359.69,xslant=0.02]  {$$};
\draw  [color={rgb, 255:red, 0; green, 0; blue, 0 }  ,draw opacity=1 ][line width=1.5]   (450, 99.67) circle [x radius= 10.06, y radius= 10.06]   ;
\draw (444.03,97.1) node [anchor=north west][inner sep=0.75pt]  [font=\tiny,rotate=-359.69,xslant=0.02]  {$$};
\draw  [color={rgb, 255:red, 0; green, 0; blue, 0 }  ,draw opacity=1 ][line width=1.5]   (543, 153.67) circle [x radius= 10.06, y radius= 10.06]   ;
\draw (537.03,151.1) node [anchor=north west][inner sep=0.75pt]  [font=\tiny,rotate=-359.69,xslant=0.02]  {$$};
\draw  [color={rgb, 255:red, 0; green, 0; blue, 0 }  ,draw opacity=1 ][line width=1.5]   (543, 210.67) circle [x radius= 10.06, y radius= 10.06]   ;
\draw (537.03,208.1) node [anchor=north west][inner sep=0.75pt]  [font=\tiny,rotate=-359.69,xslant=0.02]  {$$};
\draw (537,67.07) node [anchor=north west][inner sep=0.75pt]    {$4$};
\draw (537,122.07) node [anchor=north west][inner sep=0.75pt]    {$7$};
\draw  [color={rgb, 255:red, 0; green, 0; blue, 0 }  ,draw opacity=1 ][line width=1.5]   (452, 156.67) circle [x radius= 10.06, y radius= 10.06]   ;
\draw (446.03,154.1) node [anchor=north west][inner sep=0.75pt]  [font=\tiny,rotate=-359.69,xslant=0.02]  {$$};
\draw (537,179.07) node [anchor=north west][inner sep=0.75pt]    {$8$};
\draw (406,84.07) node [anchor=north west][inner sep=0.75pt]  [font=\Large,color={rgb, 255:red, 208; green, 2; blue, 27 }  ,opacity=1 ]  {$\bot $};
\draw (552,82.07) node [anchor=north west][inner sep=0.75pt]  [font=\Large,color={rgb, 255:red, 208; green, 2; blue, 27 }  ,opacity=1 ]  {$\bot $};
\draw (370,146.07) node [anchor=north west][inner sep=0.75pt]  [font=\large,color={rgb, 255:red, 208; green, 2; blue, 27 }  ,opacity=1 ]  {$( 1,2,3)$};
\draw (367,198.07) node [anchor=north west][inner sep=0.75pt]  [font=\large,color={rgb, 255:red, 208; green, 2; blue, 27 }  ,opacity=1 ]  {$( 1,2,3)$};
\draw (564,140.07) node [anchor=north west][inner sep=0.75pt]  [font=\large,color={rgb, 255:red, 208; green, 2; blue, 27 }  ,opacity=1 ]  {$( 4,7,8)$};
\draw (565,197.07) node [anchor=north west][inner sep=0.75pt]  [font=\large,color={rgb, 255:red, 208; green, 2; blue, 27 }  ,opacity=1 ]  {$( 4,7,8)$};
\draw (793.33,72.73) node [anchor=north west][inner sep=0.75pt]    {$1$};
\draw (794.33,130.73) node [anchor=north west][inner sep=0.75pt]    {$2$};
\draw (795.33,183.73) node [anchor=north west][inner sep=0.75pt]    {$3$};
\draw  [color={rgb, 255:red, 0; green, 0; blue, 0 }  ,draw opacity=1 ][line width=1.5]   (799.33, 215.33) circle [x radius= 10.06, y radius= 10.06]   ;
\draw (793.36,212.77) node [anchor=north west][inner sep=0.75pt]  [font=\tiny,rotate=-359.69,xslant=0.02]  {$$};
\draw  [color={rgb, 255:red, 0; green, 0; blue, 0 }  ,draw opacity=1 ][line width=1.5]   (891.33, 103.33) circle [x radius= 10.06, y radius= 10.06]   ;
\draw (885.36,100.77) node [anchor=north west][inner sep=0.75pt]  [font=\tiny,rotate=-359.69,xslant=0.02]  {$$};
\draw  [color={rgb, 255:red, 0; green, 0; blue, 0 }  ,draw opacity=1 ][line width=1.5]   (798.33, 104.33) circle [x radius= 10.06, y radius= 10.06]   ;
\draw (792.36,101.77) node [anchor=north west][inner sep=0.75pt]  [font=\tiny,rotate=-359.69,xslant=0.02]  {$$};
\draw  [color={rgb, 255:red, 0; green, 0; blue, 0 }  ,draw opacity=1 ][line width=1.5]   (891.33, 158.33) circle [x radius= 10.06, y radius= 10.06]   ;
\draw (885.36,155.77) node [anchor=north west][inner sep=0.75pt]  [font=\tiny,rotate=-359.69,xslant=0.02]  {$$};
\draw  [color={rgb, 255:red, 0; green, 0; blue, 0 }  ,draw opacity=1 ][line width=1.5]   (891.33, 215.33) circle [x radius= 10.06, y radius= 10.06]   ;
\draw (885.36,212.77) node [anchor=north west][inner sep=0.75pt]  [font=\tiny,rotate=-359.69,xslant=0.02]  {$$};
\draw (885.33,71.73) node [anchor=north west][inner sep=0.75pt]    {$4$};
\draw (885.33,126.73) node [anchor=north west][inner sep=0.75pt]    {$7$};
\draw  [color={rgb, 255:red, 0; green, 0; blue, 0 }  ,draw opacity=1 ][line width=1.5]   (800.33, 161.33) circle [x radius= 10.06, y radius= 10.06]   ;
\draw (794.36,158.77) node [anchor=north west][inner sep=0.75pt]  [font=\tiny,rotate=-359.69,xslant=0.02]  {$$};
\draw (885.33,183.73) node [anchor=north west][inner sep=0.75pt]    {$8$};
\draw (742.33,202.73) node [anchor=north west][inner sep=0.75pt]  [font=\Large,color={rgb, 255:red, 208; green, 2; blue, 27 }  ,opacity=1 ]  {$\bot $};
\draw (900.33,86.73) node [anchor=north west][inner sep=0.75pt]  [font=\Large,color={rgb, 255:red, 208; green, 2; blue, 27 }  ,opacity=1 ]  {$\bot $};
\draw (710.33,92.73) node [anchor=north west][inner sep=0.75pt]  [font=\large,color={rgb, 255:red, 208; green, 2; blue, 27 }  ,opacity=1 ]  {$( 1,2,3)$};
\draw (711.33,148.73) node [anchor=north west][inner sep=0.75pt]  [font=\large,color={rgb, 255:red, 208; green, 2; blue, 27 }  ,opacity=1 ]  {$( 1,2,3)$};
\draw (912.33,144.73) node [anchor=north west][inner sep=0.75pt]  [font=\large,color={rgb, 255:red, 208; green, 2; blue, 27 }  ,opacity=1 ]  {$( 4,7,8)$};
\draw (913.33,201.73) node [anchor=north west][inner sep=0.75pt]  [font=\large,color={rgb, 255:red, 208; green, 2; blue, 27 }  ,opacity=1 ]  {$( 4,7,8)$};
\draw (1137.33,75.73) node [anchor=north west][inner sep=0.75pt]    {$1$};
\draw (1138.33,133.73) node [anchor=north west][inner sep=0.75pt]    {$2$};
\draw (1139.33,186.73) node [anchor=north west][inner sep=0.75pt]    {$3$};
\draw  [color={rgb, 255:red, 0; green, 0; blue, 0 }  ,draw opacity=1 ][line width=1.5]   (1143.33, 218.33) circle [x radius= 10.06, y radius= 10.06]   ;
\draw (1137.36,215.77) node [anchor=north west][inner sep=0.75pt]  [font=\tiny,rotate=-359.69,xslant=0.02]  {$$};
\draw  [color={rgb, 255:red, 0; green, 0; blue, 0 }  ,draw opacity=1 ][line width=1.5]   (1235.33, 106.33) circle [x radius= 10.06, y radius= 10.06]   ;
\draw (1229.36,103.77) node [anchor=north west][inner sep=0.75pt]  [font=\tiny,rotate=-359.69,xslant=0.02]  {$$};
\draw  [color={rgb, 255:red, 0; green, 0; blue, 0 }  ,draw opacity=1 ][line width=1.5]   (1142.33, 107.33) circle [x radius= 10.06, y radius= 10.06]   ;
\draw (1136.36,104.77) node [anchor=north west][inner sep=0.75pt]  [font=\tiny,rotate=-359.69,xslant=0.02]  {$$};
\draw  [color={rgb, 255:red, 0; green, 0; blue, 0 }  ,draw opacity=1 ][line width=1.5]   (1235.33, 161.33) circle [x radius= 10.06, y radius= 10.06]   ;
\draw (1229.36,158.77) node [anchor=north west][inner sep=0.75pt]  [font=\tiny,rotate=-359.69,xslant=0.02]  {$$};
\draw  [color={rgb, 255:red, 0; green, 0; blue, 0 }  ,draw opacity=1 ][line width=1.5]   (1235.33, 218.33) circle [x radius= 10.06, y radius= 10.06]   ;
\draw (1229.36,215.77) node [anchor=north west][inner sep=0.75pt]  [font=\tiny,rotate=-359.69,xslant=0.02]  {$$};
\draw (1229.33,74.73) node [anchor=north west][inner sep=0.75pt]    {$4$};
\draw (1229.33,129.73) node [anchor=north west][inner sep=0.75pt]    {$7$};
\draw  [color={rgb, 255:red, 0; green, 0; blue, 0 }  ,draw opacity=1 ][line width=1.5]   (1144.33, 164.33) circle [x radius= 10.06, y radius= 10.06]   ;
\draw (1138.36,161.77) node [anchor=north west][inner sep=0.75pt]  [font=\tiny,rotate=-359.69,xslant=0.02]  {$$};
\draw (1229.33,186.73) node [anchor=north west][inner sep=0.75pt]    {$8$};
\draw (1096.33,204.73) node [anchor=north west][inner sep=0.75pt]  [font=\Large,color={rgb, 255:red, 208; green, 2; blue, 27 }  ,opacity=1 ]  {$\bot $};
\draw (1244.33,205.73) node [anchor=north west][inner sep=0.75pt]  [font=\Large,color={rgb, 255:red, 208; green, 2; blue, 27 }  ,opacity=1 ]  {$\bot $};
\draw (1059.33,94.73) node [anchor=north west][inner sep=0.75pt]  [font=\large,color={rgb, 255:red, 208; green, 2; blue, 27 }  ,opacity=1 ]  {$( 1,2,3)$};
\draw (1058.33,150.73) node [anchor=north west][inner sep=0.75pt]  [font=\large,color={rgb, 255:red, 208; green, 2; blue, 27 }  ,opacity=1 ]  {$( 1,2,3)$};
\draw (1255.33,93.73) node [anchor=north west][inner sep=0.75pt]  [font=\large,color={rgb, 255:red, 208; green, 2; blue, 27 }  ,opacity=1 ]  {$( 4,7,8)$};
\draw (1256.33,148.73) node [anchor=north west][inner sep=0.75pt]  [font=\large,color={rgb, 255:red, 208; green, 2; blue, 27 }  ,opacity=1 ]  {$( 4,7,8)$};
\draw (100,352.07) node [anchor=north west][inner sep=0.75pt]    {$9$};
\draw (101,410.07) node [anchor=north west][inner sep=0.75pt]    {$10$};
\draw (102,463.07) node [anchor=north west][inner sep=0.75pt]    {$3$};
\draw  [color={rgb, 255:red, 0; green, 0; blue, 0 }  ,draw opacity=1 ][line width=1.5]   (106, 494.67) circle [x radius= 10.06, y radius= 10.06]   ;
\draw (100.03,492.1) node [anchor=north west][inner sep=0.75pt]  [font=\tiny,rotate=-359.69,xslant=0.02]  {$$};
\draw  [color={rgb, 255:red, 0; green, 0; blue, 0 }  ,draw opacity=1 ][line width=1.5]   (198, 382.67) circle [x radius= 10.06, y radius= 10.06]   ;
\draw (192.03,380.1) node [anchor=north west][inner sep=0.75pt]  [font=\tiny,rotate=-359.69,xslant=0.02]  {$$};
\draw  [color={rgb, 255:red, 0; green, 0; blue, 0 }  ,draw opacity=1 ][line width=1.5]   (105, 383.67) circle [x radius= 10.06, y radius= 10.06]   ;
\draw (99.03,381.1) node [anchor=north west][inner sep=0.75pt]  [font=\tiny,rotate=-359.69,xslant=0.02]  {$$};
\draw  [color={rgb, 255:red, 0; green, 0; blue, 0 }  ,draw opacity=1 ][line width=1.5]   (198, 437.67) circle [x radius= 10.06, y radius= 10.06]   ;
\draw (192.03,435.1) node [anchor=north west][inner sep=0.75pt]  [font=\tiny,rotate=-359.69,xslant=0.02]  {$$};
\draw  [color={rgb, 255:red, 0; green, 0; blue, 0 }  ,draw opacity=1 ][line width=1.5]   (198, 494.67) circle [x radius= 10.06, y radius= 10.06]   ;
\draw (192.03,492.1) node [anchor=north west][inner sep=0.75pt]  [font=\tiny,rotate=-359.69,xslant=0.02]  {$$};
\draw (192,351.07) node [anchor=north west][inner sep=0.75pt]    {$4$};
\draw (192,406.07) node [anchor=north west][inner sep=0.75pt]    {$7$};
\draw  [color={rgb, 255:red, 0; green, 0; blue, 0 }  ,draw opacity=1 ][line width=1.5]   (107, 440.67) circle [x radius= 10.06, y radius= 10.06]   ;
\draw (101.03,438.1) node [anchor=north west][inner sep=0.75pt]  [font=\tiny,rotate=-359.69,xslant=0.02]  {$$};
\draw (192,463.07) node [anchor=north west][inner sep=0.75pt]    {$8$};
\draw (59,481.07) node [anchor=north west][inner sep=0.75pt]  [font=\Large,color={rgb, 255:red, 208; green, 2; blue, 27 }  ,opacity=1 ]  {$\bot $};
\draw (207,482.07) node [anchor=north west][inner sep=0.75pt]  [font=\Large,color={rgb, 255:red, 208; green, 2; blue, 27 }  ,opacity=1 ]  {$\bot $};
\draw (15,372.07) node [anchor=north west][inner sep=0.75pt]  [font=\large,color={rgb, 255:red, 208; green, 2; blue, 27 }  ,opacity=1 ]  {$( 3,9,10)$};
\draw (17,427.07) node [anchor=north west][inner sep=0.75pt]  [font=\large,color={rgb, 255:red, 208; green, 2; blue, 27 }  ,opacity=1 ]  {$( 3,9,10)$};
\draw (218,370.07) node [anchor=north west][inner sep=0.75pt]  [font=\large,color={rgb, 255:red, 208; green, 2; blue, 27 }  ,opacity=1 ]  {$( 4,7,8)$};
\draw (219,425.07) node [anchor=north west][inner sep=0.75pt]  [font=\large,color={rgb, 255:red, 208; green, 2; blue, 27 }  ,opacity=1 ]  {$( 4,7,8)$};
\draw (451.33,348.73) node [anchor=north west][inner sep=0.75pt]    {$9$};
\draw (452.33,406.73) node [anchor=north west][inner sep=0.75pt]    {$10$};
\draw (453.33,459.73) node [anchor=north west][inner sep=0.75pt]    {$3$};
\draw  [color={rgb, 255:red, 0; green, 0; blue, 0 }  ,draw opacity=1 ][line width=1.5]   (457.33, 491.33) circle [x radius= 10.06, y radius= 10.06]   ;
\draw (451.36,488.77) node [anchor=north west][inner sep=0.75pt]  [font=\tiny,rotate=-359.69,xslant=0.02]  {$$};
\draw  [color={rgb, 255:red, 0; green, 0; blue, 0 }  ,draw opacity=1 ][line width=1.5]   (549.33, 379.33) circle [x radius= 10.06, y radius= 10.06]   ;
\draw (543.36,376.77) node [anchor=north west][inner sep=0.75pt]  [font=\tiny,rotate=-359.69,xslant=0.02]  {$$};
\draw  [color={rgb, 255:red, 0; green, 0; blue, 0 }  ,draw opacity=1 ][line width=1.5]   (456.33, 380.33) circle [x radius= 10.06, y radius= 10.06]   ;
\draw (450.36,377.77) node [anchor=north west][inner sep=0.75pt]  [font=\tiny,rotate=-359.69,xslant=0.02]  {$$};
\draw  [color={rgb, 255:red, 0; green, 0; blue, 0 }  ,draw opacity=1 ][line width=1.5]   (549.33, 434.33) circle [x radius= 10.06, y radius= 10.06]   ;
\draw (543.36,431.77) node [anchor=north west][inner sep=0.75pt]  [font=\tiny,rotate=-359.69,xslant=0.02]  {$$};
\draw  [color={rgb, 255:red, 0; green, 0; blue, 0 }  ,draw opacity=1 ][line width=1.5]   (549.33, 491.33) circle [x radius= 10.06, y radius= 10.06]   ;
\draw (543.36,488.77) node [anchor=north west][inner sep=0.75pt]  [font=\tiny,rotate=-359.69,xslant=0.02]  {$$};
\draw (543.33,347.73) node [anchor=north west][inner sep=0.75pt]    {$1$};
\draw (543.33,402.73) node [anchor=north west][inner sep=0.75pt]    {$2$};
\draw  [color={rgb, 255:red, 0; green, 0; blue, 0 }  ,draw opacity=1 ][line width=1.5]   (458.33, 437.33) circle [x radius= 10.06, y radius= 10.06]   ;
\draw (452.36,434.77) node [anchor=north west][inner sep=0.75pt]  [font=\tiny,rotate=-359.69,xslant=0.02]  {$$};
\draw (543.33,459.73) node [anchor=north west][inner sep=0.75pt]    {$8$};
\draw (410.33,477.73) node [anchor=north west][inner sep=0.75pt]  [font=\Large,color={rgb, 255:red, 208; green, 2; blue, 27 }  ,opacity=1 ]  {$\bot $};
\draw (558.33,478.73) node [anchor=north west][inner sep=0.75pt]  [font=\Large,color={rgb, 255:red, 208; green, 2; blue, 27 }  ,opacity=1 ]  {$\bot $};
\draw (366.33,368.73) node [anchor=north west][inner sep=0.75pt]  [font=\large,color={rgb, 255:red, 208; green, 2; blue, 27 }  ,opacity=1 ]  {$( 3,9,10)$};
\draw (368.33,423.73) node [anchor=north west][inner sep=0.75pt]  [font=\large,color={rgb, 255:red, 208; green, 2; blue, 27 }  ,opacity=1 ]  {$( 3,9,10)$};
\draw (569.33,366.73) node [anchor=north west][inner sep=0.75pt]  [font=\large,color={rgb, 255:red, 208; green, 2; blue, 27 }  ,opacity=1 ]  {$( 1,2,8)$};
\draw (570.33,421.73) node [anchor=north west][inner sep=0.75pt]  [font=\large,color={rgb, 255:red, 208; green, 2; blue, 27 }  ,opacity=1 ]  {$( 1,2,8)$};
\draw (806.67,345.4) node [anchor=north west][inner sep=0.75pt]    {$9$};
\draw (807.67,403.4) node [anchor=north west][inner sep=0.75pt]    {$10$};
\draw (808.67,456.4) node [anchor=north west][inner sep=0.75pt]    {$3$};
\draw  [color={rgb, 255:red, 0; green, 0; blue, 0 }  ,draw opacity=1 ][line width=1.5]   (812.67, 488) circle [x radius= 10.06, y radius= 10.06]   ;
\draw (806.69,485.43) node [anchor=north west][inner sep=0.75pt]  [font=\tiny,rotate=-359.69,xslant=0.02]  {$$};
\draw  [color={rgb, 255:red, 0; green, 0; blue, 0 }  ,draw opacity=1 ][line width=1.5]   (904.67, 376) circle [x radius= 10.06, y radius= 10.06]   ;
\draw (898.69,373.43) node [anchor=north west][inner sep=0.75pt]  [font=\tiny,rotate=-359.69,xslant=0.02]  {$$};
\draw  [color={rgb, 255:red, 0; green, 0; blue, 0 }  ,draw opacity=1 ][line width=1.5]   (811.67, 377) circle [x radius= 10.06, y radius= 10.06]   ;
\draw (805.69,374.43) node [anchor=north west][inner sep=0.75pt]  [font=\tiny,rotate=-359.69,xslant=0.02]  {$$};
\draw  [color={rgb, 255:red, 0; green, 0; blue, 0 }  ,draw opacity=1 ][line width=1.5]   (904.67, 431) circle [x radius= 10.06, y radius= 10.06]   ;
\draw (898.69,428.43) node [anchor=north west][inner sep=0.75pt]  [font=\tiny,rotate=-359.69,xslant=0.02]  {$$};
\draw  [color={rgb, 255:red, 0; green, 0; blue, 0 }  ,draw opacity=1 ][line width=1.5]   (904.67, 488) circle [x radius= 10.06, y radius= 10.06]   ;
\draw (898.69,485.43) node [anchor=north west][inner sep=0.75pt]  [font=\tiny,rotate=-359.69,xslant=0.02]  {$$};
\draw (898.67,344.4) node [anchor=north west][inner sep=0.75pt]    {$1$};
\draw (898.67,399.4) node [anchor=north west][inner sep=0.75pt]    {$2$};
\draw  [color={rgb, 255:red, 0; green, 0; blue, 0 }  ,draw opacity=1 ][line width=1.5]   (813.67, 434) circle [x radius= 10.06, y radius= 10.06]   ;
\draw (807.69,431.43) node [anchor=north west][inner sep=0.75pt]  [font=\tiny,rotate=-359.69,xslant=0.02]  {$$};
\draw (898.67,456.4) node [anchor=north west][inner sep=0.75pt]    {$8$};
\draw (764.67,362.4) node [anchor=north west][inner sep=0.75pt]  [font=\Large,color={rgb, 255:red, 208; green, 2; blue, 27 }  ,opacity=1 ]  {$\bot $};
\draw (913.67,475.4) node [anchor=north west][inner sep=0.75pt]  [font=\Large,color={rgb, 255:red, 208; green, 2; blue, 27 }  ,opacity=1 ]  {$\bot $};
\draw (720.67,421.4) node [anchor=north west][inner sep=0.75pt]  [font=\large,color={rgb, 255:red, 208; green, 2; blue, 27 }  ,opacity=1 ]  {$( 3,9,10)$};
\draw (722.67,475.4) node [anchor=north west][inner sep=0.75pt]  [font=\large,color={rgb, 255:red, 208; green, 2; blue, 27 }  ,opacity=1 ]  {$( 3,9,10)$};
\draw (924.67,363.4) node [anchor=north west][inner sep=0.75pt]  [font=\large,color={rgb, 255:red, 208; green, 2; blue, 27 }  ,opacity=1 ]  {$( 1,2,8)$};
\draw (925.67,418.4) node [anchor=north west][inner sep=0.75pt]  [font=\large,color={rgb, 255:red, 208; green, 2; blue, 27 }  ,opacity=1 ]  {$( 1,2,8)$};
\draw (1157.33,344.73) node [anchor=north west][inner sep=0.75pt]    {$9$};
\draw (1158.33,402.73) node [anchor=north west][inner sep=0.75pt]    {$10$};
\draw (1159.33,455.73) node [anchor=north west][inner sep=0.75pt]    {$3$};
\draw  [color={rgb, 255:red, 0; green, 0; blue, 0 }  ,draw opacity=1 ][line width=1.5]   (1163.33, 487.33) circle [x radius= 10.06, y radius= 10.06]   ;
\draw (1157.36,484.77) node [anchor=north west][inner sep=0.75pt]  [font=\tiny,rotate=-359.69,xslant=0.02]  {$$};
\draw  [color={rgb, 255:red, 0; green, 0; blue, 0 }  ,draw opacity=1 ][line width=1.5]   (1255.33, 375.33) circle [x radius= 10.06, y radius= 10.06]   ;
\draw (1249.36,372.77) node [anchor=north west][inner sep=0.75pt]  [font=\tiny,rotate=-359.69,xslant=0.02]  {$$};
\draw  [color={rgb, 255:red, 0; green, 0; blue, 0 }  ,draw opacity=1 ][line width=1.5]   (1162.33, 376.33) circle [x radius= 10.06, y radius= 10.06]   ;
\draw (1156.36,373.77) node [anchor=north west][inner sep=0.75pt]  [font=\tiny,rotate=-359.69,xslant=0.02]  {$$};
\draw  [color={rgb, 255:red, 0; green, 0; blue, 0 }  ,draw opacity=1 ][line width=1.5]   (1255.33, 430.33) circle [x radius= 10.06, y radius= 10.06]   ;
\draw (1249.36,427.77) node [anchor=north west][inner sep=0.75pt]  [font=\tiny,rotate=-359.69,xslant=0.02]  {$$};
\draw  [color={rgb, 255:red, 0; green, 0; blue, 0 }  ,draw opacity=1 ][line width=1.5]   (1255.33, 487.33) circle [x radius= 10.06, y radius= 10.06]   ;
\draw (1249.36,484.77) node [anchor=north west][inner sep=0.75pt]  [font=\tiny,rotate=-359.69,xslant=0.02]  {$$};
\draw (1249.33,343.73) node [anchor=north west][inner sep=0.75pt]    {$1$};
\draw (1249.33,398.73) node [anchor=north west][inner sep=0.75pt]    {$2$};
\draw  [color={rgb, 255:red, 0; green, 0; blue, 0 }  ,draw opacity=1 ][line width=1.5]   (1164.33, 433.33) circle [x radius= 10.06, y radius= 10.06]   ;
\draw (1158.36,430.77) node [anchor=north west][inner sep=0.75pt]  [font=\tiny,rotate=-359.69,xslant=0.02]  {$$};
\draw (1249.33,455.73) node [anchor=north west][inner sep=0.75pt]    {$8$};
\draw (1115.33,361.73) node [anchor=north west][inner sep=0.75pt]  [font=\Large,color={rgb, 255:red, 208; green, 2; blue, 27 }  ,opacity=1 ]  {$\bot $};
\draw (1265.33,363.73) node [anchor=north west][inner sep=0.75pt]  [font=\Large,color={rgb, 255:red, 208; green, 2; blue, 27 }  ,opacity=1 ]  {$\bot $};
\draw (1071.33,420.73) node [anchor=north west][inner sep=0.75pt]  [font=\large,color={rgb, 255:red, 208; green, 2; blue, 27 }  ,opacity=1 ]  {$( 3,9,10)$};
\draw (1073.33,474.73) node [anchor=north west][inner sep=0.75pt]  [font=\large,color={rgb, 255:red, 208; green, 2; blue, 27 }  ,opacity=1 ]  {$( 3,9,10)$};
\draw (1271.33,416.73) node [anchor=north west][inner sep=0.75pt]  [font=\large,color={rgb, 255:red, 208; green, 2; blue, 27 }  ,opacity=1 ]  {$( 1,2,8)$};
\draw (1269.33,475.73) node [anchor=north west][inner sep=0.75pt]  [font=\large,color={rgb, 255:red, 208; green, 2; blue, 27 }  ,opacity=1 ]  {$( 1,2,8)$};
\draw (91,624.73) node [anchor=north west][inner sep=0.75pt]    {$9$};
\draw (92,682.73) node [anchor=north west][inner sep=0.75pt]    {$4$};
\draw (93,735.73) node [anchor=north west][inner sep=0.75pt]    {$5$};
\draw  [color={rgb, 255:red, 0; green, 0; blue, 0 }  ,draw opacity=1 ][line width=1.5]   (97, 767.33) circle [x radius= 10.06, y radius= 10.06]   ;
\draw (91.03,764.77) node [anchor=north west][inner sep=0.75pt]  [font=\tiny,rotate=-359.69,xslant=0.02]  {$$};
\draw  [color={rgb, 255:red, 0; green, 0; blue, 0 }  ,draw opacity=1 ][line width=1.5]   (189, 655.33) circle [x radius= 10.06, y radius= 10.06]   ;
\draw (183.03,652.77) node [anchor=north west][inner sep=0.75pt]  [font=\tiny,rotate=-359.69,xslant=0.02]  {$$};
\draw  [color={rgb, 255:red, 0; green, 0; blue, 0 }  ,draw opacity=1 ][line width=1.5]   (96, 656.33) circle [x radius= 10.06, y radius= 10.06]   ;
\draw (90.03,653.77) node [anchor=north west][inner sep=0.75pt]  [font=\tiny,rotate=-359.69,xslant=0.02]  {$$};
\draw  [color={rgb, 255:red, 0; green, 0; blue, 0 }  ,draw opacity=1 ][line width=1.5]   (189, 710.33) circle [x radius= 10.06, y radius= 10.06]   ;
\draw (183.03,707.77) node [anchor=north west][inner sep=0.75pt]  [font=\tiny,rotate=-359.69,xslant=0.02]  {$$};
\draw  [color={rgb, 255:red, 0; green, 0; blue, 0 }  ,draw opacity=1 ][line width=1.5]   (189, 767.33) circle [x radius= 10.06, y radius= 10.06]   ;
\draw (183.03,764.77) node [anchor=north west][inner sep=0.75pt]  [font=\tiny,rotate=-359.69,xslant=0.02]  {$$};
\draw (183,623.73) node [anchor=north west][inner sep=0.75pt]    {$1$};
\draw (183,678.73) node [anchor=north west][inner sep=0.75pt]    {$2$};
\draw  [color={rgb, 255:red, 0; green, 0; blue, 0 }  ,draw opacity=1 ][line width=1.5]   (98, 713.33) circle [x radius= 10.06, y radius= 10.06]   ;
\draw (92.03,710.77) node [anchor=north west][inner sep=0.75pt]  [font=\tiny,rotate=-359.69,xslant=0.02]  {$$};
\draw (183,735.73) node [anchor=north west][inner sep=0.75pt]    {$8$};
\draw (49,641.73) node [anchor=north west][inner sep=0.75pt]  [font=\Large,color={rgb, 255:red, 208; green, 2; blue, 27 }  ,opacity=1 ]  {$\bot $};
\draw (199,643.73) node [anchor=north west][inner sep=0.75pt]  [font=\Large,color={rgb, 255:red, 208; green, 2; blue, 27 }  ,opacity=1 ]  {$\bot $};
\draw (5,700.73) node [anchor=north west][inner sep=0.75pt]  [font=\large,color={rgb, 255:red, 208; green, 2; blue, 27 }  ,opacity=1 ]  {$( 4,5,9)$};
\draw (7,754.73) node [anchor=north west][inner sep=0.75pt]  [font=\large,color={rgb, 255:red, 208; green, 2; blue, 27 }  ,opacity=1 ]  {$( 4,5,9)$};
\draw (205,696.73) node [anchor=north west][inner sep=0.75pt]  [font=\large,color={rgb, 255:red, 208; green, 2; blue, 27 }  ,opacity=1 ]  {$( 1,2,8)$};
\draw (203,755.73) node [anchor=north west][inner sep=0.75pt]  [font=\large,color={rgb, 255:red, 208; green, 2; blue, 27 }  ,opacity=1 ]  {$( 1,2,8)$};
\draw (444.33,636.02) node [anchor=north west][inner sep=0.75pt]    {$9$};
\draw (445.33,694.02) node [anchor=north west][inner sep=0.75pt]    {$4$};
\draw (446.33,747.02) node [anchor=north west][inner sep=0.75pt]    {$5$};
\draw  [color={rgb, 255:red, 0; green, 0; blue, 0 }  ,draw opacity=1 ][line width=1.5]   (450.33, 778.62) circle [x radius= 10.06, y radius= 10.06]   ;
\draw (444.36,776.06) node [anchor=north west][inner sep=0.75pt]  [font=\tiny,rotate=-359.69,xslant=0.02]  {$$};
\draw  [color={rgb, 255:red, 0; green, 0; blue, 0 }  ,draw opacity=1 ][line width=1.5]   (542.33, 666.62) circle [x radius= 10.06, y radius= 10.06]   ;
\draw (536.36,664.06) node [anchor=north west][inner sep=0.75pt]  [font=\tiny,rotate=-359.69,xslant=0.02]  {$$};
\draw  [color={rgb, 255:red, 0; green, 0; blue, 0 }  ,draw opacity=1 ][line width=1.5]   (449.33, 667.62) circle [x radius= 10.06, y radius= 10.06]   ;
\draw (443.36,665.06) node [anchor=north west][inner sep=0.75pt]  [font=\tiny,rotate=-359.69,xslant=0.02]  {$$};
\draw  [color={rgb, 255:red, 0; green, 0; blue, 0 }  ,draw opacity=1 ][line width=1.5]   (542.33, 721.62) circle [x radius= 10.06, y radius= 10.06]   ;
\draw (536.36,719.06) node [anchor=north west][inner sep=0.75pt]  [font=\tiny,rotate=-359.69,xslant=0.02]  {$$};
\draw  [color={rgb, 255:red, 0; green, 0; blue, 0 }  ,draw opacity=1 ][line width=1.5]   (542.33, 778.62) circle [x radius= 10.06, y radius= 10.06]   ;
\draw (536.36,776.06) node [anchor=north west][inner sep=0.75pt]  [font=\tiny,rotate=-359.69,xslant=0.02]  {$$};
\draw (536.33,635.02) node [anchor=north west][inner sep=0.75pt]    {$1$};
\draw (536.33,690.02) node [anchor=north west][inner sep=0.75pt]    {$2$};
\draw  [color={rgb, 255:red, 0; green, 0; blue, 0 }  ,draw opacity=1 ][line width=1.5]   (451.33, 724.62) circle [x radius= 10.06, y radius= 10.06]   ;
\draw (445.36,722.06) node [anchor=north west][inner sep=0.75pt]  [font=\tiny,rotate=-359.69,xslant=0.02]  {$$};
\draw (536.33,747.02) node [anchor=north west][inner sep=0.75pt]    {$3$};
\draw (402.33,653.02) node [anchor=north west][inner sep=0.75pt]  [font=\Large,color={rgb, 255:red, 208; green, 2; blue, 27 }  ,opacity=1 ]  {$\bot $};
\draw (552.33,655.02) node [anchor=north west][inner sep=0.75pt]  [font=\Large,color={rgb, 255:red, 208; green, 2; blue, 27 }  ,opacity=1 ]  {$\bot $};
\draw (358.33,712.02) node [anchor=north west][inner sep=0.75pt]  [font=\large,color={rgb, 255:red, 208; green, 2; blue, 27 }  ,opacity=1 ]  {$( 4,5,9)$};
\draw (360.33,766.02) node [anchor=north west][inner sep=0.75pt]  [font=\large,color={rgb, 255:red, 208; green, 2; blue, 27 }  ,opacity=1 ]  {$( 4,5,9)$};
\draw (558.33,708.02) node [anchor=north west][inner sep=0.75pt]  [font=\large,color={rgb, 255:red, 208; green, 2; blue, 27 }  ,opacity=1 ]  {$( 1,2,3)$};
\draw (556.33,767.02) node [anchor=north west][inner sep=0.75pt]  [font=\large,color={rgb, 255:red, 208; green, 2; blue, 27 }  ,opacity=1 ]  {$( 1,2,3)$};
\draw (803.67,636.02) node [anchor=north west][inner sep=0.75pt]    {$9$};
\draw (804.67,694.02) node [anchor=north west][inner sep=0.75pt]    {$4$};
\draw (805.67,747.02) node [anchor=north west][inner sep=0.75pt]    {$5$};
\draw  [color={rgb, 255:red, 0; green, 0; blue, 0 }  ,draw opacity=1 ][line width=1.5]   (809.67, 778.62) circle [x radius= 10.06, y radius= 10.06]   ;
\draw (803.69,776.06) node [anchor=north west][inner sep=0.75pt]  [font=\tiny,rotate=-359.69,xslant=0.02]  {$$};
\draw  [color={rgb, 255:red, 0; green, 0; blue, 0 }  ,draw opacity=1 ][line width=1.5]   (901.67, 666.62) circle [x radius= 10.06, y radius= 10.06]   ;
\draw (895.69,664.06) node [anchor=north west][inner sep=0.75pt]  [font=\tiny,rotate=-359.69,xslant=0.02]  {$$};
\draw  [color={rgb, 255:red, 0; green, 0; blue, 0 }  ,draw opacity=1 ][line width=1.5]   (808.67, 667.62) circle [x radius= 10.06, y radius= 10.06]   ;
\draw (802.69,665.06) node [anchor=north west][inner sep=0.75pt]  [font=\tiny,rotate=-359.69,xslant=0.02]  {$$};
\draw  [color={rgb, 255:red, 0; green, 0; blue, 0 }  ,draw opacity=1 ][line width=1.5]   (901.67, 721.62) circle [x radius= 10.06, y radius= 10.06]   ;
\draw (895.69,719.06) node [anchor=north west][inner sep=0.75pt]  [font=\tiny,rotate=-359.69,xslant=0.02]  {$$};
\draw  [color={rgb, 255:red, 0; green, 0; blue, 0 }  ,draw opacity=1 ][line width=1.5]   (901.67, 778.62) circle [x radius= 10.06, y radius= 10.06]   ;
\draw (895.69,776.06) node [anchor=north west][inner sep=0.75pt]  [font=\tiny,rotate=-359.69,xslant=0.02]  {$$};
\draw (895.67,635.02) node [anchor=north west][inner sep=0.75pt]    {$1$};
\draw (895.67,690.02) node [anchor=north west][inner sep=0.75pt]    {$2$};
\draw  [color={rgb, 255:red, 0; green, 0; blue, 0 }  ,draw opacity=1 ][line width=1.5]   (810.67, 724.62) circle [x radius= 10.06, y radius= 10.06]   ;
\draw (804.69,722.06) node [anchor=north west][inner sep=0.75pt]  [font=\tiny,rotate=-359.69,xslant=0.02]  {$$};
\draw (895.67,747.02) node [anchor=north west][inner sep=0.75pt]    {$3$};
\draw (765.67,764.02) node [anchor=north west][inner sep=0.75pt]  [font=\Large,color={rgb, 255:red, 208; green, 2; blue, 27 }  ,opacity=1 ]  {$\bot $};
\draw (911.67,655.02) node [anchor=north west][inner sep=0.75pt]  [font=\Large,color={rgb, 255:red, 208; green, 2; blue, 27 }  ,opacity=1 ]  {$\bot $};
\draw (726.67,655.02) node [anchor=north west][inner sep=0.75pt]  [font=\large,color={rgb, 255:red, 208; green, 2; blue, 27 }  ,opacity=1 ]  {$( 4,5,9)$};
\draw (729.67,711.02) node [anchor=north west][inner sep=0.75pt]  [font=\large,color={rgb, 255:red, 208; green, 2; blue, 27 }  ,opacity=1 ]  {$( 4,5,9)$};
\draw (917.67,708.02) node [anchor=north west][inner sep=0.75pt]  [font=\large,color={rgb, 255:red, 208; green, 2; blue, 27 }  ,opacity=1 ]  {$( 1,2,3)$};
\draw (915.67,767.02) node [anchor=north west][inner sep=0.75pt]  [font=\large,color={rgb, 255:red, 208; green, 2; blue, 27 }  ,opacity=1 ]  {$( 1,2,3)$};
\draw (1156.67,635.02) node [anchor=north west][inner sep=0.75pt]    {$6$};
\draw (1157.67,693.02) node [anchor=north west][inner sep=0.75pt]    {$4$};
\draw (1158.67,746.02) node [anchor=north west][inner sep=0.75pt]    {$5$};
\draw  [color={rgb, 255:red, 0; green, 0; blue, 0 }  ,draw opacity=1 ][line width=1.5]   (1162.67, 777.62) circle [x radius= 10.06, y radius= 10.06]   ;
\draw (1156.69,775.06) node [anchor=north west][inner sep=0.75pt]  [font=\tiny,rotate=-359.69,xslant=0.02]  {$$};
\draw  [color={rgb, 255:red, 0; green, 0; blue, 0 }  ,draw opacity=1 ][line width=1.5]   (1254.67, 665.62) circle [x radius= 10.06, y radius= 10.06]   ;
\draw (1248.69,663.06) node [anchor=north west][inner sep=0.75pt]  [font=\tiny,rotate=-359.69,xslant=0.02]  {$$};
\draw  [color={rgb, 255:red, 0; green, 0; blue, 0 }  ,draw opacity=1 ][line width=1.5]   (1161.67, 666.62) circle [x radius= 10.06, y radius= 10.06]   ;
\draw (1155.69,664.06) node [anchor=north west][inner sep=0.75pt]  [font=\tiny,rotate=-359.69,xslant=0.02]  {$$};
\draw  [color={rgb, 255:red, 0; green, 0; blue, 0 }  ,draw opacity=1 ][line width=1.5]   (1254.67, 720.62) circle [x radius= 10.06, y radius= 10.06]   ;
\draw (1248.69,718.06) node [anchor=north west][inner sep=0.75pt]  [font=\tiny,rotate=-359.69,xslant=0.02]  {$$};
\draw  [color={rgb, 255:red, 0; green, 0; blue, 0 }  ,draw opacity=1 ][line width=1.5]   (1254.67, 777.62) circle [x radius= 10.06, y radius= 10.06]   ;
\draw (1248.69,775.06) node [anchor=north west][inner sep=0.75pt]  [font=\tiny,rotate=-359.69,xslant=0.02]  {$$};
\draw (1248.67,634.02) node [anchor=north west][inner sep=0.75pt]    {$1$};
\draw (1248.67,689.02) node [anchor=north west][inner sep=0.75pt]    {$2$};
\draw  [color={rgb, 255:red, 0; green, 0; blue, 0 }  ,draw opacity=1 ][line width=1.5]   (1163.67, 723.62) circle [x radius= 10.06, y radius= 10.06]   ;
\draw (1157.69,721.06) node [anchor=north west][inner sep=0.75pt]  [font=\tiny,rotate=-359.69,xslant=0.02]  {$$};
\draw (1248.67,746.02) node [anchor=north west][inner sep=0.75pt]    {$3$};
\draw (1118.67,763.02) node [anchor=north west][inner sep=0.75pt]  [font=\Large,color={rgb, 255:red, 208; green, 2; blue, 27 }  ,opacity=1 ]  {$\bot $};
\draw (1264.67,654.02) node [anchor=north west][inner sep=0.75pt]  [font=\Large,color={rgb, 255:red, 208; green, 2; blue, 27 }  ,opacity=1 ]  {$\bot $};
\draw (1079.67,654.02) node [anchor=north west][inner sep=0.75pt]  [font=\large,color={rgb, 255:red, 208; green, 2; blue, 27 }  ,opacity=1 ]  {$( 4,5,6)$};
\draw (1082.67,710.02) node [anchor=north west][inner sep=0.75pt]  [font=\large,color={rgb, 255:red, 208; green, 2; blue, 27 }  ,opacity=1 ]  {$( 4,5,6)$};
\draw (1270.67,707.02) node [anchor=north west][inner sep=0.75pt]  [font=\large,color={rgb, 255:red, 208; green, 2; blue, 27 }  ,opacity=1 ]  {$( 1,2,3)$};
\draw (1268.67,766.02) node [anchor=north west][inner sep=0.75pt]  [font=\large,color={rgb, 255:red, 208; green, 2; blue, 27 }  ,opacity=1 ]  {$( 1,2,3)$};
\draw (135.33,6.67) node [anchor=north west][inner sep=0.75pt]  [font=\huge] [align=left] {$\displaystyle G_{1}$};
\draw (479.33,9.33) node [anchor=north west][inner sep=0.75pt]  [font=\huge] [align=left] {$\displaystyle G_{2}$};
\draw (827.33,11) node [anchor=north west][inner sep=0.75pt]  [font=\huge] [align=left] {$\displaystyle G_{3}$};
\draw (1175.33,13.67) node [anchor=north west][inner sep=0.75pt]  [font=\huge] [align=left] {$\displaystyle G_{4}$};
\draw (140.67,291) node [anchor=north west][inner sep=0.75pt]  [font=\huge] [align=left] {$\displaystyle G_{5}$};
\draw (484.67,293.67) node [anchor=north west][inner sep=0.75pt]  [font=\huge] [align=left] {$\displaystyle G_{6}$};
\draw (832.67,295.33) node [anchor=north west][inner sep=0.75pt]  [font=\huge] [align=left] {$\displaystyle G_{7}$};
\draw (1178,298) node [anchor=north west][inner sep=0.75pt]  [font=\huge] [align=left] {$\displaystyle G_{8}$};
\draw (136.67,575) node [anchor=north west][inner sep=0.75pt]  [font=\huge] [align=left] {$\displaystyle G_{9}$};
\draw (480.67,577.67) node [anchor=north west][inner sep=0.75pt]  [font=\huge] [align=left] {$\displaystyle G_{10}$};
\draw (828.67,579.33) node [anchor=north west][inner sep=0.75pt]  [font=\huge] [align=left] {$\displaystyle G_{11}$};
\draw (1176.67,582) node [anchor=north west][inner sep=0.75pt]  [font=\huge] [align=left] {$\displaystyle G_{12}$};
\draw    (110.06,96.56) -- (182.94,95.77) ;
\draw    (108.71,101.72) -- (184.29,145.61) ;
\draw    (106.46,104.38) -- (186.54,199.95) ;
\draw    (108.38,198.88) -- (186.62,103.45) ;
\draw    (112.06,206.78) -- (182.94,207.56) ;
\draw    (112.06,153.34) -- (182.94,151) ;
\draw    (110.66,158.8) -- (184.34,202.53) ;
\draw    (110.57,201.39) -- (184.43,155.94) ;
\draw    (110.49,148.26) -- (184.51,101.08) ;
\draw    (460.06,99.56) -- (532.94,98.77) ;
\draw    (458.71,104.72) -- (534.29,148.61) ;
\draw    (456.46,107.38) -- (536.54,202.95) ;
\draw    (457.39,202.89) -- (536.61,106.44) ;
\draw    (462.06,156.34) -- (532.94,154) ;
\draw    (460.66,161.8) -- (534.34,205.53) ;
\draw    (459.56,205.37) -- (534.44,158.97) ;
\draw    (460.49,151.26) -- (534.51,104.08) ;
\draw    (808.4,104.23) -- (881.27,103.44) ;
\draw    (807.04,109.39) -- (882.63,153.28) ;
\draw    (804.8,112.05) -- (884.87,207.62) ;
\draw    (805.72,207.56) -- (884.94,111.11) ;
\draw    (810.39,161) -- (881.27,158.66) ;
\draw    (808.99,166.47) -- (882.68,210.2) ;
\draw    (807.89,210.03) -- (882.78,163.63) ;
\draw    (808.82,155.92) -- (882.84,108.74) ;
\draw    (1152.4,107.23) -- (1225.27,106.44) ;
\draw    (1151.04,112.39) -- (1226.63,156.28) ;
\draw    (1148.8,115.05) -- (1228.87,210.62) ;
\draw    (1149.72,210.56) -- (1228.94,114.11) ;
\draw    (1154.39,164) -- (1225.27,161.66) ;
\draw    (1152.99,169.47) -- (1226.68,213.2) ;
\draw    (1151.89,213.03) -- (1226.78,166.63) ;
\draw    (1152.82,158.92) -- (1226.84,111.74) ;
\draw    (115.06,383.56) -- (187.94,382.77) ;
\draw    (113.71,388.72) -- (189.29,432.61) ;
\draw    (111.46,391.38) -- (191.54,486.95) ;
\draw    (112.39,486.89) -- (191.61,390.44) ;
\draw    (117.06,440.34) -- (187.94,438) ;
\draw    (115.66,445.8) -- (189.34,489.53) ;
\draw    (114.56,489.37) -- (189.44,442.97) ;
\draw    (115.49,435.26) -- (189.51,388.08) ;
\draw    (466.4,380.23) -- (539.27,379.44) ;
\draw    (465.04,385.39) -- (540.63,429.28) ;
\draw    (462.8,388.05) -- (542.87,483.62) ;
\draw    (463.72,483.56) -- (542.94,387.11) ;
\draw    (468.39,437) -- (539.27,434.66) ;
\draw    (466.99,442.47) -- (540.68,486.2) ;
\draw    (465.89,486.03) -- (540.78,439.63) ;
\draw    (466.82,431.92) -- (540.84,384.74) ;
\draw    (821.73,376.89) -- (894.6,376.11) ;
\draw    (820.37,382.05) -- (895.96,425.95) ;
\draw    (818.13,384.72) -- (898.2,480.28) ;
\draw    (819.06,480.22) -- (898.28,383.78) ;
\draw    (823.73,433.67) -- (894.61,431.33) ;
\draw    (822.32,439.14) -- (896.01,482.86) ;
\draw    (821.22,482.7) -- (896.11,436.3) ;
\draw    (822.16,428.59) -- (896.18,381.41) ;
\draw    (1172.4,376.23) -- (1245.27,375.44) ;
\draw    (1171.04,381.39) -- (1246.63,425.28) ;
\draw    (1168.8,384.05) -- (1248.87,479.62) ;
\draw    (1169.72,479.56) -- (1248.94,383.11) ;
\draw    (1174.39,433) -- (1245.27,430.66) ;
\draw    (1172.99,438.47) -- (1246.68,482.2) ;
\draw    (1171.89,482.03) -- (1246.78,435.63) ;
\draw    (1172.82,427.92) -- (1246.84,380.74) ;
\draw    (106.06,656.23) -- (178.94,655.44) ;
\draw    (104.71,661.39) -- (180.29,705.28) ;
\draw    (102.46,664.05) -- (182.54,759.62) ;
\draw    (103.39,759.56) -- (182.61,663.11) ;
\draw    (108.06,713) -- (178.94,710.66) ;
\draw    (106.66,718.47) -- (180.34,762.2) ;
\draw    (105.56,762.03) -- (180.44,715.63) ;
\draw    (106.49,707.92) -- (180.51,660.74) ;
\draw    (459.4,667.52) -- (532.27,666.73) ;
\draw    (458.04,672.68) -- (533.63,716.57) ;
\draw    (455.8,675.34) -- (535.87,770.91) ;
\draw    (456.72,770.85) -- (535.94,674.4) ;
\draw    (461.39,724.29) -- (532.27,721.96) ;
\draw    (459.99,729.76) -- (533.68,773.49) ;
\draw    (458.89,773.32) -- (533.78,726.93) ;
\draw    (459.82,719.21) -- (533.84,672.03) ;
\draw    (818.73,667.52) -- (891.6,666.73) ;
\draw    (817.37,672.68) -- (892.96,716.57) ;
\draw    (815.13,675.34) -- (895.2,770.91) ;
\draw    (816.06,770.85) -- (895.28,674.4) ;
\draw    (820.73,724.29) -- (891.61,721.96) ;
\draw    (819.32,729.76) -- (893.01,773.49) ;
\draw    (818.22,773.32) -- (893.11,726.93) ;
\draw    (819.16,719.21) -- (893.18,672.03) ;
\draw    (1171.73,666.52) -- (1244.6,665.73) ;
\draw    (1170.37,671.68) -- (1245.96,715.57) ;
\draw    (1168.13,674.34) -- (1248.2,769.91) ;
\draw    (1169.06,769.85) -- (1248.28,673.4) ;
\draw    (1173.73,723.29) -- (1244.61,720.96) ;
\draw    (1172.32,728.76) -- (1246.01,772.49) ;
\draw    (1171.22,772.32) -- (1246.11,725.93) ;
\draw    (1172.16,718.21) -- (1246.18,671.03) ;
\draw    (819.73,778.62) -- (891.6,778.62) ;
\draw    (1172.73,777.62) -- (1244.6,777.62) ;
\draw    (460.4,778.62) -- (532.27,778.62) ;
\draw    (107.06,767.33) -- (178.94,767.33) ;
\draw    (1173.4,487.33) -- (1245.27,487.33) ;
\draw    (822.73,488) -- (894.6,488) ;
\draw    (467.4,491.33) -- (539.27,491.33) ;
\draw    (116.06,494.67) -- (187.94,494.67) ;
\draw    (1153.4,218.33) -- (1225.27,218.33) ;
\draw    (809.4,215.33) -- (881.27,215.33) ;
\draw    (461.06,210.67) -- (532.94,210.67) ;

\end{tikzpicture}